\documentclass[longauth]{aa} % for the long lists of affiliations 
\usepackage{graphicx}
\usepackage{txfonts}
\usepackage[]{hyperref}
\usepackage{breakurl}
\newcommand{\urlwofont}[1]{\urlstyle{same}\url{#1}}

\usepackage{lscape}

\def\CaII{Ca\,{\sc ii}}
\def\CI{C\,{\sc i}}

\def\FeII{Fe\,{\sc ii}}
\def\HI{H\,{\sc i}}
\def\HeI{He\,{\sc i}}

\def\MgI{Mg\,{\sc i}}
\def\NaID{Na\,{\sc i}\,D}

\def\OI{O\,{\sc i}}

\def\SiII{Si\,{\sc ii}}
\def\ScII{Sc\,{\sc ii}}
\def\BaII{Ba\,{\sc ii}}

\begin{document}

   \title{Core-collapse supernova subtypes in luminous infrared galaxies}

   \titlerunning{CCSNe in LIRGs}
   
   \author{E.~Kankare\inst{1}
          \and
          A.~Efstathiou\inst{2}
          \and
          R.~Kotak\inst{1}
          \and
          E.~C.~Kool\inst{3}
          \and
          T.~Kangas\inst{4}
          \and          
          D.~O'Neill\inst{5}
          \and
          S.~Mattila\inst{1}
          \and
          P.~V\"ais\"anen\inst{6,7}
          \and
          R.~Ramphul\inst{6,8}
          \and
          M.~Mogotsi\inst{6}
          \and
          S.~D.~Ryder\inst{9,10}
          \and
          S.~Parker\inst{11,12}
          \and
          T.~Reynolds\inst{1}
          \and
          M.~Fraser\inst{13}
          \and
          A.~Pastorello\inst{14}
          \and
          E.~Cappellaro\inst{14}
          \and
          P.~A.~Mazzali\inst{15,16}
          \and
          P.~Ochner\inst{17,14}
          \and
          L.~Tomasella\inst{14}
          \and
          M.~Turatto\inst{14}
          \and          
          J.~Kotilainen\inst{18,1}
          \and
          H.~Kuncarayakti\inst{1,18}
          \and
          M.~A.~P\'erez-Torres\inst{19,20}
          \and
          Z.~Randriamanakoto\inst{6}
          \and
          C.~Romero-Ca\~nizales\inst{21}
          \and
          M.~Berton\inst{18,22}
          \and
          R.~Cartier\inst{23}
          \and
          T.-W. Chen\inst{3}
          \and
          L.~Galbany\inst{24}
          \and
          M.~Gromadzki\inst{25}
          \and
          C.~Inserra\inst{26}
          \and
          K.~Maguire\inst{27}
          \and
          S.~Moran\inst{1}
          \and
          T.~E.~M\"uller-Bravo\inst{28}
          \and
          M.~Nicholl\inst{29,30}
          \and
          A.~Reguitti\inst{13,31,32}
          \and
          D.~R.~Young\inst{5}}

          \institute{Tuorla Observatory, Department of Physics and Astronomy, University of Turku, FI-20014 Turku, Finland\\
          \email{erkki.kankare@utu.fi}
          \and          
          School of Sciences, European University Cyprus, Diogenes Street, Engomi, 1516 Nicosia, Cyprus
          \and
          The Oskar Klein Centre, Department of Astronomy, Stockholm University, AlbaNova, SE-10691 Stockholm, Sweden
          \and
          Space Telescope Science Institute, 3700 San Martin Drive, Baltimore, MD 21218, USA
          \and
          Astrophysics Research Centre, School of Mathematics and Physics, Queen's University Belfast, Belfast BT7 1NN, UK
          \and
          South African Astronomical Observatory, P.O. Box 9, Observatory 7935, Cape Town, South Africa
          \and
          Southern African Large Telescope, P.O. Box 9, Observatory 7935, Cape Town, South Africa
          \and
          Department of Astronomy, University of Cape Town, Private Bag X3, Rondebosch 7701, South Africa
          \and
          Department of Physics and Astronomy, Macquarie University, NSW 2109, Australia
          \and
          Macquarie University Research Centre for Astronomy, Astrophysics \& Astrophotonics, Sydney, NSW 2109, Australia
          \and
          Parkdale Observatory, 225 Warren Road, RDl Oxford, Canterbury 7495, New Zealand
          \and
          Backyard Observatory Supernova Search (BOSS)
          \and
          School of Physics, O'Brien Centre for Science North, University College Dublin, Belfield, Dublin 4, Ireland
          \and
          INAF -- Osservatorio Astronomico, Vicolo Osservatorio 5, I-35122 Padova, Italy
          \and
          Astrophysics Research Institute, Liverpool John Moores University, IC2, Liverpool Science Park, 146 Brownlow Hill, Liverpool L3 5RF, UK
          \and
          Max-Planck-Institut f\"ur Astrophysik, Karl-Schwarzschild-Str. 1, D-85748 Garching, Germany         
          \and
          Dipartimento di Fisica e Astronomia, Universit\'a di Padova, Vicolo Osservatorio 2, I-35122 Padova, Italy
          \and
          Finnish Centre for Astronomy with ESO (FINCA), University of Turku, Vesilinnantie 5, FI-20014 Turku, Finland
          \and
          Instituto de Astrof\'isica de Andaluc\'ia, Glorieta de las Astronom\'ia, s/n, E-18008 Granada, Spain
          \and
          Departamento de F\'isica Teorica, Facultad de Ciencias, Universidad de Zaragoza, Spain
          \and
          Institute of Astronomy and Astrophysics, Academia Sinica, 11F of Astronomy-Mathematics Building, AS/NTU No. 1, Sec. 4, Roosevelt Rd, Taipei 10617, Taiwan, R.O.C
          \and
          Aalto University Mets\"ahovi Radio Observatory, Mets\"ahovintie 114, FI-02540 Kylm\"al\"a, Finland
          \and
          Cerro Tololo Inter-American Observatory, NSF's National Optical-Infrared Astronomy Research Laboratory, Casilla 603, La Serena, Chile
          \and
          Departamento de F\'isica Te\'orica y del Cosmos, Universidad de Granada, E-18071 Granada, Spain
          \and
          Astronomical Observatory, University of Warsaw, Al. Ujazdowskie 4, 00-478 Warszawa, Poland
          \and
          School of Physics \& Astronomy, Cardiff University, Queens Buildings, The Parade, Cardiff, CF24 3AA, UK
          \and
          School of Physics, Trinity College Dublin, The University of Dublin, Dublin 2, Ireland
          \and
          School of Physics and Astronomy, University of Southampton, Southampton, Hampshire, SO17 1BJ, UK
          \and
          Birmingham Institute for Gravitational Wave Astronomy and School of Physics and Astronomy, University of Birmingham, Birmingham B15 2TT, UK
          \and
          Institute for Astronomy, University of Edinburgh, Royal Observatory, Blackford Hill, EH9 3HJ, UK
          \and
          Departamento de Ciencias F\'isicas, Universidad Andr\'es Bello, Avda. Rep\'ublica 252, Santiago, Chile
          \and
          Millennium Institute of Astrophysics, Nuncio Monsenor S\'otero Sanz 100, Providencia, Santiago, Chile
          }

%   \date{Received September 15, 1996; accepted March 16, 1997}

% \abstract{}{}{}{}{} 
% 5 {} token are mandatory
  
\abstract{The fraction of core-collapse supernovae (CCSNe) occurring in the central regions of galaxies is not well-constrained at present. This is partly because large-scale transient surveys operate at optical wavelengths, making it challenging to detect transient sources that occur in regions susceptible to high extinction factors. Here we present the discovery and follow-up observations of two CCSNe that occurred in the luminous infrared galaxy (LIRG), NGC~3256. The first, SN~2018ec, was discovered using the ESO HAWK-I/GRAAL adaptive optics seeing enhancer, and was classified as a Type Ic with a host galaxy extinction of $A_{V} = 2.1^{+0.3}_{-0.1}$ mag. The second, AT 2018cux, was discovered during the course of follow-up observations of SN~2018ec, and is consistent with a sub-luminous Type IIP classification with an $A_{V} = 2.1 \pm 0.4$ mag of host extinction. A third CCSN, PSN~J10275082-4354034 in NGC~3256, has previously been reported in 2014, and we recovered the source in late time archival \textit{Hubble Space Telescope} imaging. Based on template light-curve fitting, we favour a Type IIn classification for it with modest host galaxy extinction of $A_{V} = 0.3^{+0.4}_{-0.3}$ mag. We also extend our study with follow-up data of the recent Type IIb SN~2019lqo and Type Ib SN~2020fkb that occurred in the LIRG system Arp 299 with host extinctions of $A_{V} = 2.1^{+0.1}_{-0.3}$ and $A_{V} = 0.4^{+0.1}_{-0.2}$ mag, respectively. Motivated by the above, we inspected, for the first time, a sample of 29 CCSNe located within a projected distance of 2.5\,kpc from the host galaxy nuclei in a sample of 16 LIRGs. We find that, if star formation within these galaxies is modelled assuming a global starburst episode and normal IMF, there is evidence of a correlation between the starburst age and the CCSN subtype. We infer that the two subgroups of 14 H-poor (Type IIb/Ib/Ic/Ibn) and 15 H-rich (Type II/IIn) CCSNe have different underlying progenitor age distributions, with the H-poor progenitors being younger at 3\,$\sigma$ significance. However, we do note that the available sample sizes of CCSNe and host LIRGs are so far small, and the statistical comparisons between subgroups do not take into account possible systematic or model errors related to the estimated starburst ages.}

   
   \maketitle  
   
\section{Introduction}

The local ($\leq$12 Mpc) core-collapse supernova (CCSN) population already clearly shows that optical transient survey programmes are not discovering all SNe in normal spiral galaxies due to host galaxy extinction \citep{mattila12}, consistent also with statistical studies of galaxy disk opacities \citep[e.g.][]{kankare09}. Furthermore, recent surveys operating at longer wavelengths have also discovered SNe missed by optical transient searches in nearby star-forming galaxies \citep[e.g., SPIRITS -- the SPitzer InfraRed Intensive Transients Survey;][]{jencson17,jencson18,jencson19}. The effect of a missing population of CCSNe becomes even more prominent in luminous $(10^{11} L_{\odot} < L_{\mathrm{IR}} < 10^{12} L_{\odot})$ and ultraluminous $(L_{\mathrm{IR}} > 10^{12} L_{\odot})$ infrared galaxies (LIRGs and ULIRGs, respectively) which are highly obscured and star-forming galaxies; it is also very common that these objects are closely paired or merging galaxies \citep{sanders88}. Radio observations have shown the existence of a rich population of radio SNe (and SN remnants) in ULIRGs \citep[e.g.][]{parra07,varenius19}, LIRGs \citep[e.g.][]{perez-torres09,ulvestad09}, and starburst galaxies \citep[e.g.][]{fenech08,mattila13}, which have remained undetected by optical surveys. In particular, LIRGs are relatively rare in the local Universe; however, they dominate the cosmic star formation history and the resulting CCSN rate beyond $z \sim 1$ \citep[e.g.][]{magnelli11}, both which peak at $z \sim 2$ \citep{madau14}. 

The Supernovae UNmasked By Infra-Red Detection (SUNBIRD) collaboration \citep{mattila07,kankare08,kankare12,kool18,kool19} has carried out systematic survey programmes to discover and study CCSNe in LIRGs using adaptive optics (AO) instruments at the Very Large Telescope (VLT), Gemini and Keck facilities. The AO observations carried out in high resolution and in the near-IR are an ideal combination to discover obscured CCSNe in crowded and dusty environments of these galaxies. For example, the AO discovery of SN 2008cs, with a derived host galaxy line-of-sight extinction of $A_{V} \approx 16$ mag, shows that there is a population of very highly reddened CCSNe in these galaxies \citep{kankare08}. Furthermore, SNe with very small projected nuclear distances but low host galaxy extinctions can also be missed by regular ground-based observations due to the high background luminosity of the LIRG host \citep[e.g. SN 2010cu;][]{kankare12}. Therefore, it is not surprising that optical transient surveys miss a significant fraction of CCSNe in LIRGs even in the local Universe \citep[e.g.][]{mattila12,kool18}. 

Previously, various statistical studies have been carried out on the properties of CCSNe in normal galaxies. Statistical studies have shown that Type Ibc SNe are spatially more closely associated to H$\alpha$ regions than Type IIP SNe, and that within Type Ibc SNe, Type Ic events show the closest H$\alpha$ association \citep[e.g.][]{anderson12}. This has been interpreted as Type Ibc SN progenitors having shorter stellar life times than Type IIP SNe, which is expected from single stellar evolution models. Similar results have also been found e.g. by \citet{crowther13}, \citet{aramyan16}, and \citet{audcent-ross20}. Furthermore, \citet{kangas17} corroborated these previous results with their statistical study combining spatial distribution of massive stars to those of different CCSN classes in local galaxies. Other statistical studies on the explosion site have also suggested that Type Ic SNe occur in more metal rich environments and have more massive progenitors than Type Ib SNe \citep{leloudas11,galbany18,kuncarayakti18}. 

\citet{anderson13} suggested that normal galaxies tend to produce similar SN types if they host multiple SNe, and speculated that this is connected to the episodic nature of the starburst event and the resulting age range distribution of possible SN progenitor stars. LIRGs with their very high SN rates (SNR) offer laboratories to investigate statistical SN characteristics, with the advantage that the recent and more strongly episodic star formation of the galaxy can be characterised more accurately. Only very recently CCSN discoveries in central regions of LIRGs made a statistically significant sample \citep{kool18}. In the sections that follow, we report discoveries of recent CCSNe in LIRGs NGC 3256 and Arp 299, and discuss the starburst age in high CCSN rate LIRGs with a connection to CCSN subtypes in these galaxies.

\section{NGC 3256}

NGC 3256 is an ongoing LIRG merger at a redshift of $z=0.009354$ \citep{wong06} with $L_{\mathrm{IR}} = 10^{11.61} L_{\odot}$ \citep{sanders03} scaled to a Tully-Fisher (TF) luminosity distance of $D_{l} = 37.4$ Mpc \citep{tully88}. The adopted TF distance is consistent with the redshift based distance of $38.6 \pm 2.7$ Mpc ($H_{0}$ = 70 km s$^{-1}$ Mpc$^{-1}$, $\Omega_{M} = 0.3$, $\Omega_{\Lambda} = 0.7$) corrected for the influence of Virgo Cluster and the Great Attractor infall \citep{mould00}. The IR luminosity of NGC 3256 suggests a rate of $\sim$1.1 CCSN yr$^{-1}$ based on the empirical relation of \citet{mattila01}. There is also some evidence of an obscured AGN in the system \citep{kotilainen96,emonts14,ohyama15}. While the expected intrinsic SN rate of NGC 3256 is very high, only one spectroscopically confirmed SN has been previously reported in this nearby LIRG. This is the Type II SN 2001db, discovered using near-IR observations and found to have a significant line-of-sight extinction of $A_{V} \approx 5.5$ mag \citep{maiolino02}. The Galactic extinction towards the LIRG is $A_{V} = 0.334$ mag \citep{schlafly11}.

\subsection{SN 2018ec}

The High-Acuity Wide field \textit{K}-band Imager \citep[HAWK-I;][]{kissler08} on the 8.2-m VLT UT4 consists of four HAWAII 2RG 2048$\times$2048 pix near-IR arrays with a total field of view of 7.5\arcmin $\times$ 7.5\arcmin\ and a pixel scale of 0\farcs106/pix. We observed in the \textit{Ks}-band two LIRGs, i.e. NGC 3256 and IRAS 08355-4944, in the European Southern Observatory (ESO) Science Verification (SV) run of HAWK-I with the GRound Layer Adaptive optics Assisted by Lasers module \citep[GRAAL;][]{paufique10}. The Tip-Tilt Star - free mode SV observations of NGC 3256 were carried out on 2018 January 3.4 UT (FWHM $\sim$ 0\farcs4) and 2018 January 6.4 UT (FWHM $\sim$ 0\farcs3). Previously, it has not been possible to target NGC 3256 with full AO facilities due to the lack of a suitable natural guide star ($m \lesssim 17$ mag) in the field.

The HAWK-I/GRAAL data were reduced in a standard manner for near-IR images, using {\sc iraf}\footnote{{\sc iraf} is distributed by the National Optical Astronomy Observatories, which are operated by the Association of Universities for Research in Astronomy, Inc., under cooperative agreement with the National Science Foundation.} based tasks. We discovered a new transient source (SN 2018ec) in NGC 3256 by using an archival ESO New Technology Telescope (NTT) Son of ISAAC \citep[SOFI;][]{moorwood98} \textit{Ks}-band image from 2003 January 23.2 UT as a reference. We reported the discovery in \citet{kankare18a}, and spectroscopically classified it as a reddened Type Ic \citep{berton18} via the extended Public ESO Spectroscopic Survey for Transient Objects \citep[ePESSTO;][]{smartt15}. The discovery image is shown in Fig.~\ref{fig:18ec_field}. To our knowledge, no optical large-scale sky survey has reported a detection of SN 2018ec, however, this might be in part due to the less-monitored southern declination of NGC 3256. 

{\begin{figure*}[!t]
\includegraphics[width=\linewidth]{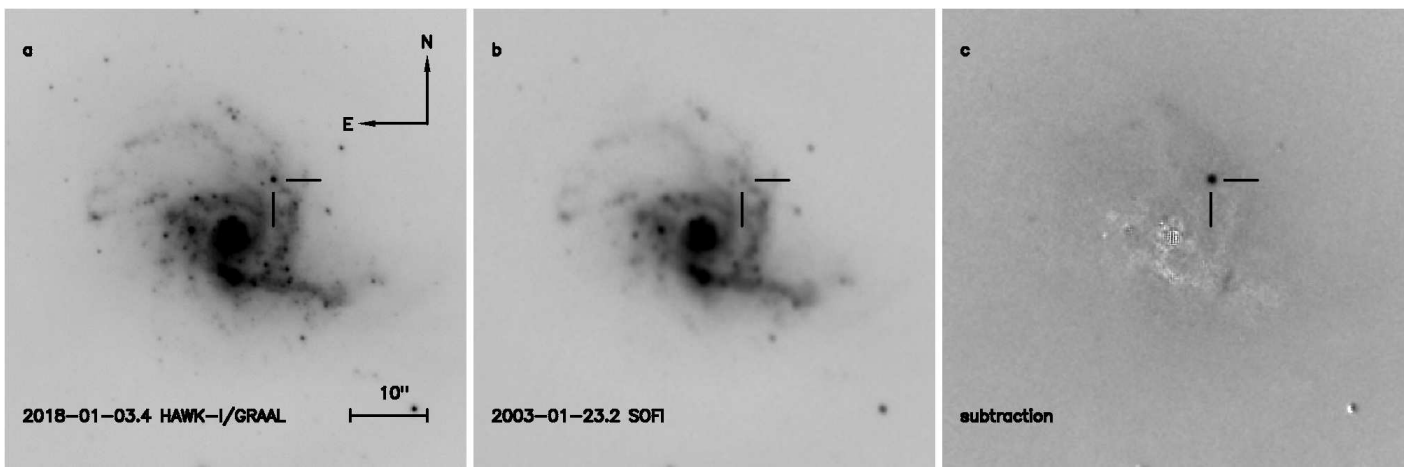}
\caption{a) 1\arcmin\ $\times$ 1\arcmin\ subsection of the HAWK-I/GRAAL \textit{Ks}-band discovery image of SN 2018ec in NGC 3256, b) SOFI reference image, and c) a subtraction between the images. The AO enhanced HAWK-I/GRAAL image with FWHM $\sim$ 0\farcs4 has a superior image quality compared to the normal ground-based seeing SOFI image with FWHM $\sim$ 0\farcs8. The location of SN 2018ec is shown with tick marks in all the panels; the orientation and image scale are indicated in the first panel.}
\label{fig:18ec_field}
\end{figure*}

Astrometry of the HAWK-I/GRAAL image was derived using 17 stars selected from the Two Micron All-Sky Survey (2MASS), yielding RA = 10$^{\mathrm{h}}$27$^{\mathrm{m}}$50$\fs$77 and Dec = $-43\degr$54$\arcmin$06\farcs3 (equinox J2000.0) for the location of SN 2018ec. This is 5\farcs2 W and 7\farcs7 N from the \textit{Ks}-band nucleus of NGC 3256, and corresponds to a projected distance of 1.7 kpc. Typical for LIRGs, NGC 3256 hosts a large population of super star clusters \citep[see e.g.][]{trancho07}, and such structures can very likely contribute to the underlying emission seen at the location of SN 2018ec and other CCSNe in NGC 3256 in the pre-explosion reference images. Spectrophotometric follow-up data of SN 2018ec were obtained with NTT as a part of ePESSTO using the ESO Faint Object Spectrograph and Camera 2 \citep[EFOSC2;][]{buzzoni84} and SOFI. The data were reduced in a standard manner using the {\sc pessto} pipeline \citep{smartt15} based on standard {\sc iraf} tasks. 

All images were template subtracted using the {\sc isis2.2} package \citep{alard98,alard00}. We used \textit{H} and \textit{Ks}-band reference images from FLAMINGOS-2 at Gemini South observed on 2017 March 26.3 and 2017 March 12.1 UT, respectively, while the \textit{J}-band reference image was taken with SOFI at NTT on 2001 April 9.1 UT. The \textit{g}, \textit{r}, \textit{i}, \textit{z}-band archival reference images were observed with Dark Energy Camera \citep[DECam;][]{flaugher15} at the 4-m Blanco Telescope on 2017 February 20.3, 2017 February 21.3, 2017 March 18.2, and 2017 February 9.3 UT, respectively. The \textit{JHKs} images were calibrated using 2MASS magnitudes of stars in the field of SN 2018ec. The \textit{gri} photometry was calibrated using the AAVSO (American Association of Variable Star Observers) Photometric All-Sky Survey magnitudes for stars in the large field of DECam images which were then used to yield magnitudes for 15 sequence stars close to SN 2018ec; these are shown in Fig.~\ref{fig:field} and the magnitudes reported in Table~\ref{table:field18ec}. The \textit{z}-band sequence stars were calibrated using standard star field observations carried out with NTT. The point spread function (PSF) photometry of the SN was carried out using the {\sc quba} pipeline \citep{valenti11}. The photometry is reported in Table~\ref{table:phot18ec} with the \textit{griz} magnitudes in the AB system and \textit{JHK} magnitudes in the Vega system.

\cite{ryder18} reported radio observations of the field of SN 2018ec with the Australia Telescope Compact Array (ATCA) on 2018 January 23.8 UT. The observations yielded 3\,$\sigma$ upper limits at the location of SN 2018ec of $<$10 mJy/beam and $<$3.6 mJy/beam at 5.5 and 9.0 GHz, respectively. This corresponds to $< 6.0 \times 10^{27}$ erg s$^{-1}$ Hz$^{-1}$ at 9 GHz; the limit is above the peak of normal Type Ic SNe around $10^{26}$ to $10^{27}$ erg s$^{-1}$ Hz$^{-1}$ \citep[e.g.][and references therein]{romero-canizales14}. The limits are not particularly constraining, due to the high background radio emission of the host galaxy. Further radio limits were reported by \citet{nayana18} based on observations with the Giant Metrewave Radio Telescope (GMRT) at 1.39 GHz on 2018 January 20.9 UT, yielding a 3\,$\sigma$ upper limit of 2.1 mJy ($< 3.5 \times 10^{27}$ erg s$^{-1}$ Hz$^{-1}$) at the SN 2018ec location. 

SN 2018ec is found to be spectroscopically a normal Type Ic SN, and in our analysis we find similarity to e.g. SN 2007gr \citep{hunter09}, see below. The line-of-sight host galaxy extinction of SN 2018ec was estimated using a $\chi^{2}$ fit of broad-band light curves of SN 2018ec with those of SN 2007gr (including an extrapolation of the optical late-time light curves of SN 2007gr based on the data beyond +70 d using linear fits). The fit was carried out simultaneously in all the bands with follow-up data of SN 2018ec. The method is the same as that used e.g. in \citet{kankare14a}. The free parameters of the fit are the host galaxy line-of-sight extinction $A_{V}$, the discovery epoch $t_{0}$ relative to a suitable reference (the estimated explosion date or light-curve maximum), and fixed constant shift $C$ applied to all the bands representing the intrinsic differences in the brightness of SNe (and any systematic difference in the distance estimates of the two SNe). The Johnson-Cousins \textit{UBVRI} light curves of SN 2007gr \citep{hunter09} were converted into the \textit{griz} system with the transformations of \citet{jester05}. We adopted the well-established \citet{cardelli89} extinction law for the fitting. The errors of the data points in the analysed light curves are considered to be Gaussian. If the light curves are well sampled and the comparison SN is well suited for the comparison, the probability density functions of the fitted parameters follow approximately Gaussian distributions. The largest deviations from this typically occur within the error of $t_{0}$ if the light-curve follow up was initiated post-maximum. The reported errors corresponds to 68.3\,\% confidence intervals estimated based on the probability density functions; systematic errors related e.g. to the uncertainty of the host extinction of the comparison SN are not included in the reported values. Based on the aforementioned comparison with SN 2007gr we conclude that SN 2018ec was discovered $19^{+13}_{-3}$ days after optical maximum, has a host galaxy extinction of $A_{V} = 2.1^{+0.3}_{-0.1}$ mag, and is $C = 0.6^{+0.4}_{-0.2}$ mag brighter than SN 2007gr. The absolute magnitude light curves are shown in Fig.~\ref{fig:18ec_lc}. Based on the fit we estimate that SN 2018ec peaked at $M_{r} \approx -18$ mag; this is at the brighter end of the peak magnitude distribution of normal Type Ic SNe but not unusual \citep[e.g.][]{taddia18}. 

\begin{figure}[!t]
\includegraphics[width=\linewidth]{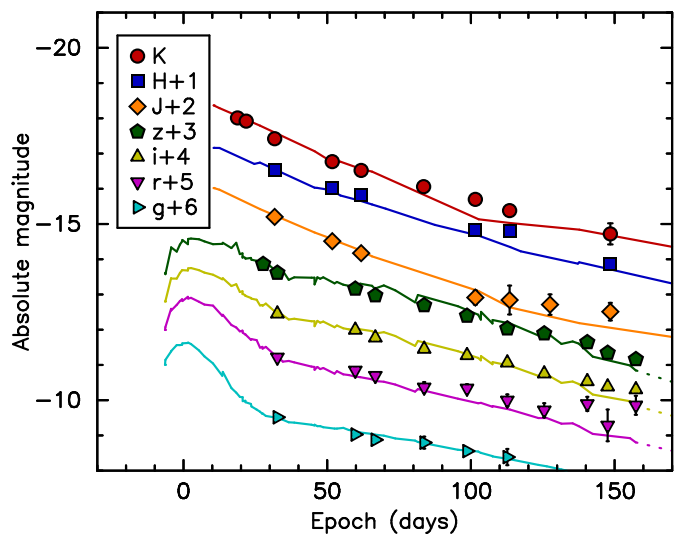}
\caption{Absolute magnitude light curves of SN 2018ec after correcting for host extinction ($A_{V} = 2.1$ mag). The light curves of the Type Ic SN 2007gr are shown with solid lines and shifted vertically by $-0.58$ mag (optical light curves extrapolated with linear fits for comparison are indicated with dotted curves). The estimated errors for most light curve points are smaller than the symbol size. The epoch 0 is the \textit{B}-band maximum estimated for SN 2007gr by \citet{hunter09}.}
\label{fig:18ec_lc}
\end{figure}

The spectroscopic sequence of SN 2018ec is shown in Fig.~\ref{fig:18ec_spect}, illustrating also the similarity to the normal Type Ic SN 2007gr \citep{hunter09}. Following a more detailed classification scheme by \citet{prentice17}, SN 2018ec appears to be a Ic-6/7. The most prominent SN feature is the \CaII\ near-IR triplet. The typically strong \OI\ $\lambda$7772 line in Type Ib/c SNe is blended with the telluric A-band feature. No clear \HeI\ features arising from the SN are evident, supporting a Type Ic classification. Actually there are broad P Cygni profiles close to the rest wavelengths of \HeI\ $\lambda$5876 and $\lambda$10830 lines, but these features can also be associated with \NaID\ and \CI\ $\lambda\lambda$10683,10691, respectively. These lines have an absorption minimum at $\sim$8500 km s$^{-1}$ around +33 d, which declines to $\sim$7000 km s$^{-1}$ within $\sim$1 month, see Fig.~\ref{fig:18ec_velocity}. As minor differences compared to SN 2007gr, SN 2018ec does not show early \CI\ $\lambda\lambda$11330,11753 features, or \SiII\ $\lambda$6355 absorption feature around +32 d. Furthermore, the \MgI\ $\lambda$15687 emission feature is not visible in the +62 d spectrum of SN 2018ec. 

\begin{figure*}[!t]
\includegraphics[width=\linewidth]{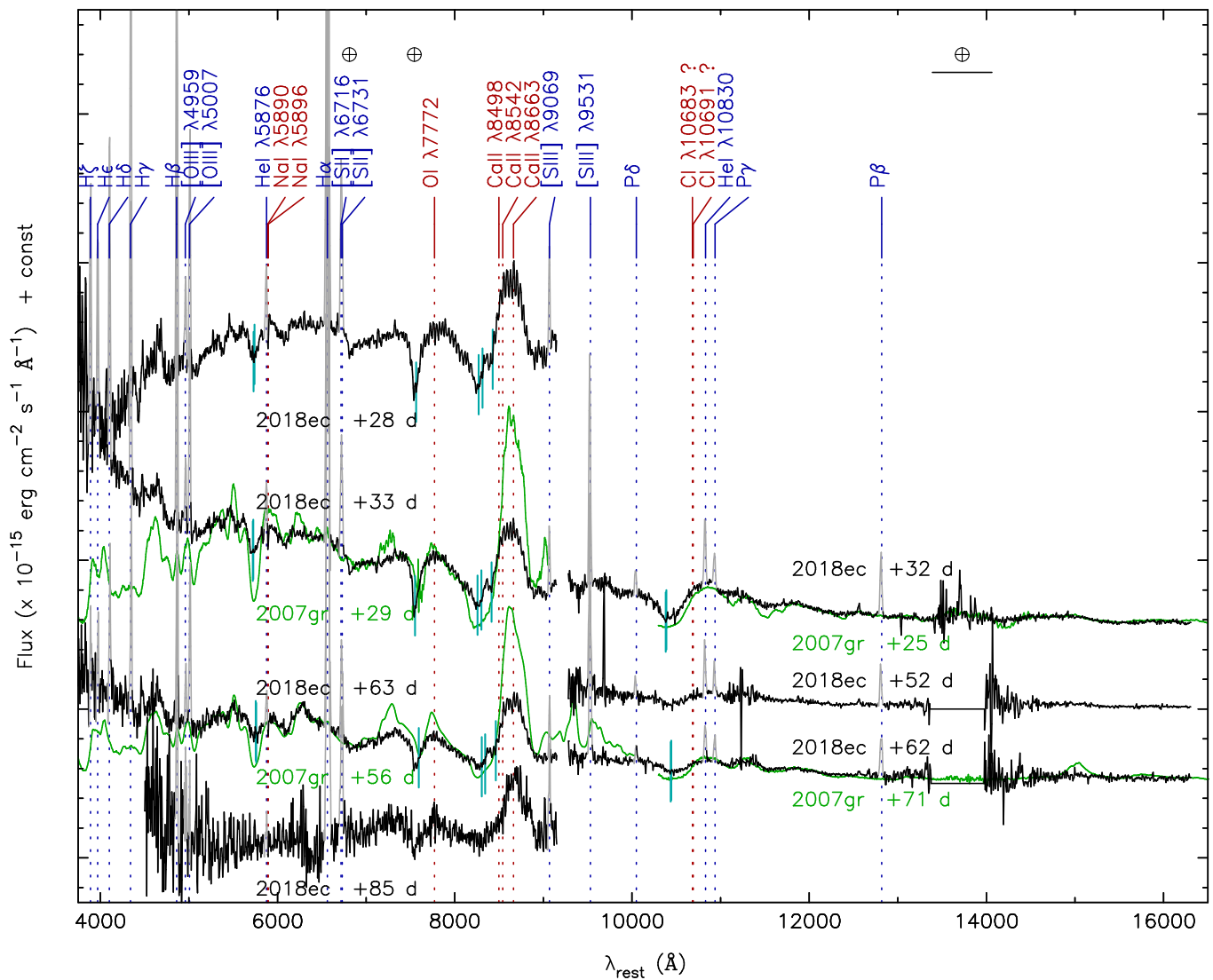}
\caption{Spectral time series of SN 2018ec. The spectra have been redshift corrected to rest frame. The spectra have been corrected for both Galactic $A_{V} = 0.334$ mag and estimated host galaxy extinction of $A_{V} = 2.1$ mag. The epochs are provided respective to the estimated light curve maximum. The main narrow-line emission features arising from an incomplete host galaxy subtraction are marked in grey. The most prominent identified spectral features have been labelled arising either from the SN (red) or from the host galaxy (blue). The Doppler shifted position of the SN lines are indicated with cyan vertical lines in selected epochs as suggested by the velocity of the Na I absorption minimum. The main telluric bands have been marked with a $\oplus$ symbol. For comparison selected spectra of normal Type Ic SN 2007gr are overlaid (green). The continua of the optical spectra of SN 2018ec are likely to be contaminated by the complex host background that was not successfully completely subtracted. Therefore, the optical SN features of SN 2018ec above and below the continuum level appear also less prominent compared to those of SN 2007gr. The spectra have been shifted vertically for clarity.}
\label{fig:18ec_spect}
\end{figure*}

\begin{figure}[!t]
\includegraphics[width=\linewidth]{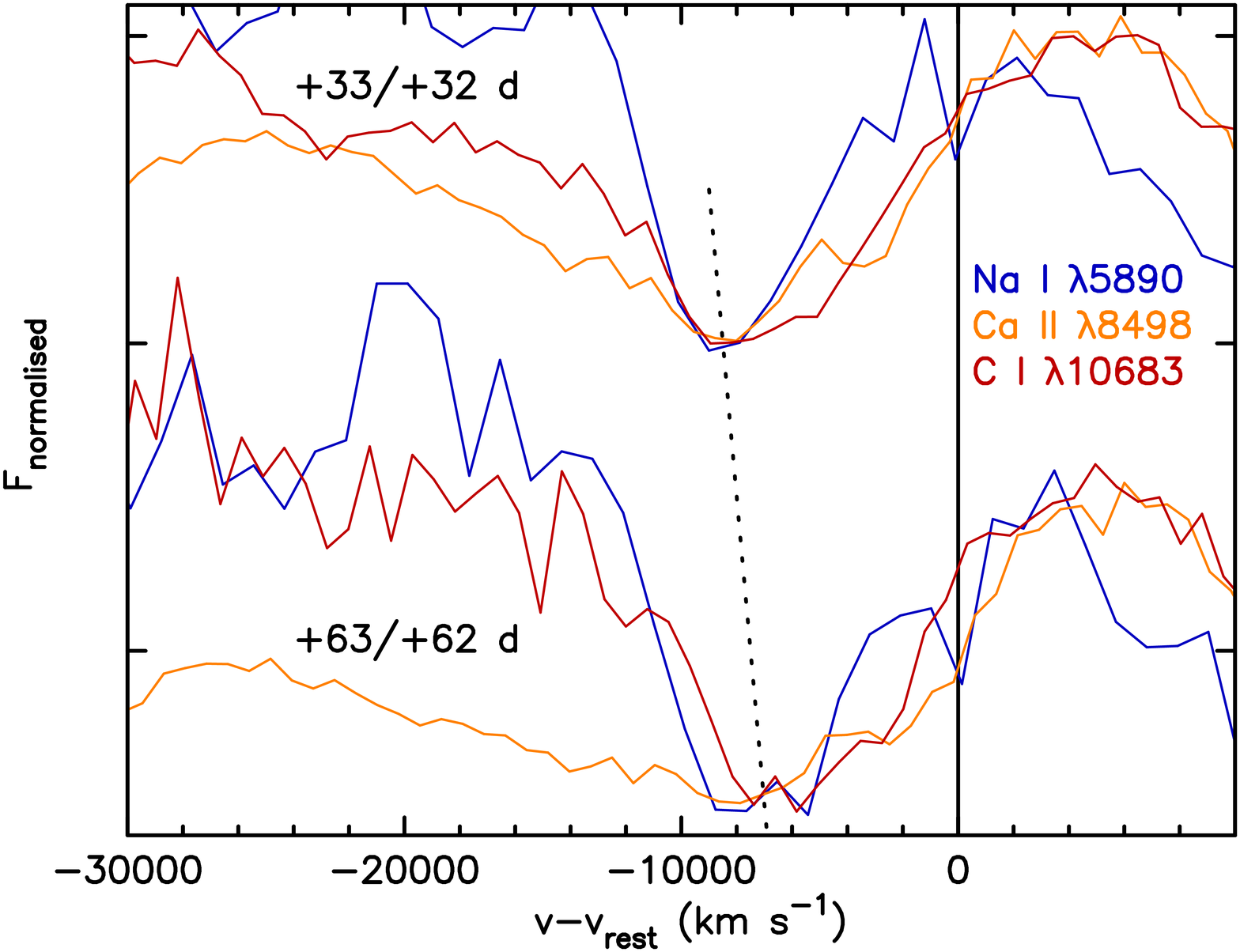}
\caption{Prominent P Cygni features of SN 2018ec at two selected epochs. The zero velocity is set to the rest wavelength of the shortest wavelength component of each line blend, i.e. at 5890 \AA\ for the \NaID\ (blue), 8498 \AA\ for the \CaII\ near-IR triplet (orange), and 10683 \AA\ for a \CI\ blend (red). Velocities at absorption minima shown by the features are surprisingly similar with $\sim$8500 km s$^{-1}$ around $\sim$1 month from the estimated maximum light and $\sim$7000 km s$^{-1}$ at $\sim$2 months from light curve peak, with the latter epoch showing somewhat more scatter, which can (partly) arise from the lower signal-to-noise of the spectra. The spectra have been binned by a factor of 4 for clarity and the line profiles are normalised to match the flux level of both their absorption minimum and emission peak within a given epoch.}
\label{fig:18ec_velocity}
\end{figure}

\subsection{AT 2018cux}

During the ePESSTO follow up of SN 2018ec the EFOSC2 data led to a serendipitous discovery of another transient AT 2018cux in NGC 3256 using images obtained on 2018 March 24.2 UT in comparison to images from 2018 January 17.2 UT as a reference. The discovery was reported in \cite{kankare18b} and the \textit{i}-band discovery image is shown in Fig.~\ref{fig:18cux_field}. We yielded coordinates RA = 10$^{\mathrm{h}}$27$^{\mathrm{m}}$51$\fs$41 and Dec = $-43\degr$54$\arcmin$18\farcs0 (equinox J2000.0) for AT 2018cux from the EFOSC2 images processed by the {\sc pessto} pipeline. This corresponds to 1.7" E and 4.0" S of the host galaxy main \textit{Ks}-band nucleus. This translates to a projected distance of 0.8 kpc. The location of AT 2018cux is relatively close, but not coincident, to the southern nucleus of NGC 3256, located at RA = 10$^{\mathrm{h}}$27$^{\mathrm{m}}$51$\fs$22 and Dec = $-43\degr$54$\arcmin$19\farcs2 (equinox J2000.0) based on radio observations \citep{neff03}. However, this southern nucleus is heavily obscured at optical and near-IR wavelengths by dust and thus the association of AT 2018cux observed in optical wavelengths to this component is uncertain. The PSF photometry of AT 2018cux was carried out similar to that of SN 2018ec and is listed in Table~\ref{table:phot18cux}. Late-time long-slit spectroscopy of AT 2018cux was obtained on 2018 July 3.7 UT with the Southern African Large Telescope \citep[SALT;][]{buckley06} using the Robert Stobie Spectrograph \citep[RSS;][]{burgh03,kobulnicky03}. The spectrum was reduced in a standard manner with basic {\sc iraf} tasks.

{\begin{figure*}[!t]
\includegraphics[width=\linewidth]{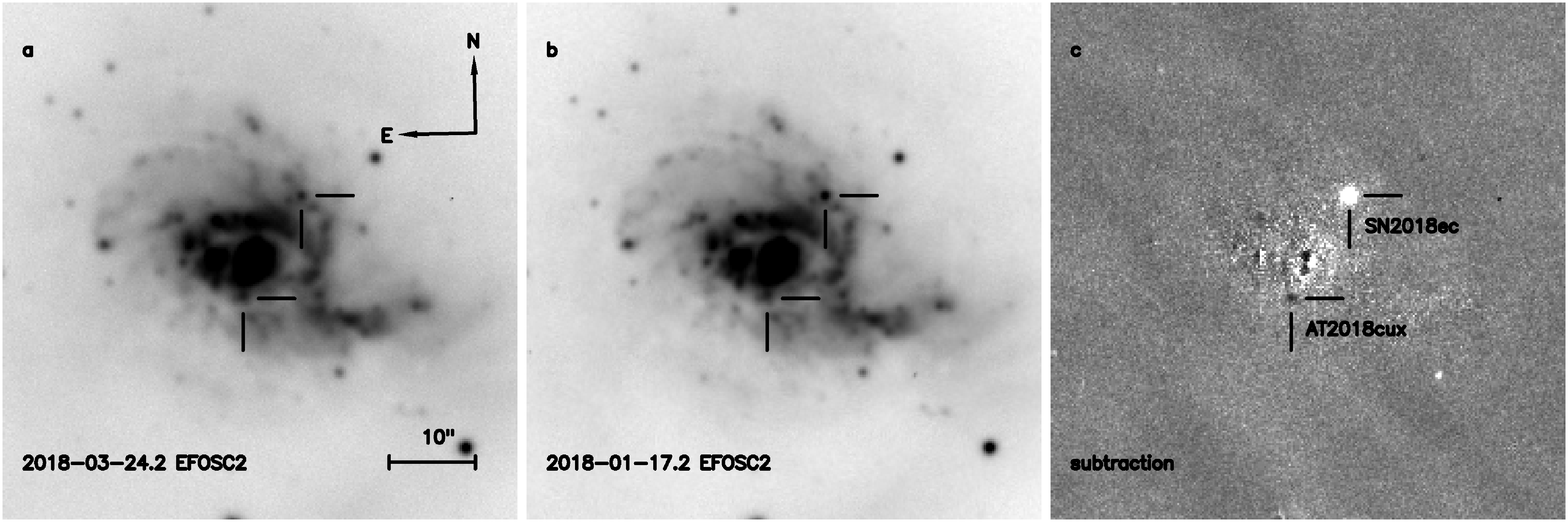}
\caption{a) 1\arcmin\ $\times$ 1\arcmin\ subsection of the EFOSC2 \textit{i}-band discovery image (FWHM $\sim$ 0\farcs9) of AT 2018cux in NGC 3256, b) pre-discovery EFOSC2 reference image (FWHM $\sim$ 0\farcs8), and c) a subtraction between the images. The locations of SN 2018ec and AT 2018cux are shown with tick marks in all the panels. Since the reference image contains increased flux from SN 2018ec compared to that of the AT 2018cux discovery image, the colour scale of the two events is reverse in the subtraction. The image scale and orientation are indicated in the left panel.}
\label{fig:18cux_field}
\end{figure*}

We used a similar approach as previously described for SN 2018ec to estimate the host galaxy extinction for AT 2018cux. The transient shows flat and plateau-like light curves, with colours that are only somewhat reddened in combination with relatively faint magnitudes, which suggest a subluminous Type IIP SN \citep[e.g.][]{spiro14,muller-bravo20}. This is also supported by our one low signal-to-noise spectrum of the event, see the paragraph below. Therefore, SN 2005cs \citep{pastorello09} was adopted as a canonical reference example of a normal subluminous Type IIP SN, with the optical light curves transferred into the \textit{griz} system with the \citet{jester05} conversions. Based on the comparison fit, AT 2018cux was found to be most consistent with a host galaxy extinction of $A_{V} = 2.1 \pm 0.4$ mag, a discovery $6^{+7}_{-3}$ d after the explosion, and a $0.6^{+0.4}_{-0.3}$ mag fainter plateau, see Fig.~\ref{fig:18cux_lc}. With the estimated line-of-sight extinction, at around +50 d from the explosion, the plateau magnitudes reach $M \approx -14.5$ mag in \textit{riz} bands. However, the faintest subluminous Type IIP events like SNe 1999br and 2001dc have plateau magnitudes of $M_{V} \approx -13.6$ and $-14.3$ mag, respectively \citep{pastorello04}. Similarly, the absolute magnitudes of AT 2018cux are generally brighter than those of faint transients that are interpreted as SN impostors, i.e. non-terminal outbursts of massive stars \citep[see e.g.][]{smith11}. AT 2018cux is also brighter compared to transients such as SN 2008S that have been associated both with a SN impostor or an electron-capture SN origin \citep{prieto08,botticella09,smith09}.

\begin{figure}[!t]
\includegraphics[width=\linewidth]{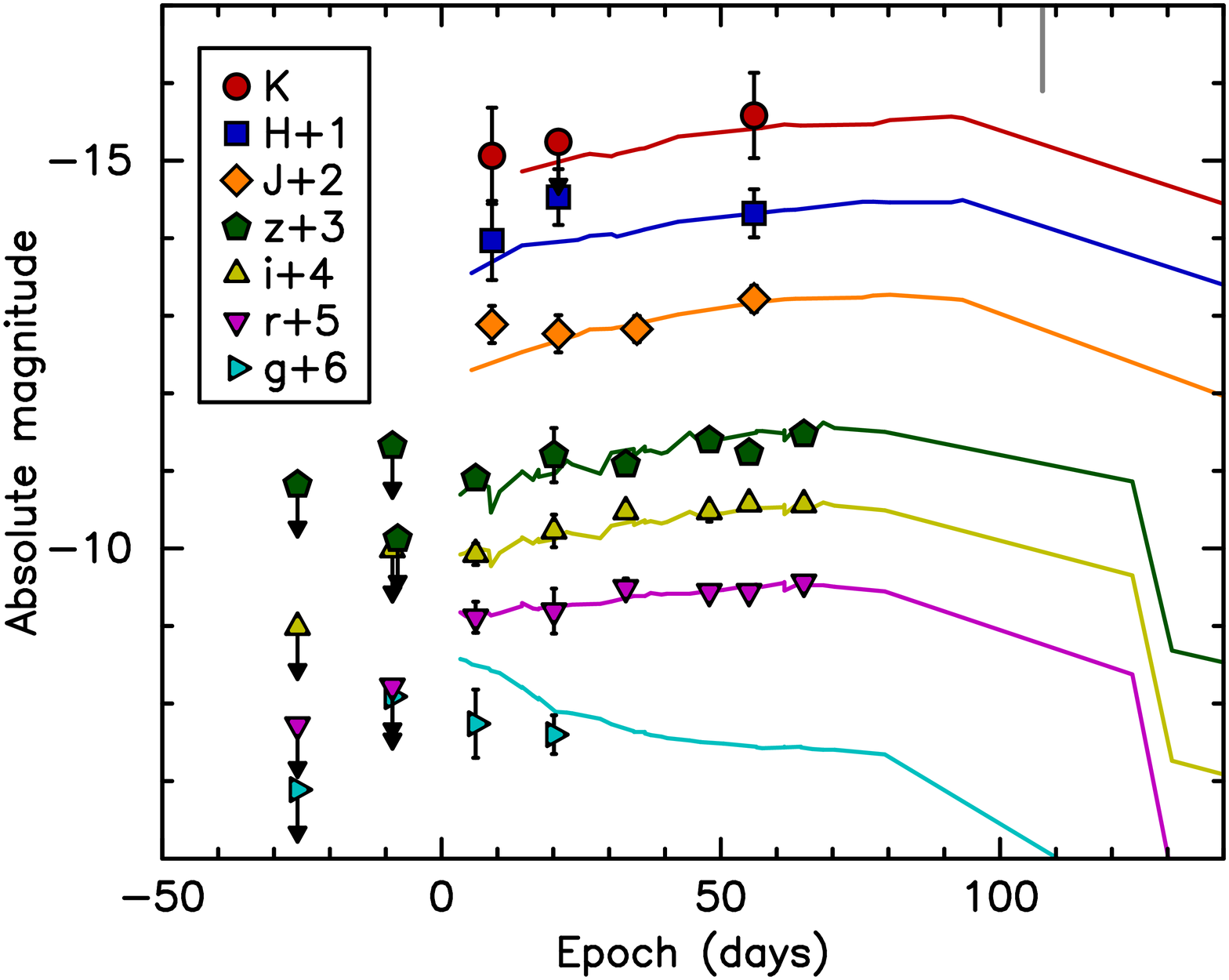}
\caption{Absolute magnitude light curves of AT 2018cux including an estimated host galaxy extinction of $A_{V} = 2.1$ mag. The light curves of the subluminous Type IIP SN 2005cs \citep{pastorello09} are shown with solid lines and shifted vertically by +0.57 mag. The grey vertical line indicates the epoch of the SALT spectrum. The epoch 0 is set to the estimated explosion date of SN 2005cs.}
\label{fig:18cux_lc}
\end{figure}

Based on the aforementioned estimated explosion epoch of AT 2018cux, the SALT spectrum of the site of AT 2018cux was obtained at $+108^{+7}_{-3}$ d. In Fig.~\ref{fig:18cux_spect} the SALT spectrum is shown corrected for extinction, overlaid with an arbitrarily shifted +106 d spectrum of SN 2005cs at the end of the plateau phase, consistent with a subluminous Type IIP SN classification. In particular, quite typical features for sub-luminous Type IIP SNe are the flux attenuation below $\sim$5500 \AA, and the absorption feature around 6150$-$6300 \AA\ consistent with a blend of \ScII\ and \BaII\ lines. 

\begin{figure}[!t]
\includegraphics[width=\linewidth]{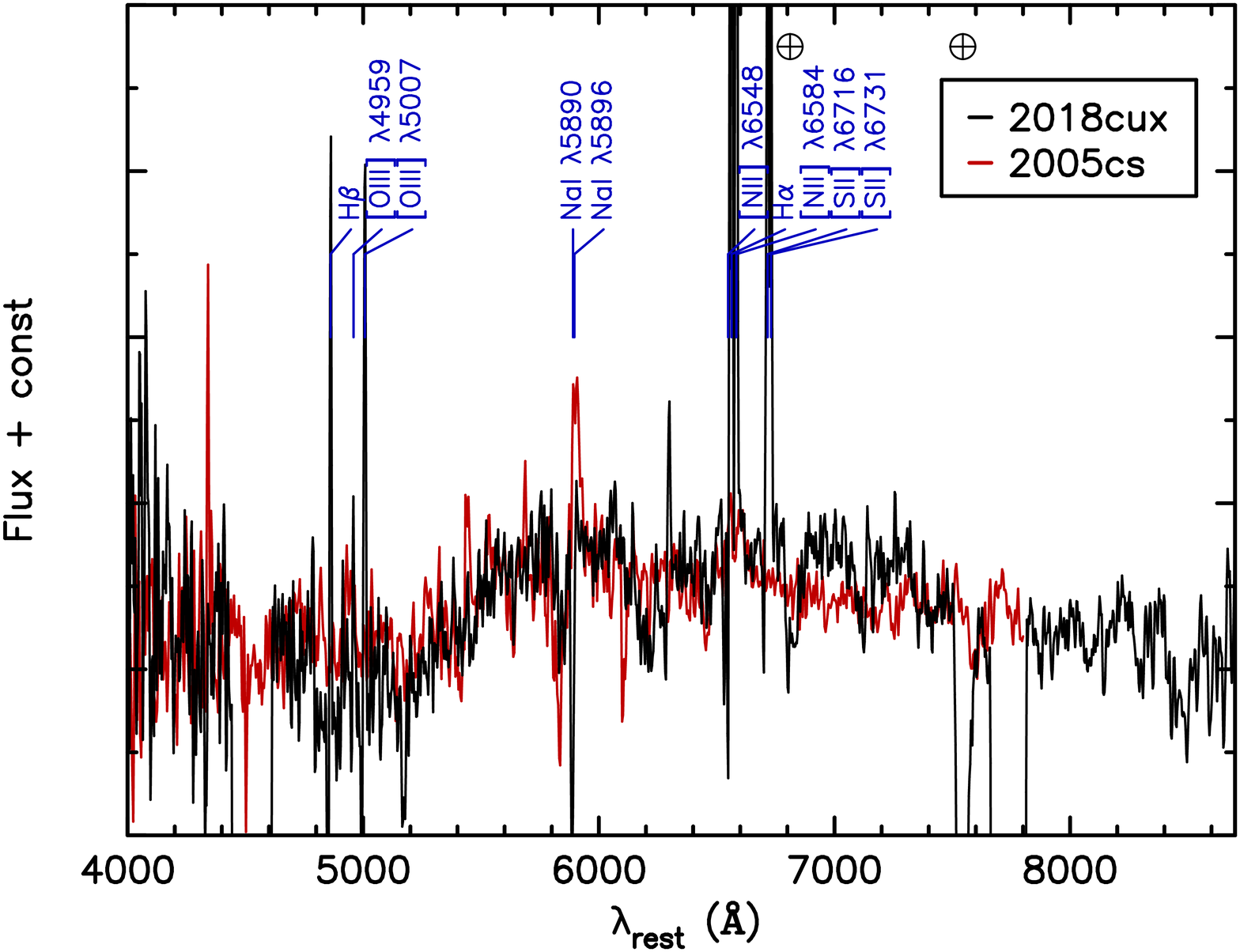}
\caption{SALT spectrum (black) at the location of AT 2018cux corrected for host galaxy extinction of $A_{V} = 2.1$ mag. The overlaid (red) +106 d spectrum of the subluminous Type IIP SN 2005cs \citep{pastorello09} is shifted with an arbitrary added constant and shows overall similarity to that of the AT 2018cux observations. The spectra have been redshift corrected to rest frame. The most prominent host lines of NGC 3256 are indicated. The main telluric bands are marked with a $\oplus$ symbol.}
\label{fig:18cux_spect}
\end{figure}

\subsection{PSN J10275082-4354034}

In addition to SN 2001db, SN 2018ec, and AT 2018cux in NGC 3256, an SN candidate PSN J10275082-4354034 (hereafter, shortly referred to as PSN102750) has been reported\footnote{Reported at the Central Bureau for Astronomical Telegrams Transient Objects Confirmation Page: \urlwofont{http://www.cbat.eps.harvard.edu/unconf/followups/J10275082-4354034.html}} in 2014 in the same LIRG at RA = 10$^{\mathrm{h}}$27$^{\mathrm{m}}$50$\fs$82 and Dec = $-43\degr$54$\arcmin$03\farcs4, discovered by Peter Aldous at Geraldine Observatory\footnote{\urlwofont{https://geraldineobservatory.co.nz/}}. Based on the coordinates the projected distance of the transient from the \textit{Ks}-band nucleus is 2.1 kpc. An unfiltered discovery magnitude on 2014 May 7.45 UT of 15.6 mag was reported. Unfortunately, no spectroscopic classification of the transient was carried out by anyone to our knowledge. 

Follow-up imaging of PSN102750 was carried out by S. Parker. This included one epoch of luminance filter observations using the 50-cm T30 telescope\footnote{\urlwofont{https://www.itelescope.net/t30/}} with FLI-PL6303E CCD camera at the Siding Springs Observatory, and two epochs of unfiltered observations with a 35-cm Celestron C14 reflector and SBIG ST-10 CCD camera at the Parkdale Observatory. The data reduction included basic bias and dark subtraction steps and flat fielding. The {\sc quba} pipeline was used to carry out the PSF photometry of the SN candidate by calibrating the images directly into \textit{R}-band, reported in Table~\ref{table:photPSN}. 

Furthermore, we recovered PSN102750 in \textit{Hubble Space Telescope} (HST) archival images obtained with the Wide Field Camera 3 (WFC3). NGC 3256 was observed in the \textit{F467M} (Str\"omgren \textit{b}) and \textit{F621M} filters on 2014 June 10 and 2014 November 13, respectively\footnote{Programme 13333, PI: Rich.}. Fig.~\ref{fig:hst} clearly shows the late-time detection of PSN102750 in comparison to an archival image obtained with a similar filter. Photometry on the images with the transient was carried out using {\sc dolphot}\footnote{\urlwofont{http://americano.dolphinsim.com/dolphot/}}, an HST dedicated photometry package. The individual charge transfer efficiency (CTE) corrected images were masked for bad pixels and the sky background was fitted and subtracted, before being aligned to produce the drizzle-combined image. PSFs were then fit to all the identified sources present in the images, yielding their magnitude values. The magnitudes of the SN candidate in \textit{F467M} and \textit{F621M} filters were adopted as those of \textit{B} and \textit{R}, respectively.

\begin{figure*}[!t]
\includegraphics[width=\linewidth]{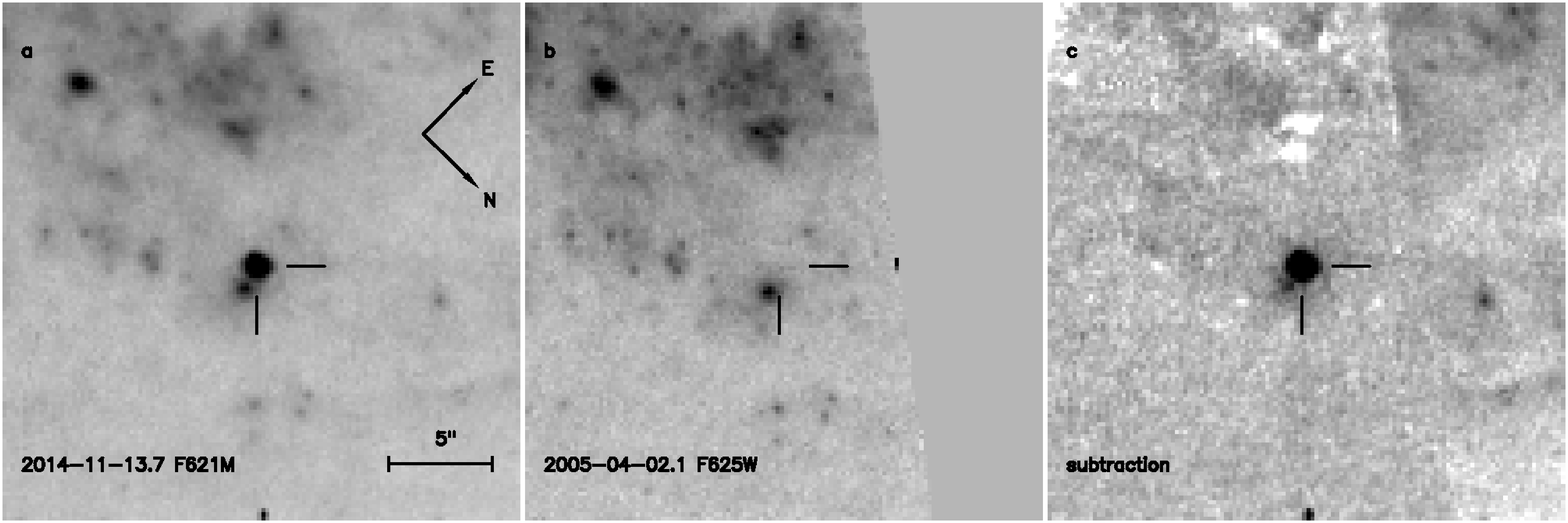}
\caption{a) Late-time HST \textit{F621M} image of PSN102750 observed 190 d after the discovery of the transient, b) pre-explosion HST \textit{F625W} image of the field of the transient, and c) a subtraction between the images. In addition to the event, the subtraction results also in some faint residuals from sources in the field likely due to the somewhat different filter widths of the two epochs of observations. The transient location is shown with tick marks. The orientation and image scale are indicated in the left panel.}
\label{fig:hst}
\end{figure*}

The resulting light curve of PSN102750 was template fitted with the same method that was used for SN 2018ec and AT 2018cux, however, a selection of different CCSN subtypes was used as templates to yield a tentative classification and host galaxy extinction for the transient. The comparison events include Type IIn SNe 1998S \citep{fassia00,liu00} and 2005ip \citep{stritzinger12}, a normal Type IIP SN 2004et \citep{sahu06}, and a normal Type Ic SN 2007gr \citep{hunter09}. Furthermore, a comparison to the normal Type Ia SN 2011fe \citep{munari13} was also carried out. The fits are shown in Fig.~\ref{fig:photPSN}. PSN102750 is most consistent with a Type IIn event that generally can show major diversity in their late-time evolution as shown by the comparisons. Our best fit is with SN 1998S, with PSN102750 discovered shortly before the peak magnitude. However, a Type IIP SN cannot be fully excluded either, though the early photometry of PSN102750 shows deviations from a flat plateau. H-poor stripped-envelope events are excluded based on the blue colours of PSN102750 even if a negligible host extinction is assumed; increasing the extinction would only make the discrepancy larger. A Type Ia SN classification can be excluded primarily based on the bright late-time detection of PSN102750. Based on the light curve fit with SN 1998S as a template, we conclude that PSN102750 has a host galaxy extinction of $A_{V} = 0.3^{+0.4}_{-0.3}$ mag, is $1.6^{+0.1}_{-0.4}$ mag fainter than SN 1998S, and discovered at $-9^{+2}_{-3}$ days relative to the light curve peak of SN 1998S assuming the extinction law of \citet{cardelli89}. For comparison, the Type IIP fit using SN 2004et as a template yields $A_{V} = 0.0^{+0.3}_{-0.0}$ mag, a very small brightness difference of $0.1^{+0.3}_{-0.1}$ mag, and $t_{0} = 12^{+2}_{-4}$ days relative to the estimated explosion date. SN 2004et is at the bright end of the intrinsic magnitude distribution of normal Type IIP SNe at plateau \citep{anderson14}.

\begin{figure*}[!t]
\includegraphics[width=0.50\linewidth]{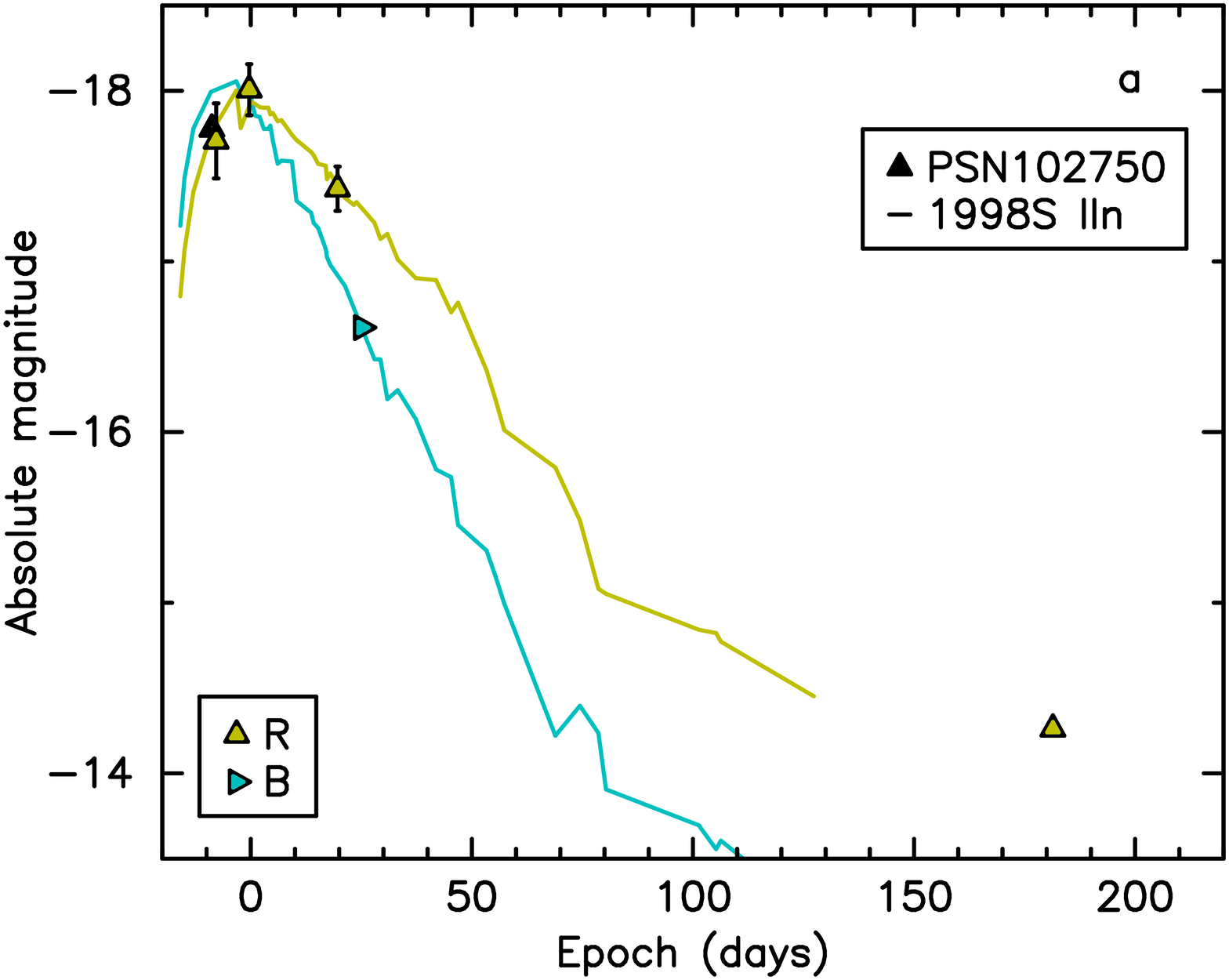}
\includegraphics[width=0.50\linewidth]{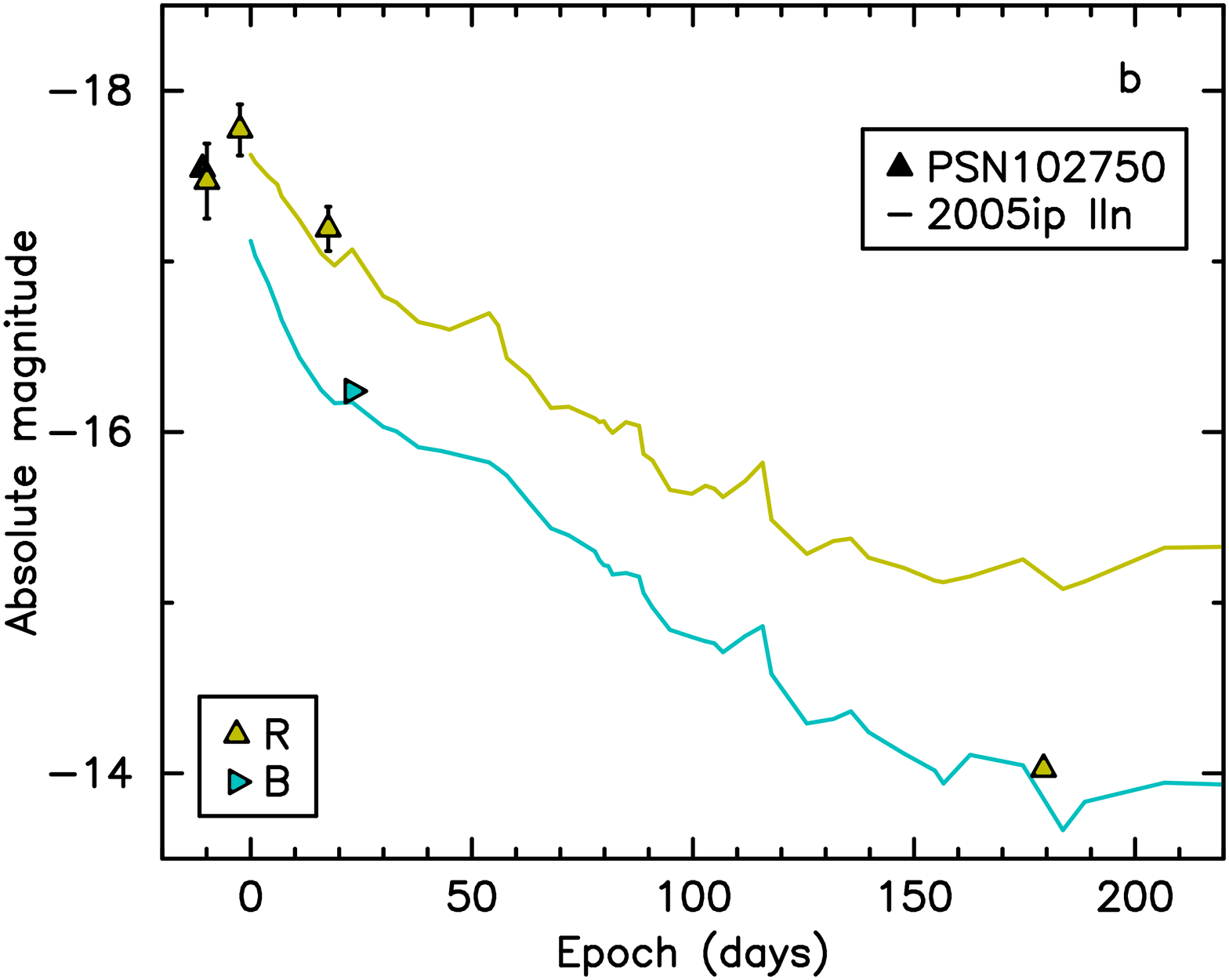}
\includegraphics[width=0.50\linewidth]{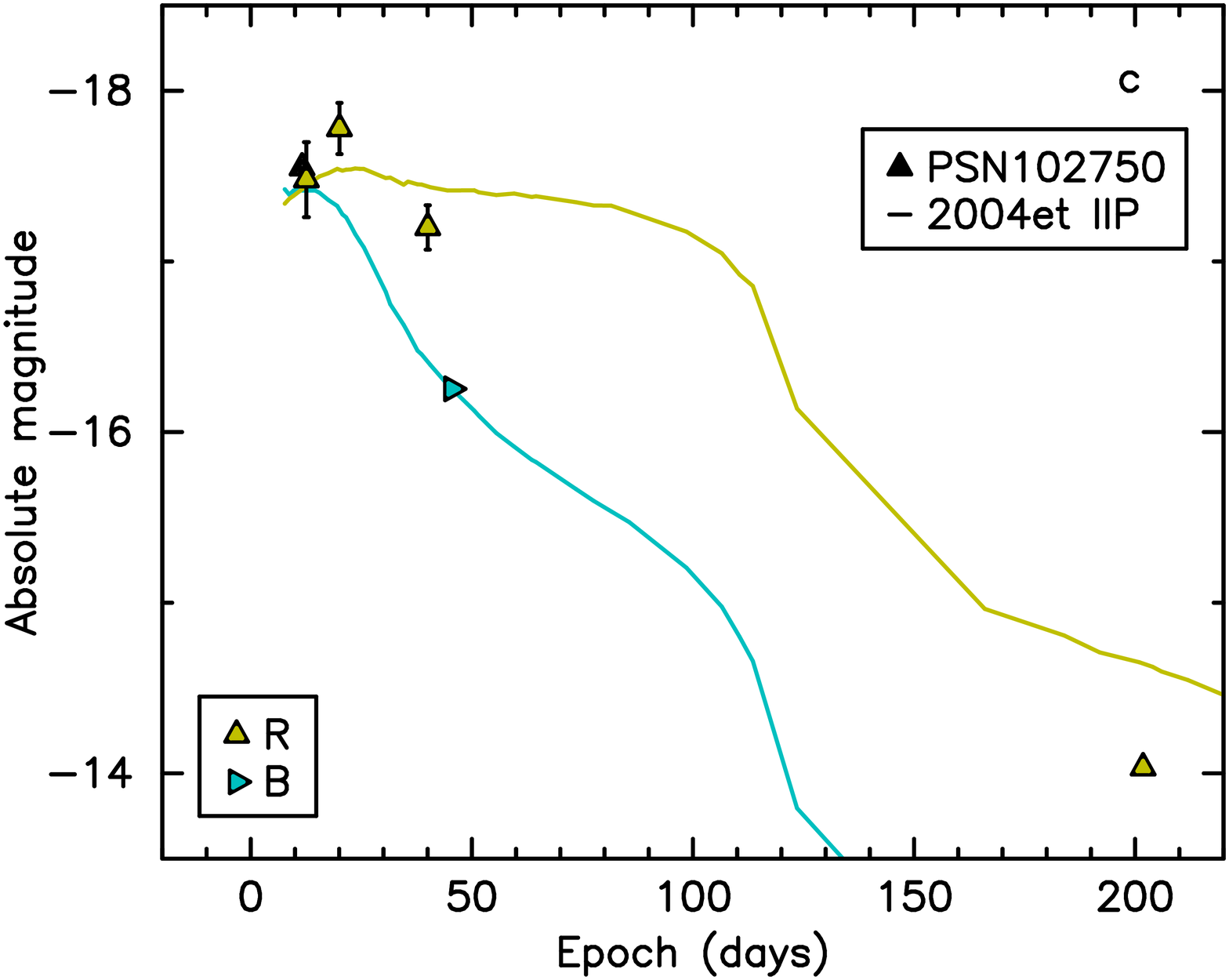}
\includegraphics[width=0.50\linewidth]{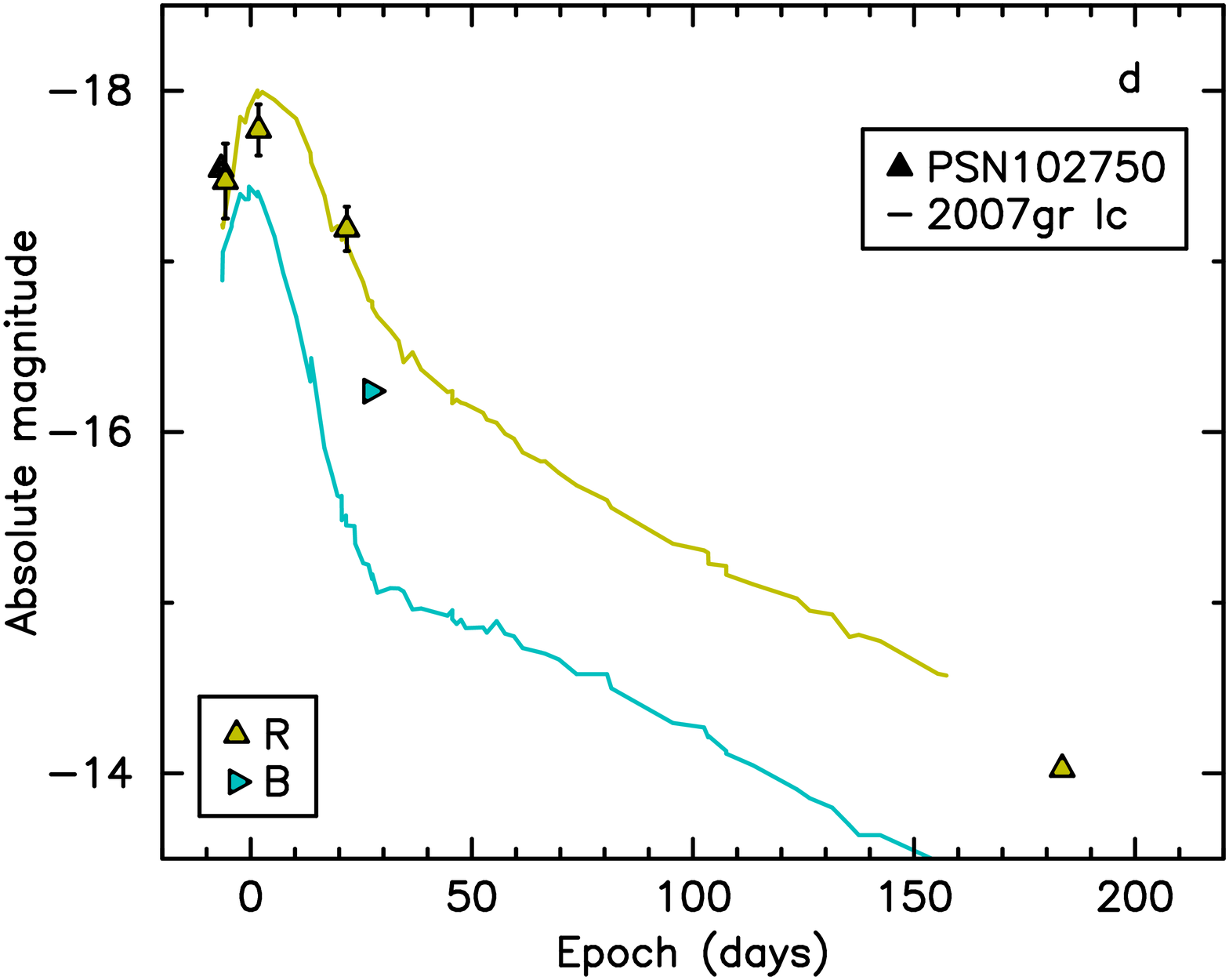}
\includegraphics[width=0.50\linewidth]{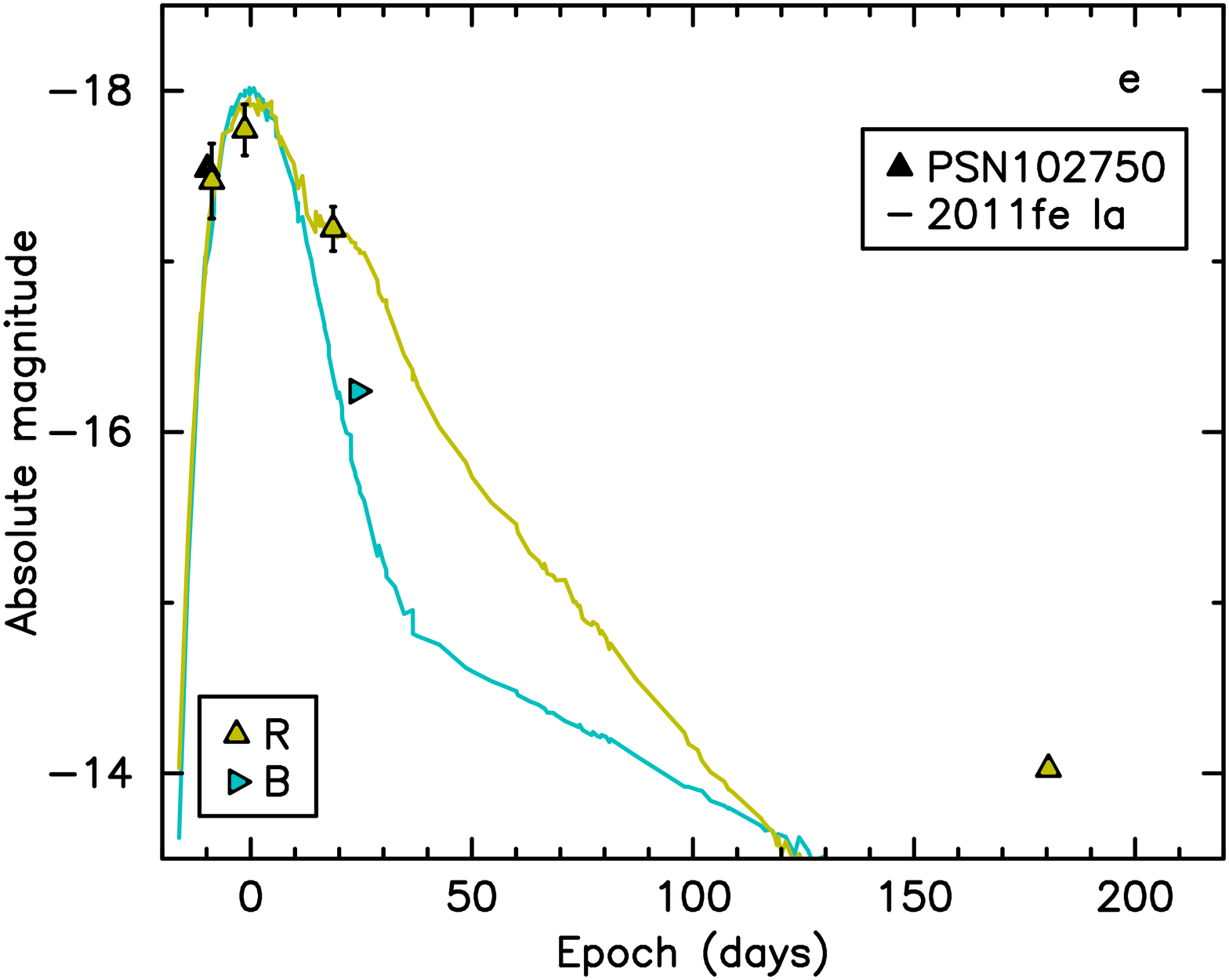}
\caption{PSN102750 light curve comparisons to those of a) Type IIn SN 1998S, b) Type IIn SN 2005ip, c) Type IIP SN 2004et, d) Type Ic SN 2007gr, and e) Type Ia SN 2011fe. The reported unfiltered discovery magnitude is also shown in the plot with a black symbol.}
\label{fig:photPSN}
\end{figure*}

\section{Arp 299}

Arp 299 is a nearby LIRG at a distance of 44.8 Mpc \citep{huo04}. The IR luminosity of the galaxy, $L_{\mathrm{IR}} = 10^{11.82} L_{\odot}$ \citep{sanders03}, suggests a rate of $\sim$1.8 CCSN yr$^{-1}$ based on the empirical relation of \citet{mattila01}. The Galactic extinction towards Arp 299 is $A_{V} = 0.046$ mag \citep{schlafly11}. The recent SNe 2019lqo and 2020fkb presented here increase the total number of spectroscopically classified CCSNe in Arp 299 to eight events, in addition to one unclassified near-IR discovered SN 1992bu \citep[see e.g.][]{mattila12}.

\subsection{SN 2019lqo}

SN 2019lqo was discovered on 2019 July 21 07:33:36 UT (JD = 2458685.81500) by the photometric instrument on board the \textit{Gaia} spacecraft, and reported on behalf of the science alerts team as Gaia19dcu \citep{hodgkin19}.The Gaia \textit{G}-band AB discovery magnitude of SN 2019lqo was 18.33$\pm$0.20 mag, with a reported non-detection of $m_{G} > 21.5$ mag on 2019 July 16 05:26:52 UT (JD = 2458680.72699). The coordinates of the transient provided by \textit{Gaia} are RA = 11$^{\mathrm{h}}$28$^{\mathrm{m}}$32$\fs$460 and Dec = $+58\degr$33$\arcmin$44\farcs82 (equinox J2000.0), pointing to component `A' as the likely host (Fig.~\ref{fig:19lqo_field}) of Arp 299. This yields a projected distance of 9\farcs3 from the nucleus of Arp299-A \citep{romero-canizales11} for SN 2019lqo, corresponding to 2.0 kpc. We classified SN 2019lqo within the Nordic Optical Telescope (NOT) Unbiased Transient Survey 2 (NUTS2) on 2019 July 24.9 UT as a young Type IIb SN \citep{bose19}. 1800 s spectroscopic observations were obtained with the Andalucia Faint Object Spectrograph and Camera (ALFOSC) using the low-resolution grism \#4 and 1\farcs0 slit. Another epoch of spectroscopy was obtained with the same setup on 2019 August 2.9 UT. 

\begin{figure*}[!t]
\includegraphics[width=\linewidth]{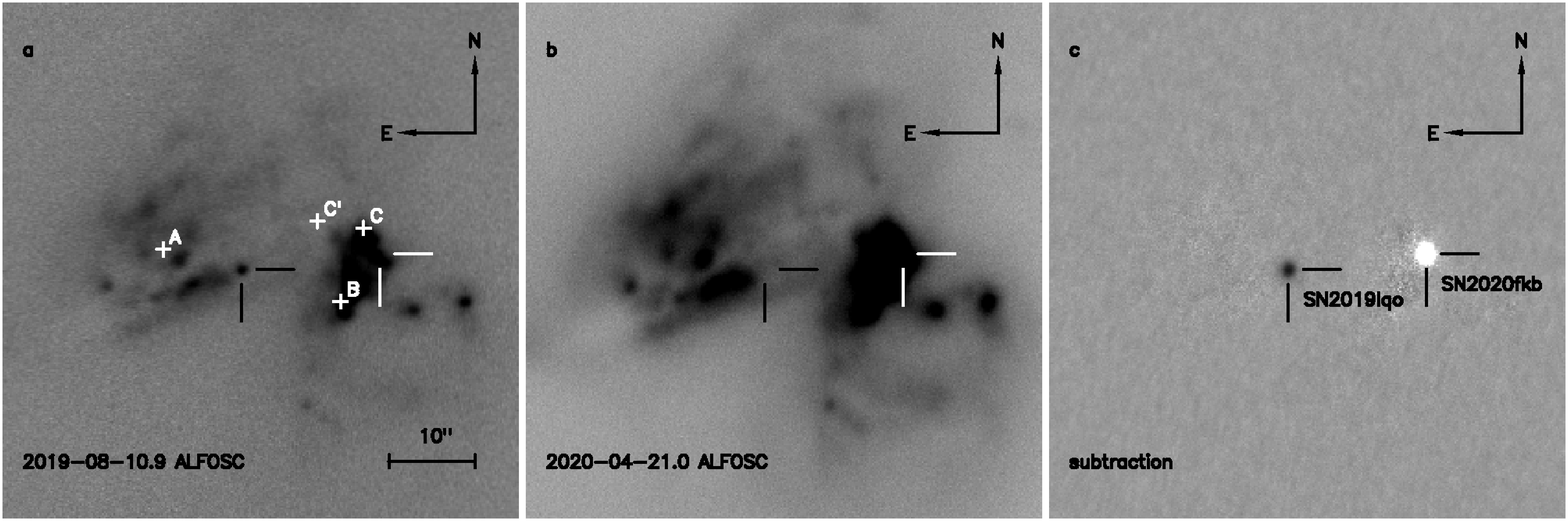}
\caption{a) 1\arcmin\ $\times$ 1\arcmin\ subsection of a NOT \textit{V}-band follow-up image of SN 2019lqo in Arp 299, b) follow-up image of SN 2020fkb, and c) a subtraction between the images. Tick marks indicate the locations of SNe 2019lqo and 2020fkb. The image scale and orientation are indicated in the left panel. The marked locations of IR bright main components (A, B, C, C$'$) of Arp 299 \citep{gehrz83} illustrate that some of these regions are obscured in optical wavelengths.}
\label{fig:19lqo_field}
\end{figure*}

SN 2019lqo was discovered shortly before the field moved into conjunction with the Sun. However, we were able to obtain multi-band light curve follow up of the event around the light curve maximum, and observed some additional data points in the tail phase of the SN when the field became observable again. All the optical ALFOSC data were reduced in a standard manner using the {\sc quba} pipeline. The NOT near-IR Camera and spectrograph (NOTCam) imaging was reduced by making use of the external {\sc notcam} package\footnote{\urlwofont{http://www.not.iac.es/instruments/notcam/guide/observe.html}} for {\sc iraf}. The {\sc quba} pipeline PSF photometry of SN 2019lqo was calibrated with the sequence star magnitudes obtained from \citet{kankare14b}. The resulting photometry is listed in Table~\ref{table:phot19lqo}. 

The first Zwicky Transient Facility \citep[ZTF;][]{bellm19} detections of SN 2019lqo (with an internal name ZTF19abgbbzy) are reported on the Transient Name Server; additional ZTF limits and detections are also available e.g., via the public MARS broker\footnote{\urlwofont{https://mars.lco.global/109717499/}}. Intriguingly, the first reported ZTF detection of $m_{r} = 18.86$ mag was obtained 24.7 hrs before the reported \textit{Gaia} non-detection of $>$21.5 mag. While a SN light curve can evolve extremely rapidly during a very short shock breakout phase, the most likely case here is that a combination of a crowded LIRG background, a faint SN, and a non-optimal \textit{Gaia} sky scanning angle made the SN non-detectable, and resulted in the reported non-detection with a nominal limit. Furthermore, ZTF reports an \textit{r}-band non-detection of $>$19.5 mag following their initial detection, and 2 days before the \textit{Gaia} discovery epoch. However, the magnitude limits reported by ZTF are difference image estimates over the whole CCD quadrant \citep{masci19}. The location of SN 2019lqo is quite crowded with a luminous galaxy background making image subtraction challenging and sky condition dependent. Therefore, it is not surprising if a limit yielded globally for the field is overly optimistic for the location of SN 2019lqo if the subtraction has not been optimal. We proceeded to download the science data product files \citep{masci19} from the ZTF data release 3 (DR3)\footnote{\urlwofont{https://www.ztf.caltech.edu/page/dr3}} covering the follow up of SN 2019lqo and corresponding template reference images. We carried out image subtractions using the {\sc isis2.2} package and our resulting PSF photometry, calibrated against the Sloan Digital Sky Survey (SDSS), for SN 2019lqo is listed in Table~\ref{table:phot_19lqo} for epochs for which we yielded a $>$ 3\,$\sigma$ detection. This includes our measurements of $m_{r} = 19.04 \pm 0.09$ mag for the first ZTF epoch and $18.77 \pm 0.12$ mag for the epoch following the first detection for which a non-detection was previously reported. 

Similar to the SNe in NGC 3256, the line-of-sight extinction of SN 2019lqo was estimated with a simultaneous $\chi^{2}$ comparison of \textit{UBVRIJHK} light curves to those of well-sampled Type IIb SNe 2011dh \citep{arcavi11,ergon14,ergon15} and 1993J \citep{pressberger93,ripero93,richmond94,richmond96}. The fitting was carried out with the \citet{cardelli89} extinction law. The resulting absolute magnitude light curves are shown in Fig.~\ref{fig:19lqo_lc}, including photometry based on publicly available data. The early public \textit{gGr}-band detections of SN 2019lqo suggest that the SN had a relatively long rise time and was discovered possibly during a post-shock cooling phase, more similar to that of SN 1993J than the more rapidly evolving SN 2011dh, but this is fairly poorly constrained. The fit with SN 1993J as a template is superior compared to that of SN 2011dh as a reference; this fit suggests that SN 2019lqo has a host galaxy extinction of $A_{V} = 2.1^{+0.1}_{-0.3}$ mag, is $0.2^{+0.1}_{-0.3}$ mag brighter than SN 1993J, and peaked in \textit{R}-band around JD = 2458702 $\pm$ 1 (i.e. around 2019 August 6). There is uncertainty in the host galaxy extinction of SN 1993J and an average value of $E(B-V)=0.2$ mag was adopted here for the comparison. The rise of the SN 2019lqo light curve appears to be relatively long compared to those of SNe 1993J and 2011dh, and also the post-maximum decline is slightly slower in comparison to these prototypical Type IIb SNe. Intriguingly SN 2019lqo is spectroscopically more similar to SN 2011dh with e.g., prominent absorption features of \HI\ and \CaII\ near-IR triplet lines compared to SN 1993J (Fig.~\ref{fig:19lqo_spect}). Therefore, it seems that SN 2019lqo bridges the observational characteristics of these two canonical Type IIb SNe.

\begin{figure*}[!t]
\includegraphics[width=0.50\linewidth]{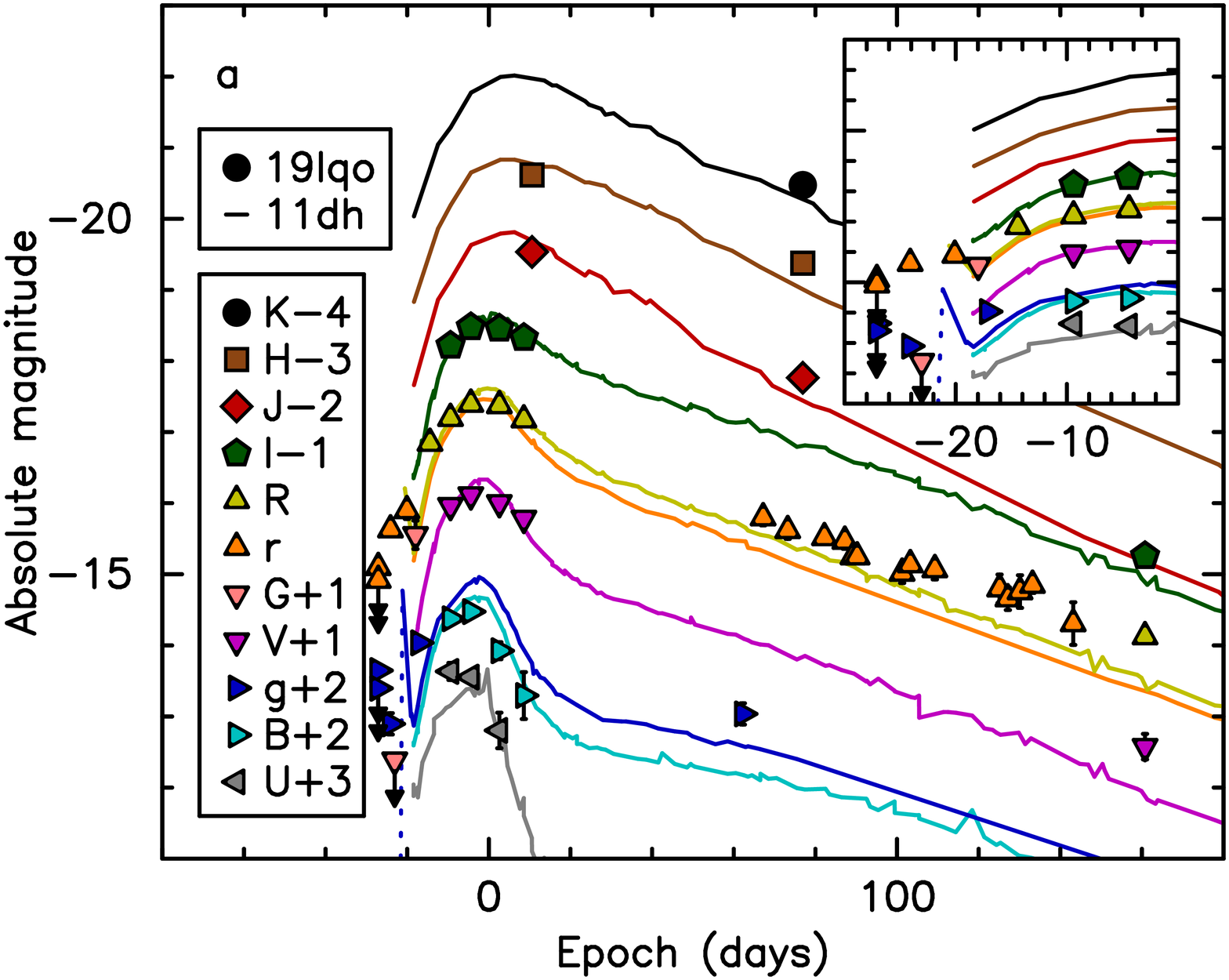}
\includegraphics[width=0.50\linewidth]{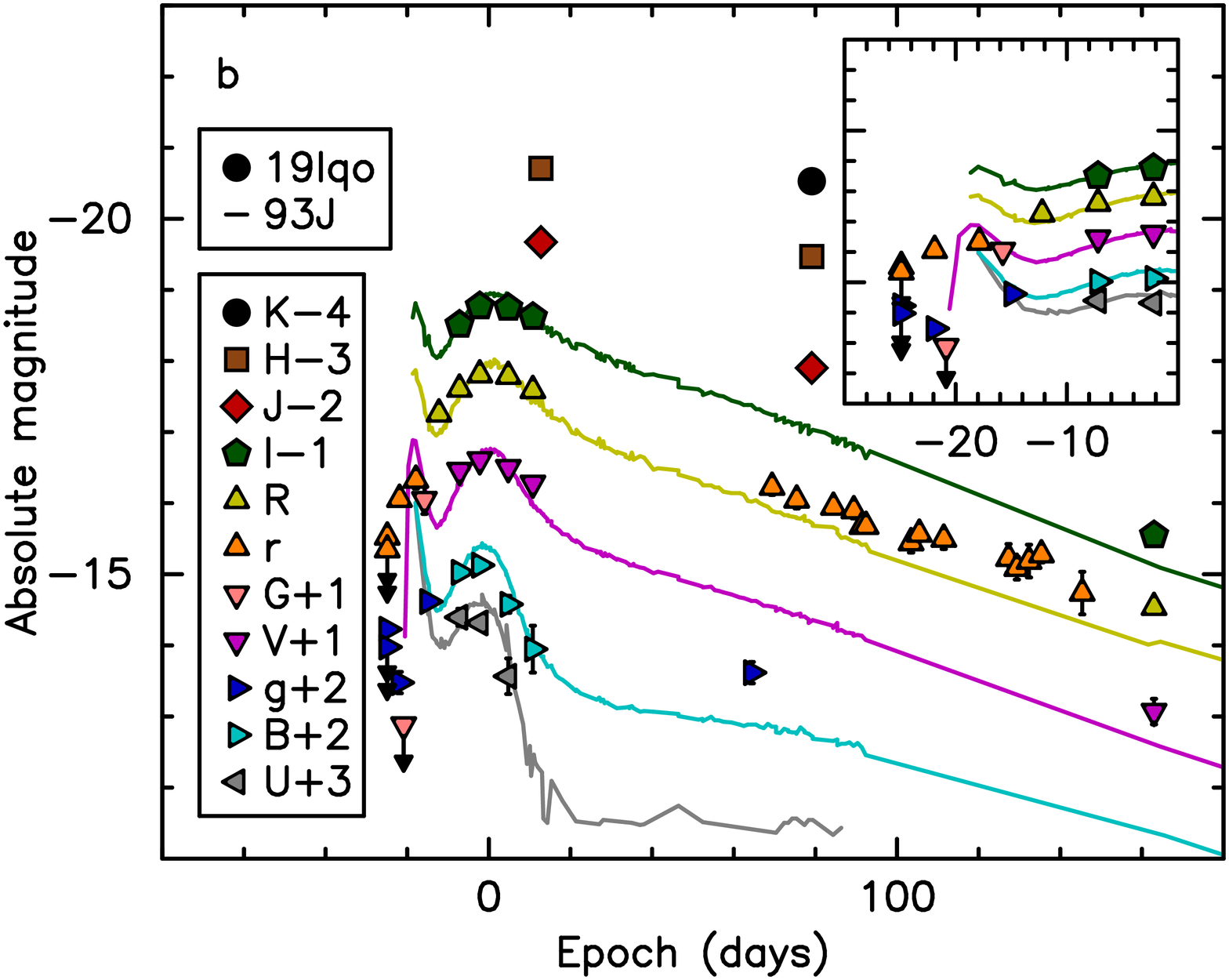}
\caption{SN 2019lqo light curve fits to Type IIb SNe a) 2011dh and b) 1993J. The inset zoom-in panels show the pre-maximum light curves suggesting a relatively long shock cooling phase. As previously noted, the \textit{Gaia} \textit{G}-band upper limit is probably optimistic. The epoch 0 is set to the estimated \textit{R}-band peak.}
\label{fig:19lqo_lc}
\end{figure*}

\begin{figure}[!t]
\includegraphics[width=\linewidth]{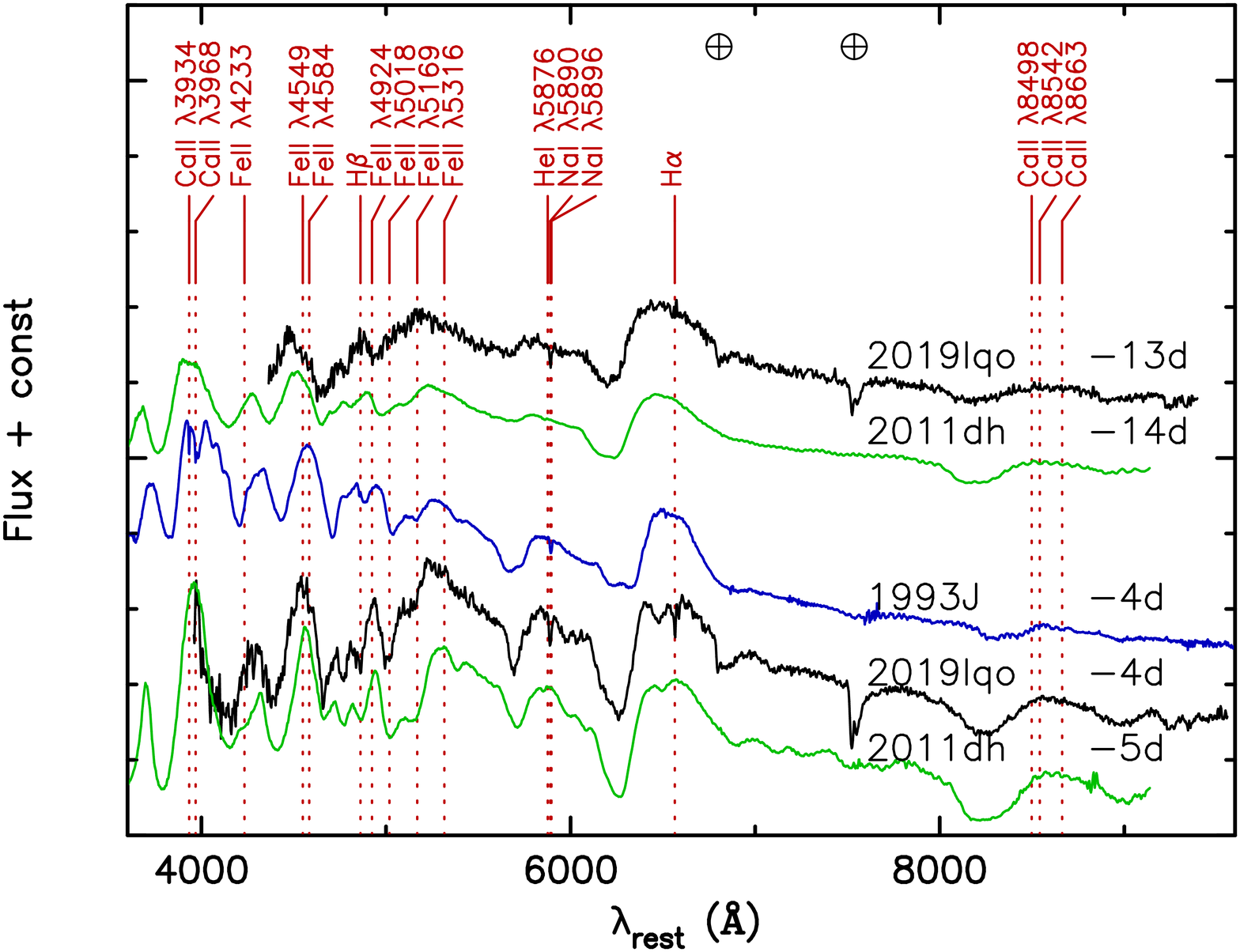}
\caption{The spectra of SN 2019lqo compared to those of Type IIb SNe 2011dh \citep{ergon14} and 1993J \citep{matheson00}. The most prominent SN lines are indicated based on \citet{ergon14}. The
spectra have been extinction and redshift corrected, and shifted vertically for clarity. The main telluric band wavelengths for SN 2019lqo are indicated with a $\oplus$ symbol.}
\label{fig:19lqo_spect}
\end{figure}

\subsection{SN 2020fkb}

SN 2020fkb in Arp 299 was discovered by the ZTF survey, reported by \citet{pignata20}, with an internal name ZTF18aarlpzd. The transient was first detected on 2020 March 28 06:19:45 UT at $m_{g} = 17.83 \pm 0.08$ mag with a reported non-detection on 2020 March 7 09:49:20 UT at $m_{r} = 19.34$ mag. The SN was classified by \citet{tomasella20} on 2020 April 2.8 UT as a young Type Ib, roughly a week before maximum, using the 1.82 m Copernico Telescope with the Asiago Faint Object Spectrograph and Camera (AFOSC). We obtained further optical spectrophotometric follow up of SN 2020fkb using AFOSC, and the NOT with ALFOSC; optical imaging with the Asiago 67/92 cm Schmidt telescope with the KAF-16803 CCD; optical and near-IR imaging with the Liverpool Telescope \citep[LT;][]{steele04} using the IO:O and IO:I \citep{barnsley16} instruments; and near-IR imaging using the NOT with NOTCam. 

The standard reduction of the Asiago observatory data was carried out using the {\sc foscgui} pipeline\footnote{\urlwofont{https://sngroup.oapd.inaf.it/foscgui.html}}, and basic instrumental pipeline reduction products of the LT data were obtained for the analysis. A selection of pre-explosion images were used as image subtraction templates from the NOT, the Asiago telescope, and the Pan-STARRS1 \citep{chambers16} data release 2 archive\footnote{\urlwofont{https://panstarrs.stsci.edu/}}. The resulting PSF photometry of SN 2020fkb is listed in Tables~\ref{table:phot20fkb} and \ref{table:phot_20fkb}.

The host galaxy extinction of SN 2020fkb was estimated using light curves of a normal Type Ib SN 2004gq \citep{bianco14,stritzinger18a} with an excellent multiband coverage and a low host galaxy extinction estimate of $A_{V} = 0.26$ mag \citep{stritzinger18b}. The fit suggests that SN 2020fkb peaked around JD = 2458949 $\pm$ 1 (i.e. around 2020 April 9), has a host galaxy extinction of $A_{V} = 0.4^{+0.1}_{-0.2}$ mag, and is $0.1^{+0.2}_{-0.1}$ mag fainter than SN 2004gq; the best match is shown in Fig.~\ref{fig:20fkb_lc}. For example, in \textit{r}-band SN 2020fkb peaked at $M_{r} \approx -17.2$ mag; this is not an unusual peak magnitude for a normal Type Ib SN, but is placed at the faint end of the \textit{r}-band distribution of these SNe around $-17$ to $-18$ mag \citep[e.g.][]{taddia18}. The aforementioned pre-explosion non-detection limit, at an epoch corresponding to $-33$ d from maximum light, is not strongly constraining the explosion epoch of SN 2020fkb, as normal Type Ib SNe have typical \textit{r}-band rise times of $\sim$21 d \citep{taddia15} with a distribution around a few days.

The spectral time series of SN 2020fkb (Table~\ref{table:spec20fkb}) is shown in Fig.~\ref{fig:20fkb_spect}, compared to a selection of SN 2004gq spectra. The evolution of SN 2020fkb is quite normal for a Type Ib SN. The spectra show broad and evolving P Cygni profiles of \HeI\ features, in particular the $\lambda\lambda$5876,6678,7065 lines. Furthermore, at pre-maximum the spectra show the \CaII\ H\&K doublet, and broad features arising likely from \FeII\ blends. The \HeI\ $\lambda$5876 line Doppler velocity of the P Cygni absorption minimum close to light curve peak in our $-4$ d spectrum of SN 2020fkb is $\sim$9900 km s$^{-1}$; this is consistent with typical velocities of this feature at the corresponding phase of normal Type Ib SNe \citep{taddia18}, whereas the photometrically similar SN 2004gq shows in fact overall larger velocities. The wavelength coverage of the +20 d spectrum onwards extends further redwards and the \CaII\ near-IR triplet is prominently visible. The +72 d spectrum is not completely nebular, but some nebular features have appeared; the doublets [\OI] $\lambda\lambda$6300,6304 and [\CaII] $\lambda\lambda$7291,7324 are clearly present with a [\OI]/[\CaII] flux ratio of $\sim$0.6. There is also strong indication of the [\OI] $\lambda$5577 emission line, however, the \MgI] $\lambda$4571 feature seen in the nebular spectra of many normal Type Ib SNe \citep[see e.g.][]{kuncarayakti15} is not clearly present at this phase. 

\begin{figure}[!t]
\includegraphics[width=\linewidth]{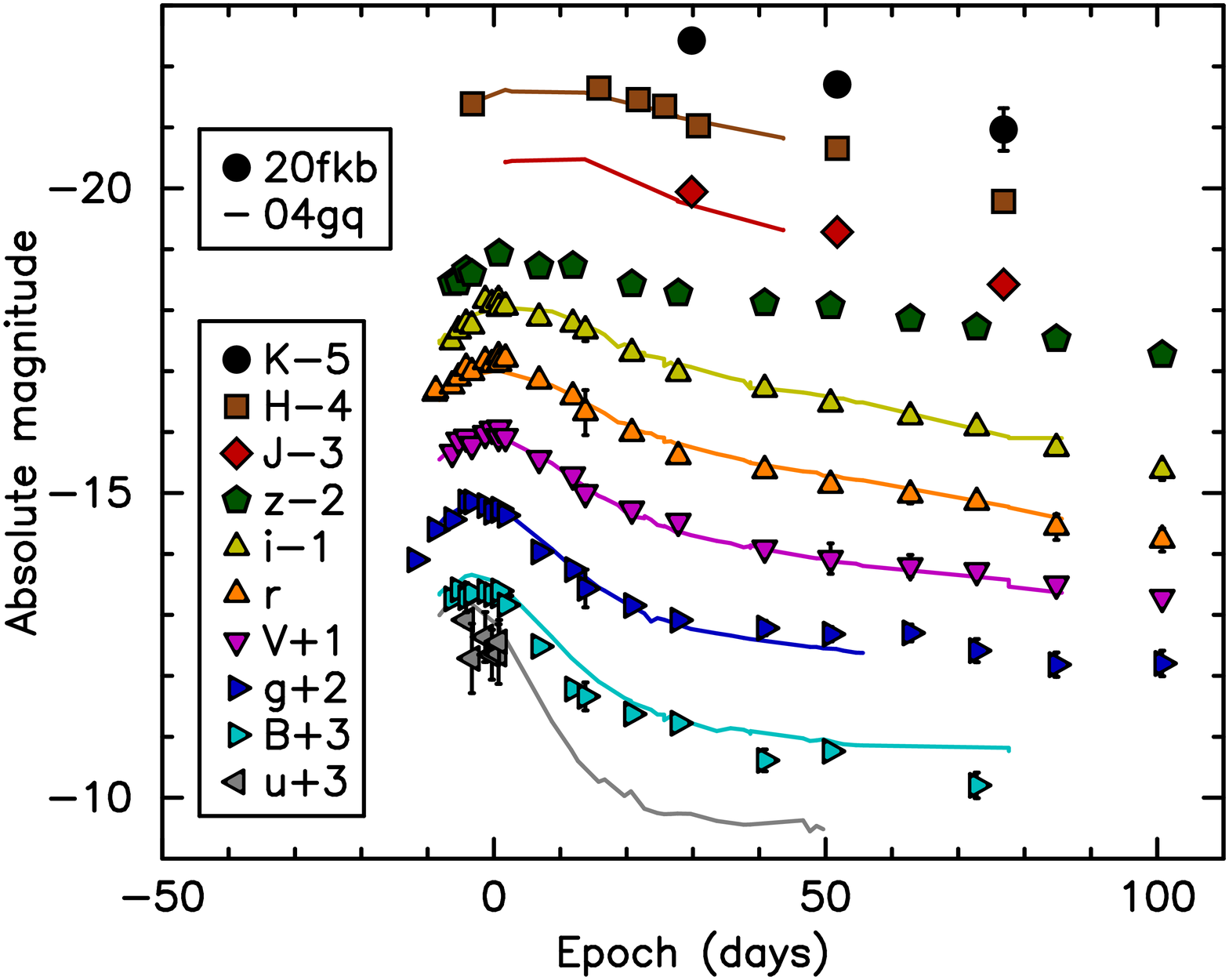}
\caption{SN 2020fkb absolute magnitude light curve (points) with an estimated host galaxy extinction of $A_{V} = 0.4$ mag yielded with a fit to Type Ib SN 2004gq, shown with solid lines and shifted vertically by $+0.13$ mag. The epoch 0 is set to the estimated \textit{V}-band maximum.}
\label{fig:20fkb_lc}
\end{figure}

\begin{figure*}[!t]
\includegraphics[width=\linewidth]{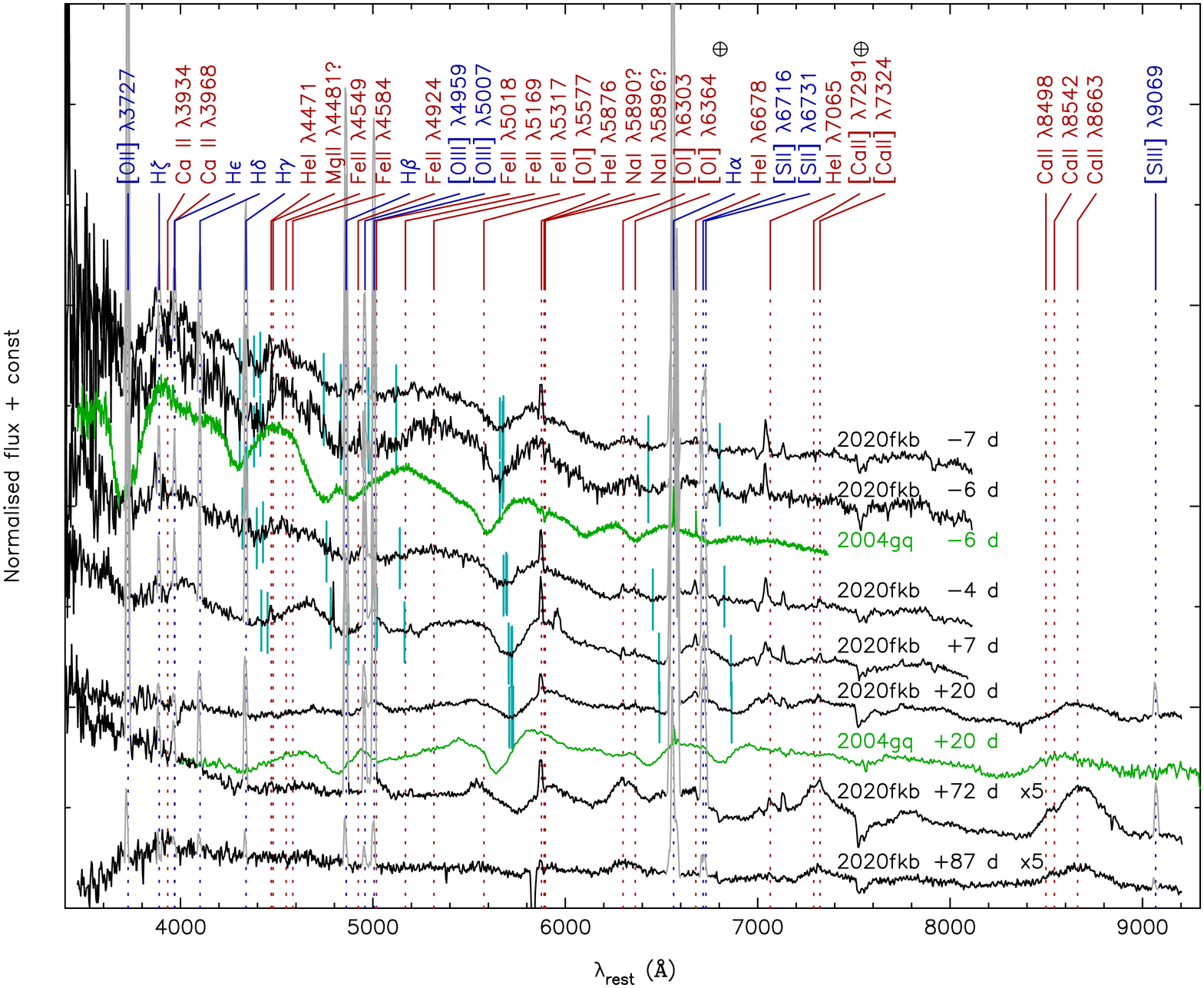}
\caption{Spectral time series of SN 2020fkb redshift corrected to rest frame. The spectra have been corrected for both Galactic and estimated host galaxy extinctions, and vertically shifted and normalised for clarity. The epochs are provided respective to the estimated light curve maximum. The most prominent spectral features are labelled arising either from the host galaxy (blue) or from the SN (red). The Doppler shifted positions of the \HeI\ and \FeII\ lines are indicated with cyan vertical lines in early epochs as suggested by the velocity of the \HeI\ $\lambda$5876 absorption minimum. Furthermore, the most prominent narrow host emission features are marked in grey, and the telluric band wavelengths with a $\oplus$ symbol. The spectra of SN 2020fkb likely contain some host continuum contamination at the blue end, in particular in the late epochs. Selected spectra of normal Type Ib SN 2004gq \citep{modjaz14} are shown for comparison (green) which shows very similar spectrophotometric evolution, but somewhat larger line velocities.}
\label{fig:20fkb_spect}
\end{figure*}

\section{Starburst age in high CCSN rate LIRGs}

The numerous CCSN subtypes are expected to originate from different populations of massive stars, that have differences e.g. in their life times. Motivated by this, we aimed to model the basic starburst parameters in all LIRGs that have hosted nuclear CCSNe to investigate, whether there is a the trend between CCSN subtypes and starburst age, $t_{\mathrm{SB}}$. Our SN sample comprises the classified CCSNe in LIRGs, at small projected distances from the host nucleus, listed in \citet{kool18} and \citet{kool19}, expanded by the new events presented in this manuscript. The sample events are discovered either at optical or IR wavelengths (i.e., we do not consider in our sample SN candidates detected only in radio). In addition to the new CCSNe in NGC 3256 and Arp 299, we also included in our analysis the recent Type II SN 2020cuj in NGC 1614 that has a limited data set, see Appendix A.1. As the star formation in LIRGs is heavily concentrated in the nuclear regions ranging in size from $\sim$100 to $\sim$1000 pc \citep[e.g.][]{soifer01}, the contribution of this region to the total CCSN rate is typically $>$50\,\%. For example, 2/3 of the young massive clusters in the LIRG Arp 299-A reside within $\sim$2.2 kpc of the nuclear regions \citep{randriamanakoto19}. Furthermore, the panchromatic luminosity of these galaxies is dominated by their central regions. While the SED fitting of the LIRGs is carried out here for the whole galaxy (largely due to the limitations of the available photometry), the luminosity and star formation characteristics are nonetheless dominated by the central regions of these systems. We set the CCSN sample limit conservatively at $\leq$2.5 kpc projected distance limit from the centres of their host galaxies. Our initial sample thus consists of 16 LIRGS ($18 \leq D_{l} \leq 150$ Mpc), and 29 SNe (12 II, 3 IIn, 13 IIb/Ib/Ic, and 1 Ibn). We modelled the SEDs of these sample LIRGs to study the distribution of the CCSN subtypes and the host starburst age in these galaxies. We note that we were not able to model the nearby and extended IC 2163/NGC 2207 system \citep[host of SPIRITS\,14buu and SPIRITS\,15c;][]{jencson17} due to the vastly differing spatial coverage across the range of wavelengths we considered; thus this LIRG is not part of our sample. Furthermore, we did not include Type II SN 2004gh \citep{folatelli04} in our analysis as the SED of its host galaxy MCG -04-25-06 has a very limited wavelength coverage available, which would not enable robust modelling. 

It is possible that due to chance alignment some of the CCSNe with small projected distances are in fact located at much larger distances from the host nucleus. While this cannot be accurately constrained, the host galaxy extinction provides a hint of the true location of the CCSNe in LIRGs; CCSNe that are more obscured by host galaxy extinction are also more likely to be embedded in the dusty central regions of their hosts. This is a strong motivation to estimate host galaxy extinctions for CCSNe in LIRGs, as we have done for the new events presented in this paper. The sample CCSNe are listed in Table~\ref{table:LIRGs} with the available reported host galaxy extinctions. 

We used a grid of radiative transfer models for a starburst \citep{efstathiou00, efstathiou09}, AGN \citep{efstathiou95,efstathiou13}, and a spheroidal galaxy (Efstathiou et al., 2020, submitted) or a disc galaxy (Efstathiou \& Siebenmorgen 2020, in prep.) to fit the SEDs using the SED Analysis Through Markov Chains (SATMC) Monte Carlo code \citep{johnson13}. The host galaxy model resulting for the best fit is indicated in Table~\ref{table:LIRGs}. The starburst and host galaxy models incorporate the stellar population synthesis model of \citet{bruzual93,bruzual03} and assume a solar metallicity and a Salpeter initial mass function (IMF). \citet{alonso-herrero06} studied a local sample of LIRGs and found them to have approximately solar metallicity. While adopting a different IMF can have some effect on the derived parameters \citep[see discussion in][]{herrero-illana17}, we adopt the classical Salpeter IMF, consistently with previous studies. The fitting predicts the starburst SED at different ages with the assumption that the star formation rate declines exponentially. The spheroidal galaxy model used here is an evolution of the cirrus model of \citet{efstathiou03} and is similar to that of \citet{silva98}, incorporated in the GRAphite and SILicate (GRASIL) code. The model assumes that stars, dust and molecular clouds in which young stars are embedded are mixed in a spherical cloud with a Sersic profile with index $n = 4$. The method for computing the disc model is similar to that used for the spheroidal models, apart from the difference in the geometry. The SATMC fitting includes 13 and 14 free parameters for the spherical and the disc models, respectively. The degrees of freedom (DOF) for the best fit are indicated in Table~\ref{table:LIRGs}. The SATMC derives errors for the fitting parameters, which Johnson et al. states to be the 68\,\% contained ranges from the one-dimensional confidence levels, marginalised over the other fitting parameters.

\begin{figure}[!t]
\includegraphics[width=\linewidth]{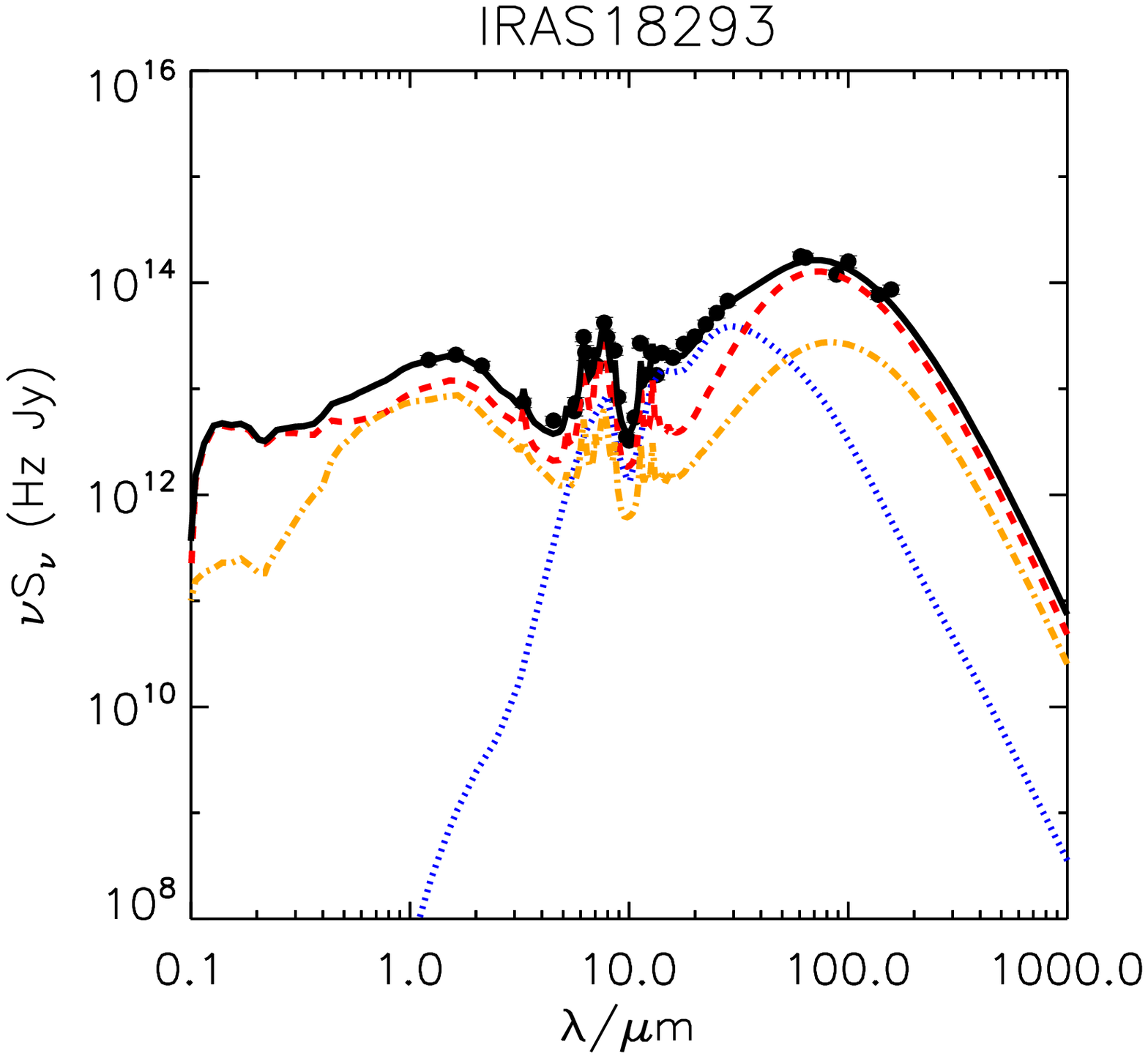}
\caption{An example SED fit (solid black curve) to the data of one of the sample LIRGs, IRAS 18293-3413 (points), with a combination of a starburst (dashed red), a spheroidal galaxy (dot-dashed orange), and an AGN (dotted blue) component.}
\label{fig:SED_IRAS18293}
\end{figure}

\begin{table*}[!t]
\caption{Confirmed and likely CCSNe at the central regions ($d_{\mathrm{proj}} \leq 2.5$ kpc) of LIRGs with SED model based global star formation properties of LIRG hosts \citep[see compilations of][and references therein]{kool18,kool19}. The horizontal line separates the two events that are excluded from further analysis with very uncertain starburst parameters of the host galaxy. Luminosity distances are obtained via NED. References for the CCSNe and their host galaxy extinction estimates (if available) are provided in parenthesis.} 
\centering
\begin{tabular}{ccccccccccccccc}
\hline
\hline
Host & $D_{l}$ & log($L_{IR}$) & $t_{\mathrm{SB}}$ & SNR$_{\mathrm{SB}}$ & SFR & $L_{\mathrm{AGN}}$ & SED & DOF & CCSN & Type & $d_{\mathrm{proj}}$ & $A_{V}^{\mathrm{host}}$ & Disc. & Ref. \\
 & (Mpc) & ($L_{\odot}$) & (Myr) & (SN yr$^{-1}$) & ($M_{\odot}$ yr$^{-1}$) & (\%) & host &  &  &  & (kpc) & (mag) & & \\ 
\hline
IRAS\,18293-3413 & 84.1 & 11.89 & $40^{+1}_{-5}$ & $2.3^{+0.4}_{-0.2}$ & $275^{+50}_{-42}$ & $49^{+13}_{-9}$ & S & 22 & AT\,2012iz & IIP & 0.6 & 1.8 & IR & (1) \\
IRAS\,18293-3413 & \dots & \dots & \dots & \dots & \dots & \dots & \dots & \dots & SN\,2013if & IIP & 0.2 & 0 & IR & (2) \\
IC\,883 & 107.1 & 11.73 & $38^{+1}_{-1}$ & $2.1^{+0.1}_{-0.1}$ & $233^{+21}_{-17}$ & $45^{+13}_{-20}$ & D & 27 & SN\,2010cu & IIP & 0.2 & 0 & IR & (3) \\
IC\,883 & \dots & \dots & \dots & \dots & \dots & \dots & \dots & \dots & SN\,2011hi & IIP & 0.4 & 6 & IR & (3) \\
NGC\,5331 & 150.0 & 11.66 & $35^{+1}_{-3}$ & $1.8^{+0.2}_{-0.3}$ & $212^{+19}_{-39}$ & $21^{+4}_{-5}$ & D & 5 & AT\,2017chi & IIn & 1.5 & 12 & IR & (1) \\
Arp\,299 & 44.8 & 11.82 & $16^{+2}_{-4}$ & $1.5^{+0.1}_{-0.2}$ & $99^{+9}_{-16}$ & 0.0 & S & 5 & SN\,1998T & Ib & 1.5 & ? & Opt & (4) \\
Arp\,299 & \dots & \dots & \dots & \dots & \dots & \dots & \dots & \dots & SN\,2005U & IIb & 1.4 & ? & IR & (4) \\
Arp\,299 & \dots & \dots & \dots & \dots & \dots & \dots & \dots & \dots & SN\,2010O & Ib & 1.2 & 1.9 & Opt & (5) \\
Arp\,299 & \dots & \dots & \dots & \dots & \dots & \dots & \dots & \dots & SN\,2010P & IIb & 0.2 & 7 & IR & (5) \\
Arp\,299 & \dots & \dots & \dots & \dots & \dots & \dots & \dots & \dots & SN\,2019lqo & IIb & 2.0 & 2.1 & Opt & (6) \\
Arp\,299 & \dots & \dots & \dots & \dots & \dots & \dots & \dots & \dots & SN\,2020fkb & Ib & 0.7 & 0.4 & Opt & (6) \\
NGC\,1614 & 66.8 & 11.66 & $25^{+2}_{-1}$ & $1.1^{+0.1}_{-0.1}$ & $94^{+11}_{-3}$ & $61^{+4}_{-4}$ & S & 33 & SN\,1996D & Ic & 2.1 & ? & Opt & (2) \\
NGC\,1614 & \dots & \dots & \dots & \dots & \dots & \dots & \dots & \dots & SN\,2020cuj & II & 2.5 & 2.8 & Opt & (5) \\
NGC\,838 & 53.3 & 11.05 & $25^{+2}_{-1}$ & $1.0^{+0.1}_{-0.1}$ & $90^{+8}_{-7}$ & $30^{+30}_{-9}$ & S & 7 & SN\,2005H & II & 0.4 & ? & Opt & (2) \\
IRAS\,17138-1017 & 82.2 & 11.49 & $28^{+7}_{-1}$ & $0.8^{+0.1}_{-0.1}$ & $69^{+37}_{-4}$ & $31^{+20}_{-4}$ & S & 22 & SN\,2008cs & IIn & 1.5 & 16 & IR & (7) \\
IRAS\,17138-1017 & \dots & \dots & \dots & \dots & \dots & \dots & \dots & \dots & SN\,2015cb & IIP & 0.6 & 4.6 & IR & (2) \\
NGC\,2388 & 61.1 & 11.28 & $25^{+1}_{-1}$ & $0.7^{+0.1}_{-0.1}$ & $63^{+5}_{-3}$ & $37^{+3}_{-2}$ & D & 8 & SN\,2015U & Ibn & 1.8 & 3.1 & Opt & (8) \\
NGC\,317B & 77.0 & 11.19 & $37^{+1}_{-1}$ & $0.7^{+0.1}_{-0.1}$ & $73^{+2}_{-4}$ & $45^{+4}_{-3}$ & S & 30 & SN\,1999gl & II & 1.9 & ? & Opt & (2) \\
NGC\,317B & \dots & \dots & \dots & \dots & \dots & \dots & \dots & \dots & SN\,2014dj & Ic & 1.5 & ? & Opt & (2) \\
NGC\,3256 & 37.4 & 11.61 & $43^{+1}_{-4}$ & $0.6^{+0.1}_{-0.1}$ & $64^{+8}_{-14}$ & $37^{+4}_{-2}$ & S & 16 & SN\,2001db & II & 2.1 & 5.5 & IR & (6) \\
NGC\,3256 & \dots & \dots & \dots & \dots & \dots & \dots & \dots & \dots & PSN102750 & IIn & 2.1 & 0.3 & Opt & (6) \\
NGC\,3256 & \dots & \dots & \dots & \dots & \dots & \dots & \dots & \dots & SN\,2018ec & Ic & 1.7 & 2.1 & IR & (6) \\
NGC\,3256 & \dots & \dots & \dots & \dots & \dots & \dots & \dots & \dots & AT\,2018cux & IIP & 0.8 & 2.1 & Opt & (6) \\
ESO\,138-G27 & 96.1 & 11.41 & $31^{+2}_{-2}$ & $0.6^{+0.1}_{-0.1}$ & $60^{+6}_{-10}$ & $46^{+2}_{-3}$ & S & 27 & SN\,2009ap & Ic & 1.2 & ? & Opt & (2) \\
NGC\,6907 & 49.2 & 11.11 & $39^{+1}_{-2}$ & $0.6^{+0.1}_{-0.2}$ & $79^{+13}_{-25}$ & $45^{+4}_{-6}$ & S & 2 & SN\,2008fq & II & 1.4 & ? & Opt & (2) \\
NGC\,2146 & 18.0 & 11.15 & $30^{+2}_{-1}$ & $0.3^{+0.1}_{-0.1}$ & $33^{+3}_{-2}$ & $82^{+3}_{-3}$ & S & 8 & SN\,2005V & Ib/c & 0.5 & ? & IR & (2) \\
MCG\,-02-01-052 & 115.2 & 11.48 & $41^{+4}_{-3}$ & $0.1^{+0.1}_{-0.1}$ & $18^{+12}_{-7}$ & $58^{+20}_{-22}$ & S & 3 & SN\,2010hp & II & 2.1 & 0.5 & IR & (9) \\
NGC\,6000 & 28.1 & 10.95 & $11^{+6}_{-2}$ & $0.1^{+0.1}_{-0.1}$ & $7^{+5}_{-3}$ & $32^{+3}_{-3}$ & S & 4 & SN\,2010as & IIb & 0.6 & 1.8 & Opt & (10) \\
%\hline
NGC\,5433 & 70.3 & 11.01 & $46^{+2}_{-11}$ & $<$0.03 & $24^{+2}_{-1}$ & $25^{+19}_{-5}$ & S & 2 & SN\,2010gk & Ic & 2.0 & ? & Opt & (2) \\ 
\hline
\end{tabular}
\label{table:LIRGs}
\tablefoot{1) \citet{kool19}, 2) \citet{kool18}, 3) \citet{kankare12}, 4) \citet{mattila12}, 5) \citet{kankare14a}, 6) This work, 7) \citet{kankare08}, 8) \citet{pastorello15}, 9) \citet{miluzio13}, 10) \citet{folatelli14}.}
\end{table*}

For the usage of the SED fitting, we collated the optical to far-IR photometric data of the sample LIRGs via the NASA/IPAC Extragalactic Database\footnote{\urlwofont{http://ned.ipac.caltech.edu/}} (NED) mainly from the GALEX (\textit{Galaxy Evolution Explorer}), SDSS, 2MASS, IRAS (the \textit{Infrared Astronomical Satellite}), and ISO (the \textit{Infrared Space Observatory}). Furthermore, we also used photometric data obtained with the \textit{Herschel Space Observatory} from \citet{chu17}. 

For three LIRGs (IC 883, IRAS 17138-1017, and IRAS 18293-3413) the \textit{Spitzer} spectra used by \citet{herrero-illana17} for their modelling were re-utilised for the re-analysis here. For five additional LIRGs in our sample (ESO 138-G27, NGC 1614, NGC 2388, NGC 317B, and NGC 838), low-resolution ($R \approx 60-127$) mid-IR spectra were available from the short-low (SL; 5.2$-$14.5\,$\mu$m) and long-low (LL; 14$-$38\,$\mu$m) modules of the Infrared Spectrograph \citep[IRS;][]{houck04} on board the \textit{Spitzer Space Telescope}. For observations that were carried out in mapping mode, we only selected those frames where the spectrograph slit included the nucleus of the galaxy. The data were processed using standard spectroscopic techniques. We began by subtracting one nod position from the other to remove the sky background. All subsequent steps (extraction, wavelength calibration, and flux calibration) were carried out using the optimal extraction mode of the Spitzer IRS Custom Extractor (SPICE) tool (v2.5.1). The absolute flux calibration was checked against the photometric data. In the case of nearby ($<$50 Mpc) LIRGs we decided not to use IRS data as the field of view of IRS is much smaller than the angular size of the galaxies. The IRS spectra were binned to steps of $\log_{10}(\lambda_{\mathrm{rest}}/\mu\mathrm{m})=0.05$ to match more similarly the resolution of the radiative transfer models; a more relaxed match was applied around the complex regions of the polycyclic aromatic hydrocarbon (PAH) and the 9.7 $\mu$m silicate features. For each sample LIRG three sets of Monte Carlo fits were carried out both for the spheroidal and disc galaxy model cases, and the individual fit with the best reduced $\chi^{2}$ value was adopted to yield the starburst parameters for the system. Our global estimates of the CCSN rate and starburst properties of the sample galaxies are reported in Table~\ref{table:LIRGs} and an example SED fit of IRAS 18293-3413 is shown in Fig.~\ref{fig:SED_IRAS18293}. The best SED fits for all the other sample galaxies are shown in Appendix Figs.~\ref{fig:sed} and \ref{fig:sed2}. We note that the LIRG sample overlaps partly with that of \citet{herrero-illana17}, with some difference in the yielded starburst values with similar simulations. The key difference to those are the more extended parameter ranges allowed in our simulations for the starburst age (5 to 50 Myr) and the e-folding time of the starburst (1 to 40 Myr). While preparing this study we noticed that the starburst age range used in the simulations of Herrero-Illana et al. had an upper limit of 29.5 Myr, which resulted in many of their older starburst age estimates to artificially cluster near this value. However, the starburst age was not the focus of that study and does not change their conclusions. Similarly any differences in the derived parameters previously reported by \citet{mattila12} and \citet{mattila18} result from the usage of different data sets for their SED fitting covering separately individual components of Arp 299 with a more limited wavelength range rather than fitting the SED of the whole system. Thus, the results for the starburst age are not directly comparable.

Fig.~\ref{fig:grid} shows our result with the host LIRG starburst age comparison to the projected distance of the CCSN, with the different CCSN subtypes highlighted. Due to the still relatively small number of discovered CCSNe in LIRGs, we combine Type IIb, Ib and Ic SNe into a single group of H-poor IIb/Ib/Ic events. We see that the H-rich CCSNe around the central regions of these galaxies are produced in LIRGs with relatively older starburst ages ($\geq$30 Myr), while the H-poor CCSNe are produced primarily in younger starburst age regions. This is consistent with canonical stellar evolution models. In younger starbursts only the most massive stars have had time to evolve to the SN phase, and explode as H-poor CCSNe; in the older starbursts the most massive stars have already exploded and the lower mass stars have had more time to evolve to the SN stage, and end their life cycles as H-rich CCSNe \citep[e.g.][]{heger03}. However, rotation, metallicity, and binarity \citep[e.g.][]{podsiadlowski92, woosley19} all have an influence on this, and can reduce the relative initial progenitor star mass and life time differences between different CCSN subtypes. 

We carried out a two-sample Anderson-Darling (AD) statistical test for the CCSN subtype distribution. The \textit{p}-value for the two basic CCSN subtypes of H-poor Type IIb/Ib/Ic/Ibn (14 SNe) and H-rich Type II/IIn (15 SNe) in these galaxies having the same underlying distribution is only 0.0027, and thus the discussed trend is statistically significant at the 3.0\,$\sigma$ level. We also considered the case of only the LIRGs with the highest yielded SN rates of $\geq$0.5 CCSN yr$^{-1}$. Then the $p$-value of the two basic CCSN subtypes having the same underlying distribution is reduced to 0.0013 (3.2\,$\sigma$). We chose the AD test over other empirical distribution statistics-based tests (e.g., a Kolmogorov-Smirnov test) because of its generally powerful ability to distinguish between distributions that have the largest differences near the minima and maxima of their cumulative distribution \citep{stephens74}. As an additional AD test, we also compared our SN samples to a `flat' larger simulated comparison sample of a sufficiently large number (10) of SNe distributed uniformly in each of the corresponding sample galaxies. The resulting \textit{p}-value for the Type IIb/Ib/Ic/Ibn SNe being uniformly distributed among these galaxies is 0.011 (2.5\,$\sigma$). However, for the SNR $\geq$ 0.5 CCSN yr$^{-1}$ sample the \textit{p}-value is reduced to 0.00060 (3.4\,$\sigma$). Type II/IIn SNe are consistent with the uniform distribution in these comparisons with a \textit{p}-value of 0.14 (1.5\,$\sigma$) and 0.075 (1.8\,$\sigma$) for the $\geq$ 0.0 and $\geq$ 0.5 CCSN yr$^{-1}$ samples, respectively. Furthermore, we also consider cases with the sample groups defined as Type II and Type IIb/Ib/Ic SNe. Generally, these groups with reduced statistics appear to slightly increase the \textit{p}-values. The full presentation of the yielded values are listed in Table~\ref{table:AD}. The \textit{p}-values for the estimated probabilities of obtaining at least as extreme a difference as seen in the comparisons were converted to the corresponding standard deviation ($\sigma$) units $z$ of a half-normal distribution by solving $p/2 = (2\pi)^{-1/2} \int_{z}^{\infty} (e^{-x^{2}/2}) dx$. A half-normal distribution is warranted, as the AD test probes the absolute difference $|t_{\mathrm{SB},1} - t_{\mathrm{SB},2}|$ between distributions. We do note that there is no standard method to take the data point errors into account with the AD (or other similar) statistical test. 

The obvious caveat of the aforementioned trends is that the results are based on a relatively small sample of only 29 CCSNe and their 16 host LIRGs. Furthermore, the model dependencies and errors associated with the starburst age estimations are not taken into account in the significance calculations. This is a strong motivation to discover and study more CCSNe in the central regions of LIRGs to improve sample statistics. Removing an individual galaxy, and thus typically 1 or 2 CCSNe, from the studied samples will not result in major differences in the yielded \textit{p}-values. However, arbitrarily removing Arp 299 from the sample, and therefore 6 Type IIb/Ib SNe of the analysed samples would result in the disappearance of the discussed trends, see Fig.~\ref{fig:test}. For completeness, jackknife resampling averages of the \textit{p}-values are also listed in Table~\ref{table:AD} with the corresponding values of standard error of the mean.

\begin{table*}[!t]
\caption{AD statistical test results of \textit{p}-values and corresponding standard deviation ($\sigma$) units for our CCSN samples having the same underlying distributions of estimated global starburst age of the host. Additional tests include comparisons to a `flat' simulated sample, including galaxies which have hosted $\geq$1 CCSN from a selection of indicated types. Two CCSN rate ranges are considered. For completeness, the table also lists the jackknife resampling mean ($\theta_{p\mathrm{-value}}$) and standard error values, and the corresponding $\sigma$ units ($\theta_{\sigma}$), which are dominated by the arbitrary removal of Arp 299 and its CCSNe from the samples.} 
\centering
\begin{tabular}{llccclccc}
\hline
\hline
 & \multicolumn{4}{c}{ SNR $\geq$ 0.0} & \multicolumn{4}{c}{ SNR $\geq$ 0.5} \\
 & \textit{p}-value & $\sigma$ & $\theta_{p\mathrm{-value}}$ & $\theta_{\sigma}$ & \textit{p}-value & $\sigma$ & $\theta_{p\mathrm{-value}}$ & $\theta_{\sigma}$ \\
\hline
IIb/Ib/Ic/Ibn vs II/IIn & 0.0027 & 3.0 & 0.017 $\pm$ 0.014 & $2.4^{+0.6}_{-0.2}$ & 0.0013 & 3.2 & 0.040 $\pm$ 0.038 & $2.1^{+1.0}_{-0.3}$\\
IIb/Ib/Ic vs II/IIn & 0.0039 & 2.9 & 0.030 $\pm$ 0.025 & $2.2^{+0.6}_{-0.3}$ & 0.0017 & 3.1 & 0.078 $\pm$ 0.076 & $1.8^{+1.3}_{-0.4}$ \\
IIb/Ib/Ic vs II & 0.0059 & 2.8 & 0.032 $\pm$ 0.025 & $2.1^{+0.6}_{-0.2}$ & 0.0035 & 2.9 & 0.083 $\pm$ 0.078 & $1.7^{+1.1}_{-0.3}$ \\
IIb/Ib/Ic/Ibn vs flat$_{\mathrm{IIb/Ib/Ic/Ibn/II/IIn}}$ & 0.011 & 2.5 & 0.058 $\pm$ 0.046 & $1.9^{+0.6}_{-0.3}$ & 0.00060 & 3.4 & 0.070 $\pm$ 0.069 & $1.8^{+1.5}_{-1.5}$ \\
IIb/Ib/Ic vs flat$_{\mathrm{IIb/Ib/Ic/II/IIn}}$ & 0.015 & 2.4 & 0.071 $\pm$ 0.054 & $1.8^{+0.6}_{-0.3}$ & 0.00072 & 3.4 & 0.089 $\pm$ 0.088 & $1.7^{+1.6}_{-0.3}$\\
IIb/Ib/Ic vs flat$_{\mathrm{IIb/Ib/Ic/II}}$ & 0.022 & 2.3 & 0.083 $\pm$ 0.060 & $1.7^{+0.6}_{-0.2}$ & 0.0011 & 3.3 & 0.094 $\pm$ 0.093 & $1.7^{+1.6}_{-0.4}$ \\
II/IIn vs flat$_{\mathrm{IIb/Ib/Ic/Ibn/II/IIn}}$ & 0.14 & 1.5 & 0.16 $\pm$ 0.02 & $1.4^{+0.1}_{-0.1}$ & 0.075 & 1.8 & 0.10 $\pm$ 0.02 & $1.6^{+0.2}_{-0.1}$ \\
II/IIn vs flat$_{\mathrm{IIb/Ib/Ic/II/IIn}}$ & 0.24 & 1.2 & 0.27 $\pm$ 0.03 & $1.1^{+0.1}_{-0.1}$ & 0.15 & 1.4 & 0.20 $\pm$ 0.03 & $1.3^{+0.1}_{-0.1}$ \\
II vs flat$_{\mathrm{IIb/Ib/Ic/II}}$ & 0.28 & 1.1 & 0.32 $\pm$ 0.03 & $1.0^{+0.1}_{-0.1}$ & 0.24 & 1.2 & 0.29 $\pm$ 0.03 & $1.1^{+0.1}_{-0.1}$ \\
\hline
\end{tabular}
\label{table:AD}
\end{table*}

Our assumption was that very high CCSN rate LIRGs can have effectively dominating starburst episodes that produce the majority of CCSNe in these galaxies. However, even if these galaxies had one recent dominating starburst episode, this would not quench the underlying and ongoing star formation and CCSN production on a lower efficiency. Similarly, the sample LIRGs could also have other more minor non-dominating starburst episodes. Therefore, these galaxies can still produce any SN subtypes, though at a more moderate rate and on average less concentrated towards the central regions of their hosts.

The starburst ages ranging from $\sim$25 to 40 Myr, as found for LIRGs hosting primarily H-rich SNe in our sample, correspond to progenitor life times for zero-age main sequence (ZAMS) mass $M_{\mathrm{ZAMS}}$ of $\sim$11 to 8 M$_{\odot}$, respectively, according to the Geneva group single star stellar evolution tracks in Solar metallicity with rotation \citep{ekstrom12}. This is consistent with both theory and the direct red supergiant (RSG) progenitor detections of nearby Type IIP SNe \citep{smartt09}. Similarly, our estimate of the global starburst age of $16^{+2}_{-4}$ Myr for Arp 299 corresponds to progenitor masses of roughly 13 to 18 M$_{\odot}$ \citep{ekstrom12}. Intriguingly, the progenitors of many canonical Type IIb SNe are typically found within this range of initial mass, and in a binary system, e.g. 1993J \citep{nomoto93,podsiadlowski93,aldering94,woosley94}, 2008ax \citep{crockett08,folatelli15}, 2011dh \citep{maund11,vandyk11,ergon14}, 2013df \citep{morales-garoffolo14,vandyk14}, and 2016gkg \citep{bersten18}. From our sample of 6 central CCSNe in Arp 299, 50\,\% are Type IIb and 50\,\% are Type Ib events. \citet{kangas17} found consistent H$\alpha$ associations between Type IIb SNe and $M_{\mathrm{ZAMS}} \sim 15 M_{\odot}$ yellow super giants, and Type Ib SNe and $M_{\mathrm{ZAMS}} \sim 15 M_{\odot}$ RSGs. Here we find results consistent with those above, and note the advantage of using LIRGs (in comparison to normal spiral galaxies) for statistical studies as the starburst age can be more strictly constrained in these galaxies. 

\begin{figure}[!t]
\includegraphics[width=\linewidth]{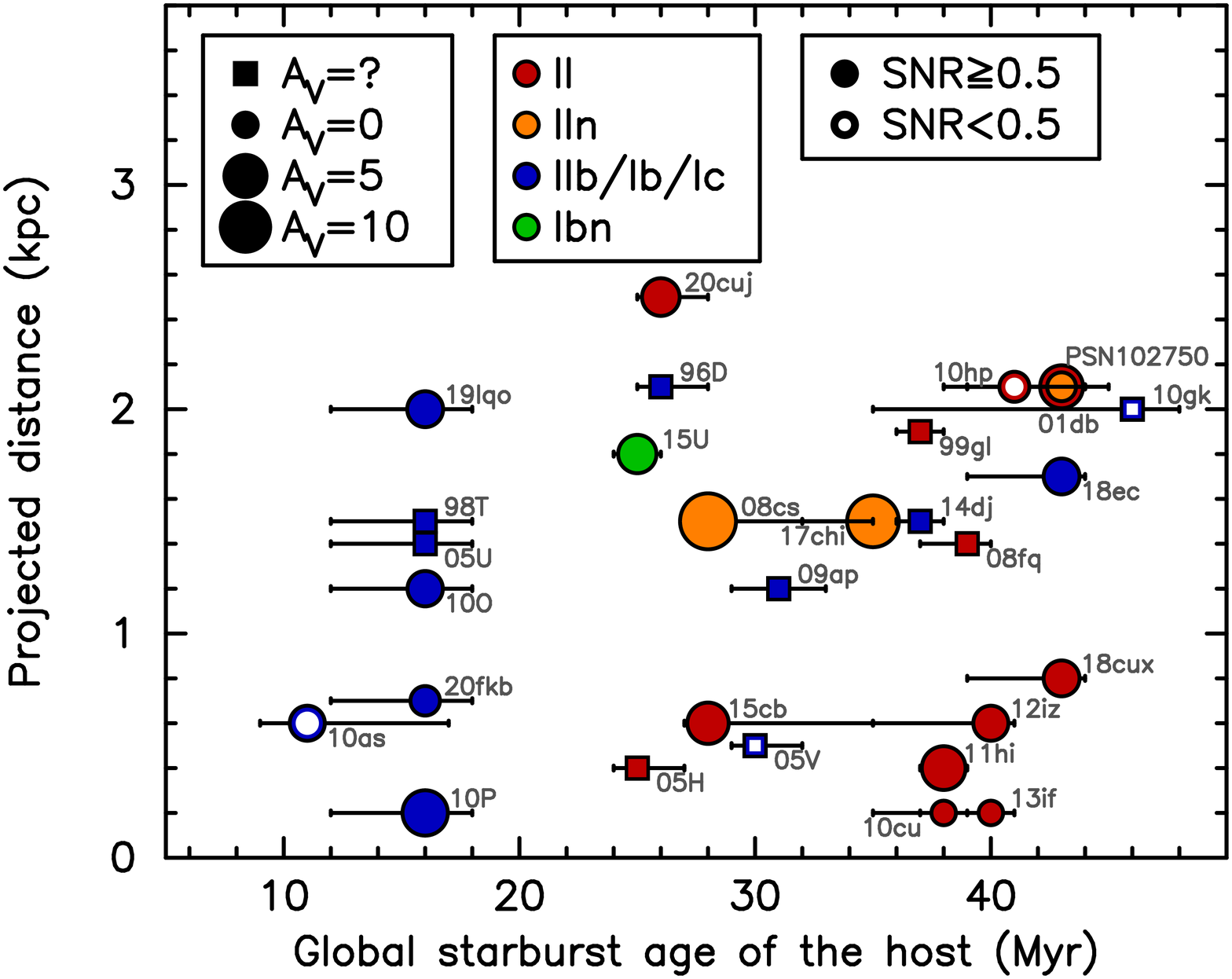}
\includegraphics[width=\linewidth]{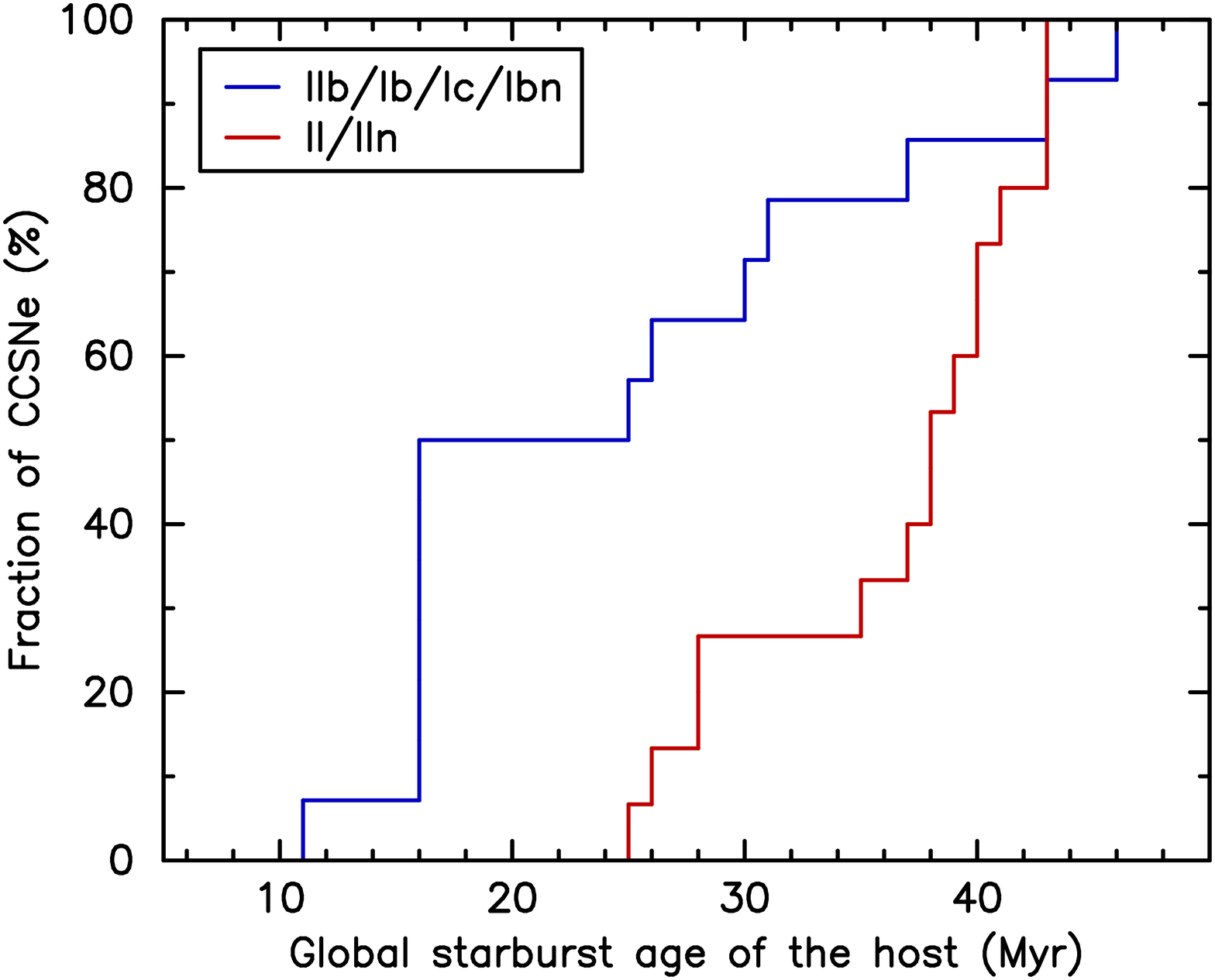}
\caption{Upper panel: Type distribution of classified CCSNe compared to our SED modelled global age of the host starburst at the central regions of our sample LIRGs. The sizes of the symbols indicate the derived host galaxy extinction of the SNe, with unknown (and likely small) extinction cases indicated with square symbols. The symbol colours indicate the CCSN subtype classifications. The events are marked with closed and open symbols based on their estimated host CCSN rate above and below 0.5 CCSN yr$^{-1}$, respectively. Lower panel: A simplified cumulative distribution presentation of the upper panel data of the Type IIb/Ib/Ic/Ibn and Type II/IIn samples of CCSNe. The data suggests that H-poor and H-rich CCSNe come from very young ($\lesssim$30 Myr) and older ($\gtrsim$30 Myr) starbursts, respectively.}
\label{fig:grid}
\end{figure}

\citet{anderson11} and \citet{anderson13} highlight the tendency of the Arp 299 system to produce Type Ib to Type IIb SNe, compared to their normal fraction of all CCSNe in the local Universe. This relative excess has been discussed to rise either from young age of the recent star formation or from a top-heavy IMF \citep{anderson11}. The recent SNe 2019lqo and 2020fkb follow the trend of Type Ib/IIb SNe in Arp 299, which we argue to be consistent with the dominating episode of recent star formation with a normal (Salpeter) IMF adopted. \citet{perez-torres09} reported the radio detection of 20 SNe or SN remnants within the $<$150 pc nuclear region of the nucleus of Arp 299-A, including three young radio SNe. They concluded that the properties of the young SNe are consistent with those of either Type IIb or normal Type II SNe. If Arp 299 produces predominantly Type Ib/IIb SNe, this would suggest that these recent SNe are predominantly Type IIb SNe. For comparison, \citet{alonso-herrero00} inferred a starburst age of $\sim$11 Myr for the Arp 299-A component that dominates the star formation in the Arp 299 system. Within errors, this is consistent with our global estimate for this LIRG. For the other less dominating nuclei (B, C, C$'$) they find indications of a somewhat younger starburst age, in particular in nuclei C$'$ and C of $\sim$4 to 5 Myr, respectively; this implies that only very massive stars in the $\sim$50 to 90 $M_{\odot}$ range are expected to explode as CCSNe. For a Salpeter IMF, the fraction of such massive stars is fairly insignificant, and thus the most recent star formation episode could not have contributed significantly to the observed CCSN rate and population in Arp 299.

\begin{figure}[!t]
\includegraphics[width=\linewidth]{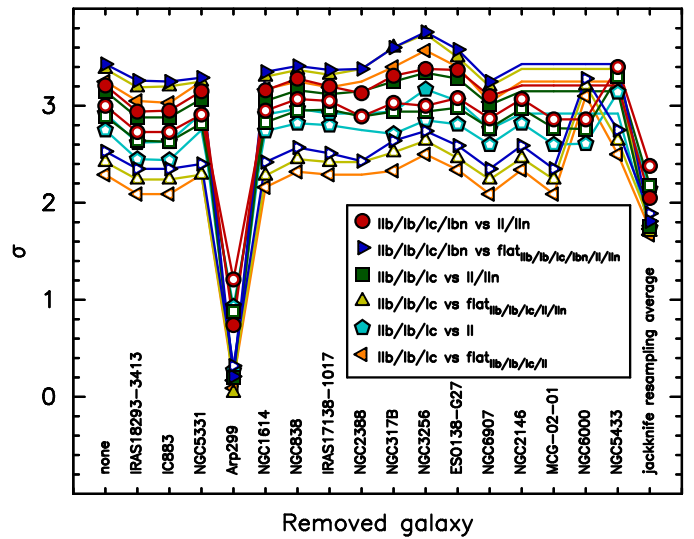}
\caption{Yielded $\sigma$ values from a selection of statistical tests for the sample with individual galaxies and their sample CCSNe removed from the analysis as a test. While removing individual galaxies and thus typically only 1 or 2 CCSNe from the sample will not have a major effect on the resulting AD test values, the arbitrary removal of Arp 299 and its 6 Type IIb/Ib SNe would make the discussed trends nonexistent. Open and closed symbols show the values for samples with the host SNR of $\geq$ 0.0 and $\geq$ 0.5 CCSN yr$^{-1}$, respectively. Data points are not shown in cases where the removal of the galaxy is irrelevant for the comparison between the analysed samples. With the large number of hosted sample CCSNe, the removal of the LIRG Arp 299 with a yielded young starburst age has also the largest weight in the jackknife resampling averages of the \textit{p}-values for each set of comparisons, and result in mean values of $\theta_{p} \geq 0.017$ ($\leq$ 2.4\,$\sigma$).}
\label{fig:test}
\end{figure}

\section{Conclusions}

We discovered a new Type Ic SN 2018ec in the LIRG NGC 3256 during the ESO science verification run of the adaptive optics seeing enhancer instrument HAWK-I/GRAAL. While carrying out follow up of this SN, we discovered another transient, AT 2018cux, in the same galaxy that was a subluminous type IIP SN. We derive the host galaxy extinctions of $A_{V} = 2.1^{+0.3}_{-0.1}$ and $2.1 \pm 0.4$ mag for SN 2018ec and AT 2018cux, respectively. A SN candidate from 2014, PSN102750, in the same galaxy was found to be consistent with H-rich CCSN with a fairly low host galaxy extinction. A best fit for PSN102750 is provided by a Type IIn using a combination of early ground-based follow-up and late-time colours from HST; however, we cannot fully rule out a Type IIP origin. We also report follow up of Type IIb SN 2019lqo and Type Ib SN 2020fkb in Arp 299 and derive a line-of-sight host extinction of $A_{V} = 2.1^{+0.1}_{-0.3}$ and $0.4^{+0.1}_{-0.2}$ mag for them, respectively. 

We find evidence that the CCSN subtypes occurring within $\sim$2.5 kpc of the nucleus of LIRGs correlate with the inferred starburst age of the host and are consistent with a normal (Salpeter) IMF. In other words, the H-rich (Type II/IIn) and H-poor (Type IIb/Ib/Ic/Ibn) CCSN progenitors have different underlying age distributions in these galaxies at a 3\,$\sigma$ significance level. However, the overall trend was drawn from a relatively small sample size of CCSNe and their host galaxies, and the significance comparisons between subgroups do not specifically take into account error ranges of the yielded starburst ages. Hence, future surveys, in particular in high spatial resolution IR, will be necessary to discover and further study CCSNe in LIRGs. 

We predict, that LIRGs that have hosted multiple discovered circumnuclear CCSNe in their central regions will continue to produce (predominantly) the subtypes that they have already shown to favour in their nuclear regions. New CCSN discoveries in these galaxies will constrain more accurately a typical projected distance limit for this trend, or alternatively a possible effective radius limit that should be adopted. At the moment, the very small number statistics of the currently available sample of SNe in these galaxies do not allow for more sophisticated constraints. Furthermore, with improved statistics, the discussed trend offers a possible new method to probe the progenitor systems of different CCSN types by using high-IR-luminosity LIRGs as laboratories.

\begin{acknowledgements}

We thank the anonymous referee for useful comments. 

We thank Marco Fiaschi for carrying out some of the Asiago observations.

EK is supported by the Turku Collegium of Science, Medicine and Technology. EK also acknowledge support from the Science and Technology Facilities Council (STFC; ST/P000312/1).

ECK acknowledges support from the G.R.E.A.T. research environment and support from The Wenner-Gren Foundations.

MF is supported by a Royal Society - Science Foundation Ireland University Research Fellowship.

EC, LT, AP, and MT are partially supported by the PRIN-INAF 2017 with the project ``Towards the SKA and CTA era: discovery, localization, and physics of transient objects''.

HK was funded by the Academy of Finland projects 324504 and 328898.

TWC acknowledges the EU Funding under Marie Sk\l{}odowska-Curie grant agreement No 842471.

LG was funded by the European Union's Horizon 2020 research and innovation programme under the Marie Sk\l{}odowska-Curie grant agreement No. 839090. This work has been partially supported by the Spanish grant PGC2018-095317-B-C21 within the European Funds for Regional Development (FEDER).

MG is supported by the Polish NCN MAESTRO grant 2014/14/A/ST9/00121.

KM acknowledges support from EU H2020 ERC grant no. 758638.

TMB was funded by the CONICYT PFCHA / DOCTORADOBECAS CHILE/2017-72180113.

MN is supported by a Royal Astronomical Society Research Fellowship.

Based on observations collected at the European Organisation for Astronomical Research in the Southern Hemisphere under ESO programmes 67.D-0438, 60.A-9475, 199.D-0143, and 1103.D-0328.

Some of the observations reported in this paper were obtained with the Southern African Large Telescope (SALT) under programme 2018-1-DDT-003 (PI: Kankare).

Polish participation in SALT is funded by grant No. MNiSW DIR/WK/2016/07.

Based on observations made with the Nordic Optical Telescope, operated by the Nordic Optical Telescope Scientific Association at the Observatorio del Roque de los Muchachos, La Palma, Spain, of the Instituto de Astrofisica de Canarias.

The data presented here were obtained in part with ALFOSC, which is provided by the Instituto de Astrofisica de Andalucia (IAA) under a joint agreement with the University of Copenhagen and NOTSA.

This work is partly based on the NUTS2 programme carried out at the NOT. NUTS2 is funded in part by the Instrument Center for Danish Astrophysics (IDA).

The Liverpool Telescope is operated on the island of La Palma by Liverpool John Moores University in the Spanish Observatorio del Roque de los Muchachos of the Instituto de Astrofisica de Canarias with financial support from the UK Science and Technology Facilities Council.

This paper is also based on observations collected at the Copernico 1.82 m and Schmidt 67/92 Telescopes operated by INAF -- Osservatorio Astronomico di Padova at Asiago, Italy. 

Based on observations obtained at the Gemini Observatory, which is operated by the Association of Universities for Research in Astronomy, Inc., under a cooperative agreement with the NSF on behalf of the Gemini partnership: the National Science Foundation (United States), the National Research Council (Canada), CONICYT (Chile), Ministerio de Ciencia, Tecnolog\'{i}a e Innovaci\'{o}n Productiva (Argentina), and Minist\'{e}rio da Ci\^{e}ncia, Tecnologia e Inova\c{c}\~{a}o (Brazil). Observations were carried out under programme GS-2017A-C-1.

This project used data obtained with the Dark Energy Camera (DECam), which was constructed by the Dark Energy Survey (DES) collaboration. Funding for the DES Projects has been provided by the DOE and NSF (USA), MISE (Spain), STFC (UK), HEFCE (UK), NCSA (UIUC), KICP (U. Chicago), CCAPP (Ohio State), MIFPA (Texas A\&M University), CNPQ, FAPERJ, FINEP (Brazil), MINECO (Spain), DFG (Germany) and the collaborating institutions in the Dark Energy Survey, which are Argonne Lab, UC Santa Cruz, University of Cambridge, CIEMAT-Madrid, University of Chicago, University College London, DES-Brazil Consortium, University of Edinburgh, ETH Z{\"u}rich, Fermilab, University of Illinois, ICE (IEEC-CSIC), IFAE Barcelona, Lawrence Berkeley Lab, LMU M{\"u}nchen and the associated Excellence Cluster Universe, University of Michigan, NOAO, University of Nottingham, Ohio State University, OzDES Membership Consortium, University of Pennsylvania, University of Portsmouth, SLAC National Lab, Stanford University, University of Sussex, and Texas A\&M University.

Based on observations obtained with the Samuel Oschin 48-inch Telescope at the Palomar Observatory as part of the Zwicky Transient Facility project. ZTF is supported by the National Science Foundation under Grant No. AST-1440341 and a collaboration including Caltech, IPAC, the Weizmann Institute for Science, the Oskar Klein Center at Stockholm University, the University of Maryland, the University of Washington, Deutsches Elektronen-Synchrotron and Humboldt University, Los Alamos National Laboratories, the TANGO Consortium of Taiwan, the University of Wisconsin at Milwaukee, and Lawrence Berkeley National Laboratories. Operations are conducted by COO, IPAC, and UW. 

Based on observations at Cerro Tololo Inter-American Observatory, National Optical Astronomy Observatory (NOAO Prop. ID 2017A-0260; and PI: Soares-Santos), which is operated by the Association of Universities for Research in Astronomy (AURA) under a cooperative agreement with the National Science Foundation.

The Pan-STARRS1 Surveys (PS1) and the PS1 public science archive have been made possible through contributions by the Institute for Astronomy, the University of Hawaii, the Pan-STARRS Project Office, the Max-Planck Society and its participating institutes, the Max Planck Institute for Astronomy, Heidelberg and the Max Planck Institute for Extraterrestrial Physics, Garching, The Johns Hopkins University, Durham University, the University of Edinburgh, the Queen's University Belfast, the Harvard-Smithsonian Center for Astrophysics, the Las Cumbres Observatory Global Telescope Network Incorporated, the National Central University of Taiwan, the Space Telescope Science Institute, the National Aeronautics and Space Administration under Grant No. NNX08AR22G issued through the Planetary Science Division of the NASA Science Mission Directorate, the National Science Foundation Grant No. AST-1238877, the University of Maryland, Eotvos Lorand University (ELTE), the Los Alamos National Laboratory, and the Gordon and Betty Moore Foundation.

Some of the data presented in this paper were obtained from the Mikulski Archive for Space Telescopes (MAST). STScI is operated by the Association of Universities for Research in Astronomy, Inc., under NASA contract NAS5-26555. 

This work is based in part on archival data obtained with the Spitzer Space Telescope, which is operated by the Jet Propulsion Laboratory, California Institute of Technology under a contract with NASA.

This research has made use of NED which is operated by the Jet Propulsion Laboratory, California Institute of Technology, under contract with the National Aeronautics and Space Administration. 

We have made use of the Weizmann Interactive Supernova Data Repository \citep[][\urlwofont{https://wiserep.weizmann.ac.il}]{yaron12}.

\end{acknowledgements}


\begin{thebibliography}{}

\bibitem[Alard \& Lupton(1998)]{alard98} Alard, C., \& Lupton, R.~H.\ 1998, \apj, 503, 325 
\bibitem[Alard(2000)]{alard00} Alard, C.\ 2000, \aaps, 144, 363 
\bibitem[Aldering et al.(1994)]{aldering94} Aldering, G., Humphreys, R.~M., \& Richmond, M.\ 1994, \aj, 107, 662
\bibitem[Alonso-Herrero et al.(2000)]{alonso-herrero00} Alonso-Herrero, A., Rieke, G.~H., Rieke, M.~J., et al.\ 2000, \apj, 532, 845
\bibitem[Alonso-Herrero et al.(2006)]{alonso-herrero06} Alonso-Herrero, A., Rieke, G.~H., Rieke, M.~J., et al.\ 2006, \apj, 650, 835
\bibitem[Anderson \& Soto(2013)]{anderson13} Anderson, J.~P., \& Soto, M.\ 2013, \aap, 550, A69
\bibitem[Anderson et al.(2011)]{anderson11} Anderson, J.~P., Habergham, S.~M., \& James, P.~A.\ 2011, \mnras, 416, 567
\bibitem[Anderson et al.(2012)]{anderson12} Anderson, J.~P., Habergham, S.~M., James, P.~A., et al.\ 2012, \mnras, 424, 1372
\bibitem[Anderson et al.(2014)]{anderson14} Anderson, J.~P., Gonz{\'a}lez-Gait{\'a}n, S., Hamuy, M., et al.\ 2014, \apj, 786, 67 
\bibitem[Aramyan et al.(2016)]{aramyan16} Aramyan, L.~S., Hakobyan, A.~A., Petrosian, A.~R., et al.\ 2016, \mnras, 459, 3130
\bibitem[Arcavi et al.(2011)]{arcavi11} Arcavi, I., Gal-Yam, A., Yaron, O., et al.\ 2011, \apjl, 742, L18
\bibitem[Audcent-Ross et al.(2020)]{audcent-ross20} Audcent-Ross, F.~M., Meurer, G.~R., Audcent, J.~R., et al.\ 2020, \mnras, 492, 848
\bibitem[Barnsley et al.(2016)]{barnsley16} Barnsley, R.~M., Jermak, H.~E., Steele, I.~A., et al.\ 2016, Journal of Astronomical Telescopes, Instruments, and Systems, 2, 015002
\bibitem[Bellm et al.(2019)]{bellm19} Bellm, E.~C., Kulkarni, S.~R., Graham, M.~J., et al.\ 2019, \pasp, 131, 018002
\bibitem[Bersten et al.(2018)]{bersten18} Bersten, M.~C., Folatelli, G., Garc{\'\i}a, F., et al.\ 2018, \nat, 554, 497
\bibitem[Berton et al.(2018)]{berton18} Berton, M., Bufano, F., Vogl, C., et al.\ 2018, The Astronomer's Telegram, 11160, 1
\bibitem[Bianco et al.(2014)]{bianco14} Bianco, F.~B., Modjaz, M., Hicken, M., et al.\ 2014, \apjs, 213, 19
\bibitem[Bose et al.(2019)]{bose19} Bose, S., Holmbo, S., Mattila, S., et al.\ 2019, Transient Name Server Classification Report 2019-1332, 1
\bibitem[Botticella et al.(2009)]{botticella09} Botticella, M.~T., Pastorello, A., Smartt, S.~J., et al.\ 2009, \mnras, 398, 1041 
\bibitem[Bruzual \& Charlot(1993)]{bruzual93} Bruzual, G., \& Charlot, S.\ 1993, \apj, 405, 538 
\bibitem[Bruzual \& Charlot(2003)]{bruzual03} Bruzual, G., \& Charlot, S.\ 2003, \mnras, 344, 1000 
\bibitem[Buckley et al.(2006)]{buckley06} Buckley, D.~A.~H., Swart, G.~P., \& Meiring, J.~G.\ 2006, \procspie, 6267, 62670Z 
\bibitem[Burgh et al.(2003)]{burgh03} Burgh, E.~B., Nordsieck, K.~H., Kobulnicky, H.~A., et al.\ 2003, \procspie, 4841, 1463 
\bibitem[Buzzoni et al.(1984)]{buzzoni84} Buzzoni, B., Delabre, B., Dekker, H., et al.\ 1984, The Messenger, 38, 9 
\bibitem[Cardelli et al.(1989)]{cardelli89} Cardelli, J.~A., Clayton, G.~C., \& Mathis, J.~S.\ 1989, \apj, 345, 245 
\bibitem[Chambers et al.(2016)]{chambers16} Chambers, K.~C., Magnier, E.~A., Metcalfe, N., et al.\ 2016, arXiv e-prints, arXiv:1612.05560
\bibitem[Chu et al.(2017)]{chu17} Chu, J.~K., Sanders, D.~B., Larson, K.~L., et al.\ 2017, \apjs, 229, 25 
\bibitem[Crockett et al.(2008)]{crockett08} Crockett, R.~M., Eldridge, J.~J., Smartt, S.~J., et al.\ 2008, \mnras, 391, L5
\bibitem[Crowther(2013)]{crowther13} Crowther, P.~A.\ 2013, \mnras, 428, 1927
\bibitem[Dahiwale \& Fremling(2020)]{dahiwale20} Dahiwale, A., \& Fremling, C.\ 2020, Transient Name Server Classification Report 2020-927, 1
\bibitem[De(2020)]{de20} De, K.\ 2020, Transient Name Server Discovery Report 2020-542, 1
\bibitem[Efstathiou \& Rowan-Robinson(1995)]{efstathiou95} Efstathiou, A., \& Rowan-Robinson, M.\ 1995, \mnras, 273, 649
\bibitem[Efstathiou et al.(2000)]{efstathiou00} Efstathiou, A., Rowan-Robinson, M., \& Siebenmorgen, R.\ 2000, \mnras, 313, 734 
\bibitem[Efstathiou \& Rowan-Robinson(2003)]{efstathiou03} Efstathiou, A., \& Rowan-Robinson, M.\ 2003, \mnras, 343, 322 
\bibitem[Efstathiou \& Siebenmorgen(2009)]{efstathiou09} Efstathiou, A., \& Siebenmorgen, R.\ 2009, \aap, 502, 541 
\bibitem[Efstathiou et al.(2013)]{efstathiou13} Efstathiou, A., Christopher, N., Verma, A., et al.\ 2013, \mnras, 436, 1873
\bibitem[Ekstr{\"o}m et al.(2012)]{ekstrom12} Ekstr{\"o}m, S., Georgy, C., Eggenberger, P., et al.\ 2012, \aap, 537, A146
\bibitem[Emonts et al.(2014)]{emonts14} Emonts, B.~H.~C., Piqueras-L{\'o}pez, J., Colina, L., et al.\ 2014, \aap, 572, A40
\bibitem[Ergon et al.(2014)]{ergon14} Ergon, M., Sollerman, J., Fraser, M., et al.\ 2014, \aap, 562, A17 
\bibitem[Ergon et al.(2015)]{ergon15} Ergon, M., Jerkstrand, A., Sollerman, J., et al.\ 2015, \aap, 580, A142
\bibitem[Fassia et al.(2000)]{fassia00} Fassia, A., Meikle, W.~P.~S., Vacca, W.~D., et al.\ 2000, \mnras, 318, 1093 
\bibitem[Fenech et al.(2008)]{fenech08} Fenech, D.~M., Muxlow, T.~W.~B., Beswick, R.~J., et al.\ 2008, \mnras, 391, 1384
\bibitem[Flaugher et al.(2015)]{flaugher15} Flaugher, B., Diehl, H.~T., Honscheid, K., et al.\ 2015, \aj, 150, 150 
\bibitem[Folatelli et al.(2004)]{folatelli04} Folatelli, G., Hamuy, M., Morrell, N., et al.\ 2004, \iaucirc, 8447
\bibitem[Folatelli et al.(2014)]{folatelli14} Folatelli, G., Bersten, M.~C., Kuncarayakti, H., et al.\ 2014, \apj, 792, 7
\bibitem[Folatelli et al.(2015)]{folatelli15} Folatelli, G., Bersten, M.~C., Kuncarayakti, H., et al.\ 2015, \apj, 811, 147
\bibitem[Galbany et al.(2018)]{galbany18} Galbany, L., Anderson, J.~P., S{\'a}nchez, S.~F., et al.\ 2018, \apj, 855, 107
\bibitem[Gehrz et al.(1983)]{gehrz83} Gehrz, R.~D., Sramek, R.~A., \& Weedman, D.~W.\ 1983, \apj, 267, 551
\bibitem[Heger et al.(2003)]{heger03} Heger, A., Fryer, C.~L., Woosley, S.~E., et al.\ 2003, \apj, 591, 288
\bibitem[Herrero-Illana et al.(2017)]{herrero-illana17} Herrero-Illana, R., P{\'e}rez-Torres, M.~{\'A}., Randriamanakoto, Z., et al.\ 2017, \mnras, 471, 1634 
\bibitem[Hodgkin et al.(2019)]{hodgkin19} Hodgkin, S.~T., Breedt, E., Delgado, A., et al.\ 2019, Transient Name Server Discovery Report 2019-1300, 1
\bibitem[Houck et al.(2004)]{houck04} Houck, J.~R., Roellig, T.~L., van Cleve, J., et al.\ 2004, \apjs, 154, 18
\bibitem[Hunter et al.(2009)]{hunter09} Hunter, D.~J., Valenti, S., Kotak, R., et al.\ 2009, \aap, 508, 371 
\bibitem[Huo et al.(2004)]{huo04} Huo, Z.~Y., Xia, X.~Y., Xue, S.~J., et al.\ 2004, \apj, 611, 208
\bibitem[Jencson et al.(2017)]{jencson17} Jencson, J.~E., Kasliwal, M.~M., Johansson, J., et al.\ 2017, \apj, 837, 167 
\bibitem[Jencson et al.(2018)]{jencson18} Jencson, J.~E., Kasliwal, M.~M., Adams, S.~M., et al.\ 2018, \apj, 863, 20 
\bibitem[Jencson et al.(2019)]{jencson19} Jencson, J.~E., Kasliwal, M.~M., Adams, S.~M., et al.\ 2019, \apj, 886, 40
\bibitem[Jester et al.(2005)]{jester05} Jester, S., Schneider, D.~P., Richards, G.~T., et al.\ 2005, \aj, 130, 873 
\bibitem[Johnson et al.(2013)]{johnson13} Johnson, S.~P., Wilson, G.~W., Tang, Y., \& Scott, K.~S.\ 2013, \mnras, 436, 2535 
\bibitem[Kangas et al.(2017)]{kangas17} Kangas, T., Portinari, L., Mattila, S., et al.\ 2017, \aap, 597, A92
\bibitem[Kankare et al.(2008)]{kankare08} Kankare, E., Mattila, S., Ryder, S., et al.\ 2008, \apjl, 689, L97 
\bibitem[Kankare et al.(2009)]{kankare09} Kankare, E., Hanski, M., Theureau, G., \& Teerikorpi, P.\ 2009, \aap, 493, 23 
\bibitem[Kankare et al.(2012)]{kankare12} Kankare, E., Mattila, S., Ryder, S., et al.\ 2012, \apjl, 744, L19 
\bibitem[Kankare et al.(2014a)]{kankare14a} Kankare, E., Fraser, M., Ryder, S., et al.\ 2014a, \aap, 572, A75 
\bibitem[Kankare et al.(2014b)]{kankare14b} Kankare, E., Mattila, S., Ryder, S., et al.\ 2014b, \mnras, 440, 1052 
\bibitem[Kankare et al.(2018a)]{kankare18a} Kankare, E., Mattila, S., Ryder, S., et al.\ 2018a, The Astronomer's Telegram, 11156, 1
\bibitem[Kankare et al.(2018b)]{kankare18b} Kankare, E., Taubenberger, S., Vogl, C., et al.\ 2018b, The Astronomer's Telegram, 11778, 1  
\bibitem[Kissler-Patig et al.(2008)]{kissler08} Kissler-Patig, M., Pirard, J.-F., Casali, M., et al.\ 2008, \aap, 491, 941 
\bibitem[Kobulnicky et al.(2003)]{kobulnicky03} Kobulnicky, H.~A., Nordsieck, K.~H., Burgh, E.~B., et al.\ 2003, \procspie, 4841, 1634 
\bibitem[Kool et al.(2018)]{kool18} Kool, E.~C., Ryder, S., Kankare, E., et al.\ 2018, \mnras, 473, 5641 
\bibitem[Kool(2019)]{kool19} Kool, E.~C.\ 2019, Ph.D. Thesis
\bibitem[Kotilainen et al.(1996)]{kotilainen96} Kotilainen, J.~K., Moorwood, A.~F.~M., Ward, M.~J., et al.\ 1996, \aap, 305, 107
\bibitem[Kuncarayakti et al.(2015)]{kuncarayakti15} Kuncarayakti, H., Maeda, K., Bersten, M.~C., et al.\ 2015, \aap, 579, A95
\bibitem[Kuncarayakti et al.(2018)]{kuncarayakti18} Kuncarayakti, H., Anderson, J.~P., Galbany, L., et al.\ 2018, \aap, 613, A35
\bibitem[Leloudas et al.(2011)]{leloudas11} Leloudas, G., Gallazzi, A., Sollerman, J., et al.\ 2011, \aap, 530, A95
\bibitem[Liu et al.(2000)]{liu00} Liu, Q.-Z., Hu, J.-Y., Hang, H.-R., et al.\ 2000, \aaps, 144, 219
\bibitem[Madau \& Dickinson(2014)]{madau14} Madau, P., \& Dickinson, M.\ 2014, \araa, 52, 415
\bibitem[Magnelli et al.(2011)]{magnelli11} Magnelli, B., Elbaz, D., Chary, R.~R., et al.\ 2011, \aap, 528, A35
\bibitem[Maiolino et al.(2002)]{maiolino02} Maiolino, R., Vanzi, L., Mannucci, F., et al.\ 2002, \aap, 389, 84 
\bibitem[Masci et al.(2019)]{masci19} Masci, F.~J., Laher, R.~R., Rusholme, B., et al.\ 2019, \pasp, 131, 995
\bibitem[Matheson et al.(2000)]{matheson00} Matheson, T., Filippenko, A.~V., Ho, L.~C., et al.\ 2000, \aj, 120, 1499
\bibitem[Mattila \& Meikle(2001)]{mattila01} Mattila, S., \& Meikle, W.~P.~S.\ 2001, \mnras, 324, 325 
\bibitem[Mattila et al.(2007)]{mattila07} Mattila, S., V{\"a}is{\"a}nen, P., Farrah, D., et al.\ 2007, \apjl, 659, L9 
\bibitem[Mattila et al.(2012)]{mattila12} Mattila, S., Dahlen, T., Efstathiou, A., et al.\ 2012, \apj, 756, 111 
\bibitem[Mattila et al.(2013)]{mattila13} Mattila, S., Fraser, M., Smartt, S.~J., et al.\ 2013, \mnras, 431, 2050
\bibitem[Mattila et al.(2018)]{mattila18} Mattila, S., P{\'e}rez-Torres, M., Efstathiou, A., et al.\ 2018, Science, 361, 482
\bibitem[Maund et al.(2011)]{maund11} Maund, J.~R., Fraser, M., Ergon, M., et al.\ 2011, \apjl, 739, L37
\bibitem[Miluzio et al.(2013)]{miluzio13} Miluzio, M., Cappellaro, E., Botticella, M.~T., et al.\ 2013, \aap, 554, A127
\bibitem[Modjaz et al.(2014)]{modjaz14} Modjaz, M., Blondin, S., Kirshner, R.~P., et al.\ 2014, \aj, 147, 99
\bibitem[Moorwood et al.(1998)]{moorwood98} Moorwood, A., Cuby, J.-G., \& Lidman, C.\ 1998, The Messenger, 91, 9 
\bibitem[Morales-Garoffolo et al.(2014)]{morales-garoffolo14} Morales-Garoffolo, A., Elias-Rosa, N., Benetti, S., et al.\ 2014, \mnras, 445, 1647
\bibitem[Mould et al.(2000)]{mould00} Mould, J.~R., Huchra, J.~P., Freedman, W.~L., et al.\ 2000, \apj, 529, 786
\bibitem[M{\"u}ller Bravo et al.(2019)]{muller-bravo19} M{\"u}ller Bravo, T., Wiseman, P., Pursiainen, M., et al.\ 2019, Transient Name Server AstroNote, 155
\bibitem[M{\"u}ller-Bravo et al.(2020)]{muller-bravo20} M{\"u}ller-Bravo, T.~E., Guti{\'e}rrez, C.~P., Sullivan, M., et al.\ 2020, \mnras, 497, 361
\bibitem[Munari et al.(2013)]{munari13} Munari, U., Henden, A., Belligoli, R., et al.\ 2013, \na, 20, 30
\bibitem[Nayana \& Chandra(2018)]{nayana18} Nayana, A.~J., \& Chandra, P.\ 2018, The Astronomer's Telegram, 11350, 1
\bibitem[Neff et al.(2003)]{neff03} Neff, S.~G., Ulvestad, J.~S., \& Campion, S.~D.\ 2003, \apj, 599, 1043
\bibitem[Nomoto et al.(1993)]{nomoto93} Nomoto, K., Suzuki, T., Shigeyama, T., et al.\ 1993, \nat, 364, 507
\bibitem[Ohyama et al.(2015)]{ohyama15} Ohyama, Y., Terashima, Y., \& Sakamoto, K.\ 2015, \apj, 805, 162
\bibitem[Parra et al.(2007)]{parra07} Parra, R., Conway, J.~E., Diamond, P.~J., et al.\ 2007, \apj, 659, 314
\bibitem[Pastorello et al.(2004)]{pastorello04} Pastorello, A., Zampieri, L., Turatto, M., et al.\ 2004, \mnras, 347, 74 
\bibitem[Pastorello et al.(2009)]{pastorello09} Pastorello, A., Valenti, S., Zampieri, L., et al.\ 2009, \mnras, 394, 2266 
\bibitem[Pastorello et al.(2015)]{pastorello15} Pastorello, A., Tartaglia, L., Elias-Rosa, N., et al.\ 2015, \mnras, 454, 4293
\bibitem[Paufique et al.(2010)]{paufique10} Paufique, J., Bruton, A., Glindemann, A., et al.\ 2010, \procspie, 77361P
\bibitem[P{\'e}rez-Torres et al.(2009)]{perez-torres09} P{\'e}rez-Torres, M.~A., Romero-Ca{\~n}izales, C., Alberdi, A., et al.\ 2009, \aap, 507, L17
\bibitem[Pignata et al.(2020)]{pignata20} Pignata, G., Bauer, F.~E., Forster, F., et al.\ 2020, Transient Name Server Discovery Report 2020-909, 1
\bibitem[Podsiadlowski et al.(1992)]{podsiadlowski92} Podsiadlowski, P., Joss, P.~C., \& Hsu, J.~J.~L.\ 1992, \apj, 391, 246
\bibitem[Podsiadlowski et al.(1993)]{podsiadlowski93} Podsiadlowski, P., Hsu, J.~J.~L., Joss, P.~C., et al.\ 1993, \nat, 364, 509
\bibitem[Prentice \& Mazzali(2017)]{prentice17} Prentice, S.~J. \& Mazzali, P.~A.\ 2017, \mnras, 469, 2672
\bibitem[Pressberger et al.(1993)]{pressberger93} Pressberger, R., Maitzen, H.~M., \& Neely, A.~W.\ 1993, \iaucirc, 5832, 2 
\bibitem[Prieto et al.(2008)]{prieto08} Prieto, J.~L., Kistler, M.~D., Thompson, T.~A., et al.\ 2008, \apjl, 681, L9 
\bibitem[Randriamanakoto et al.(2019)]{randriamanakoto19} Randriamanakoto, Z., V{\"a}is{\"a}nen, P., Ryder, S.~D., et al.\ 2019, \mnras, 482, 2530
\bibitem[Richmond et al.(1994)]{richmond94} Richmond, M.~W., Treffers, R.~R., Filippenko, A.~V., et al.\ 1994, \aj, 107, 1022 
\bibitem[Richmond et al.(1996)]{richmond96} Richmond, M.~W., Treffers, R.~R., Filippenko, A.~V., et al.\ 1996, \aj, 112, 732
\bibitem[Ripero et al.(1993)]{ripero93} Ripero, J., Garcia, F., Rodriguez, D., et al.\ 1993, \iaucirc, 5731, 1 
\bibitem[Romero-Ca{\~n}izales et al.(2011)]{romero-canizales11} Romero-Ca{\~n}izales, C., Mattila, S., Alberdi, A., et al.\ 2011, \mnras, 415, 2688
\bibitem[Romero-Ca{\~n}izales et al.(2014)]{romero-canizales14} Romero-Ca{\~n}izales, C., Herrero-Illana, R., P{\'e}rez-Torres, M.~A., et al.\ 2014, \mnras, 440, 1067
\bibitem[Ryder et al.(2018)]{ryder18} Ryder, S., Kool, E., Kankare, E., et al.\ 2018, The Astronomer's Telegram, 11224, 1
\bibitem[Sahu et al.(2006)]{sahu06} Sahu, D.~K., Anupama, G.~C., Srividya, S., et al.\ 2006, \mnras, 372, 1315
\bibitem[Sanders et al.(1988)]{sanders88} Sanders, D.~B., Soifer, B.~T., Elias, J.~H., et al.\ 1988, \apj, 325, 74
\bibitem[Sanders et al.(2003)]{sanders03} Sanders, D.~B., Mazzarella, J.~M., Kim, D.-C., Surace, J.~A., \& Soifer, B.~T.\ 2003, \aj, 126, 1607 
\bibitem[Schlafly \& Finkbeiner(2011)]{schlafly11} Schlafly, E.~F., \& Finkbeiner, D.~P.\ 2011, \apj, 737, 103 
\bibitem[Silva et al.(1998)]{silva98} Silva, L., Granato, G.~L., Bressan, A., \& Danese, L.\ 1998, \apj, 509, 103 
\bibitem[Smartt(2009)]{smartt09} Smartt, S.~J.\ 2009, \araa, 47, 63
\bibitem[Smartt et al.(2015)]{smartt15} Smartt, S.~J., Valenti, S., Fraser, M., et al.\ 2015, \aap, 579, A40 
\bibitem[Smith et al.(2009)]{smith09} Smith, N., Ganeshalingam, M., Chornock, R., et al.\ 2009, \apjl, 697, L49 
\bibitem[Smith et al.(2011)]{smith11} Smith, N., Li, W., Silverman, J.~M., Ganeshalingam, M., \& Filippenko, A.~V.\ 2011, \mnras, 415, 773 
\bibitem[Soifer et al.(2001)]{soifer01} Soifer, B.~T., Neugebauer, G., Matthews, K., et al.\ 2001, \aj, 122, 1213
\bibitem[Spiro et al.(2014)]{spiro14} Spiro, S., Pastorello, A., Pumo, M.~L., et al.\ 2014, \mnras, 439, 2873
\bibitem[Steele et al.(2004)]{steele04} Steele, I.~A., Smith, R.~J., Rees, P.~C., et al.\ 2004, \procspie, 679
\bibitem[Stephens(1974)]{stephens74} Stephens, M.~A.\ 1974, Journal of the American Statistical Association, 69, 347
\bibitem[Stritzinger et al.(2012)]{stritzinger12} Stritzinger, M., Taddia, F., Fransson, C., et al.\ 2012, \apj, 756, 173 
\bibitem[Stritzinger et al.(2018a)]{stritzinger18a} Stritzinger, M.~D., Anderson, J.~P., Contreras, C., et al.\ 2018a, \aap, 609, A134
\bibitem[Stritzinger et al.(2018b)]{stritzinger18b} Stritzinger, M.~D., Taddia, F., Burns, C.~R., et al.\ 2018b, \aap, 609, A135
\bibitem[Taddia et al.(2015)]{taddia15} Taddia, F., Sollerman, J., Leloudas, G., et al.\ 2015, \aap, 574, A60
\bibitem[Taddia et al.(2018)]{taddia18} Taddia, F., Stritzinger, M.~D., Bersten, M., et al.\ 2018, \aap, 609, A136
\bibitem[Tomasella et al.(2020)]{tomasella20} Tomasella, L., Benetti, S., Ochner, P., et al.\ 2020, The Astronomer's Telegram 13616, 1
\bibitem[Trancho et al.(2007)]{trancho07} Trancho, G., Bastian, N., Miller, B.~W., et al.\ 2007, \apj, 664, 284
\bibitem[Tully(1988)]{tully88} Tully, R.~B.\ 1988, Cambridge and New York, Cambridge University Press, 1988, 221 p.,  
\bibitem[Ulvestad(2009)]{ulvestad09} Ulvestad, J.~S.\ 2009, \aj, 138, 1529
\bibitem[Valenti et al.(2011)]{valenti11} Valenti, S., Fraser, M., Benetti, S., et al.\ 2011, \mnras, 416, 3138 
\bibitem[Van Dyk et al.(2011)]{vandyk11} Van Dyk, S.~D., Li, W., Cenko, S.~B., et al.\ 2011, \apjl, 741, L28
\bibitem[Van Dyk et al.(2014)]{vandyk14} Van Dyk, S.~D., Zheng, W., Fox, O.~D., et al.\ 2014, \aj, 147, 37
\bibitem[Varenius et al.(2019)]{varenius19} Varenius, E., Conway, J.~E., Batejat, F., et al.\ 2019, \aap, 623, A173
\bibitem[Wong et al.(2006)]{wong06} Wong, O.~I., Ryan-Weber, E.~V., Garcia-Appadoo, D.~A., et al.\ 2006, \mnras, 371, 1855
\bibitem[Woosley et al.(1994)]{woosley94} Woosley, S.~E., Eastman, R.~G., Weaver, T.~A., et al.\ 1994, \apj, 429, 300
\bibitem[Woosley(2019)]{woosley19} Woosley, S.~E.\ 2019, \apj, 878, 49
\bibitem[Yaron \& Gal-Yam(2012)]{yaron12} Yaron, O., \& Gal-Yam, A.\ 2012, \pasp, 124, 668
\bibitem[Yuan et al.(2016)]{yuan16} Yuan, F., Jerkstrand, A., Valenti, S., et al.\ 2016, \mnras, 461, 2003

\newpage 

\end{thebibliography}

\begin{appendix}

\section{Appendix}

\begin{figure}[!h]
\includegraphics[width=\linewidth]{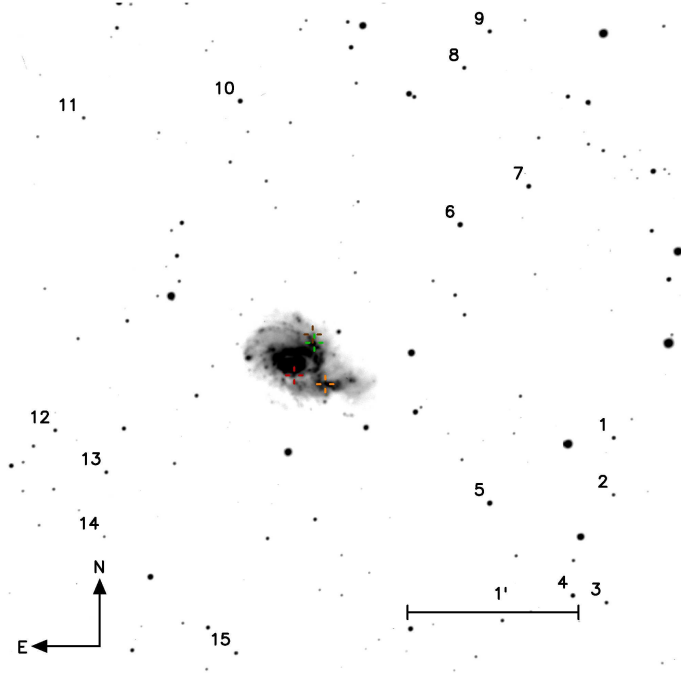}
\caption{4\arcmin\ $\times$ 4\arcmin\ ePESSTO NTT+EFOSC2 \textit{r}-band image of the field of NGC 3256 on 2018 January 16. The locations of sequence stars, SN 2018ec (green), AT 2018cux (red), PSN102750 (brown), and SN 2001db (orange) are indicated. North is up, east is left.}
\label{fig:field}
\end{figure}

\subsection{SN 2020cuj in NGC 1614}

SN 2020cuj in NGC 1614 was discovered by ZTF on 2020 February 12 03:06:02 UT with the internal name of ZTF20aaoffej \citep{de20}, and classified by \citet{dahiwale20} as a Type II SN. The SN was also detected by \textit{Gaia} (internal name Gaia20ayu). \textit{Gaia} derives coordinates RA = 04$^{\mathrm{h}}$34$^{\mathrm{m}}$00$\fs$53 and Dec = $-08\degr$34$\arcmin$43\farcs46 for the SN. We took the public photometry of SN 2020cuj obtained by these surveys, and also measured the template image subtracted magnitudes from a NTT+EFOSC2 (via ePESSTO+) image of $m_{V} = 19.28 \pm 0.06$ mag (on JD = 2458905.53), and 
a NOT+NOTCam image of $m_{K} = 15.74 \pm 0.04$ mag (on JD = 2458912.37). Based on a light curve fit with the Type IIP SN 2013ej \citep{yuan16} as a template we find that SN 2020cuj was discovered shortly after the explosion, has a host galaxy extinction of $A_{V}=2.8 \pm 0.2$ mag, and is $0.6^{+0.1}_{-0.2}$ mag brighter than SN 2013ej. The light curve fit is shown in Fig.~\ref{fig:20cuj_lc}. The late-time \textit{Gaia} detection could indicate a more linearly declining Type IIL SN, however, the \textit{Gaia} data points are not template subtracted and likely contain background contamination from the host galaxy, which can be significant at late phases. 

We note that public ZTF brokers list also another event ZTF19acyfumm in NGC 1614 at RA = 04$^{\mathrm{h}}$34$^{\mathrm{m}}$00$\fs$06 and Dec = $-08\degr$34$\arcmin$44\farcs47 discovered on 2019 November 25 07:59:00 UT. The coordinates are consistent with the host galaxy nucleus, but not with SN 2020cuj and the two events appear to be unrelated. The ePESSTO+ programme obtained an optical spectrum at the location on 2019 December 18 with the NTT and EFOSC2, however, the spectrum was reported to be red and dominated by the host galaxy without clear transient features \citep{muller-bravo19}. To our knowledge no other transient programme has carried out any observations of ZTF19acyfumm and due to the unclear nature of the event it is not part of our study.

\begin{figure}
\includegraphics[width=\linewidth]{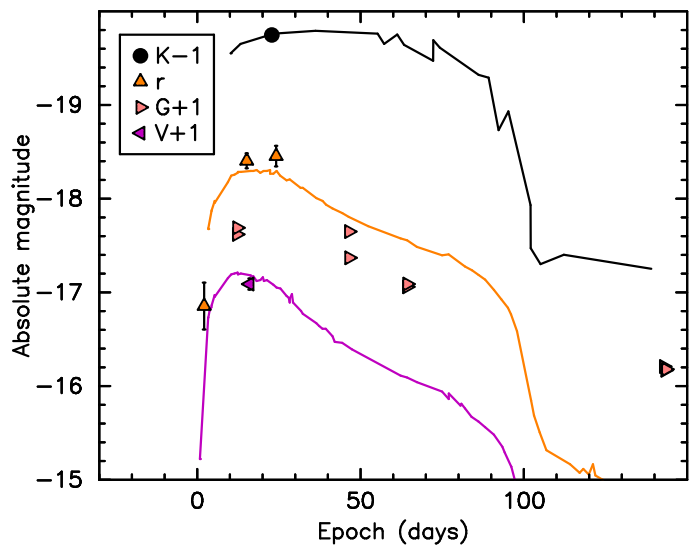}
\caption{Absolute magnitude light curves of AT 2020cuj (points) in NGC 1614 corrected for the host galaxy extinction of $A_{V} = 2.8$ mag, estimated with a fit to the light curves of the Type IIP SN 2013ej \citep[solid lines,][]{yuan16}. The epoch 0 is set to the estimated explosion date of SN 2013ej. Since the \textit{Gaia} light curve is not template subtracted, it is likely that it contains flux excess from the luminous host background, which has an effect in particular to the late-time data points.}
\label{fig:20cuj_lc}
\end{figure}

\begin{figure*}[h]
\centering
\includegraphics[width=0.35\linewidth]{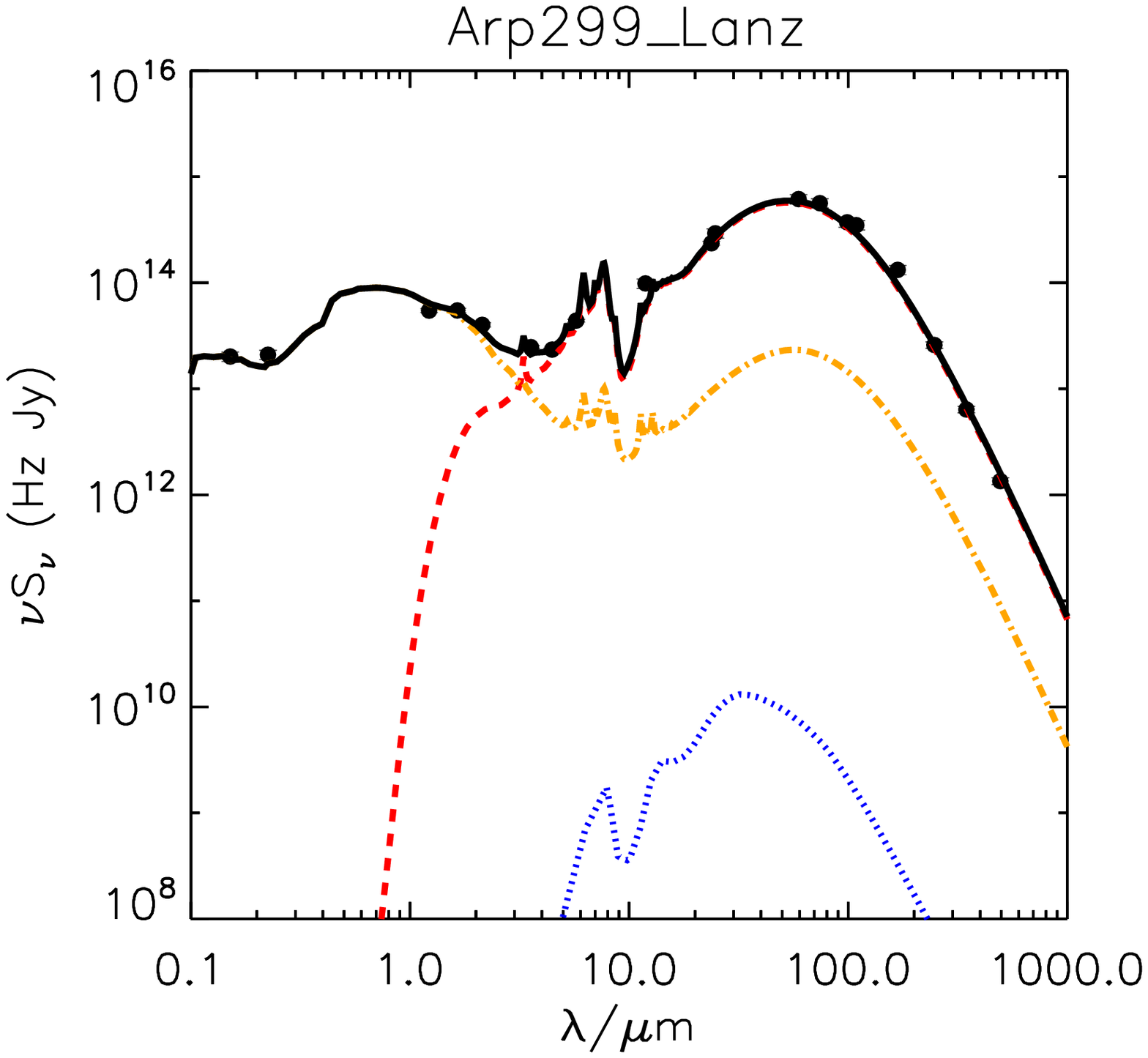}
\includegraphics[width=0.35\linewidth]{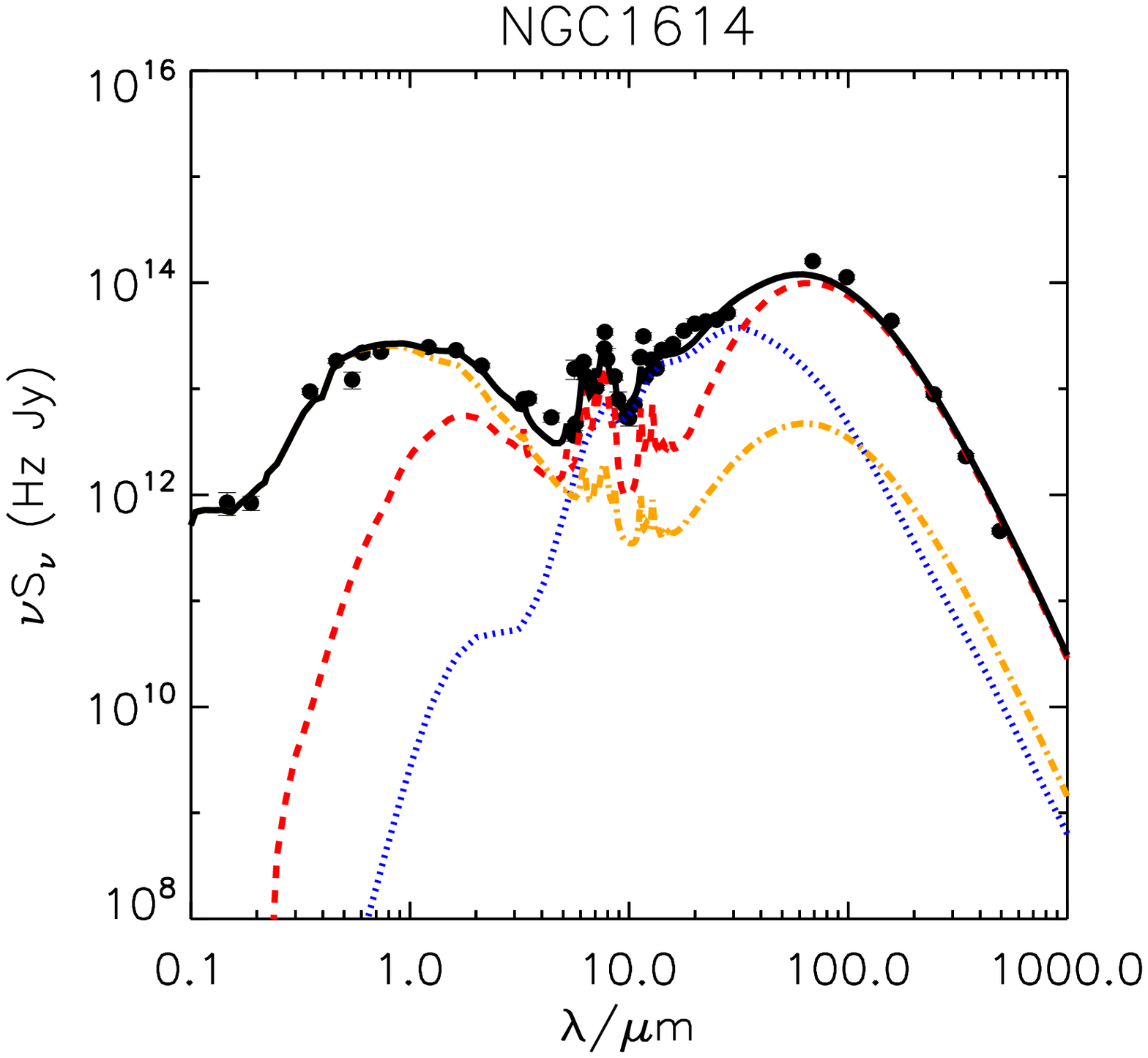}
\includegraphics[width=0.35\linewidth]{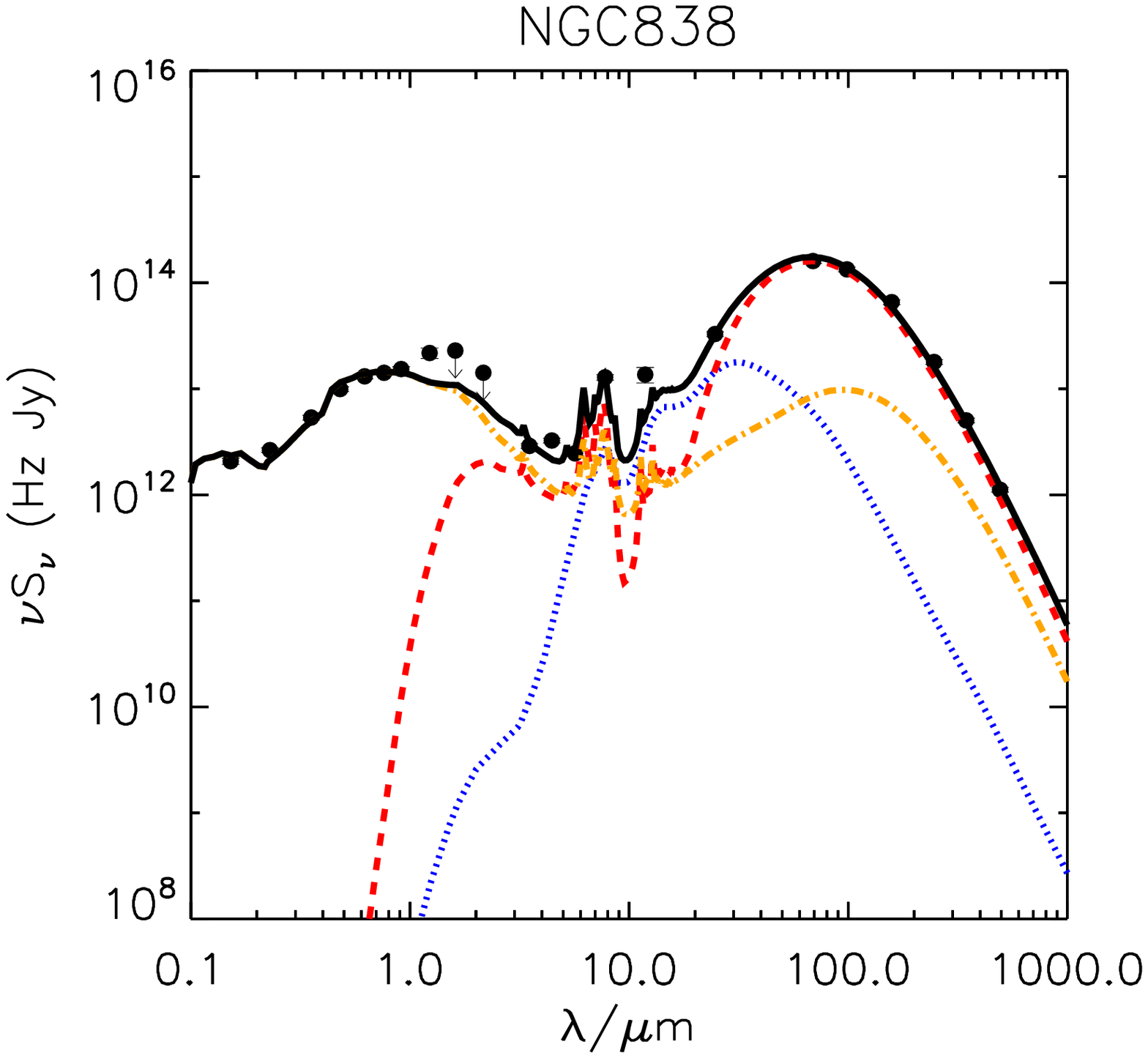}
\includegraphics[width=0.35\linewidth]{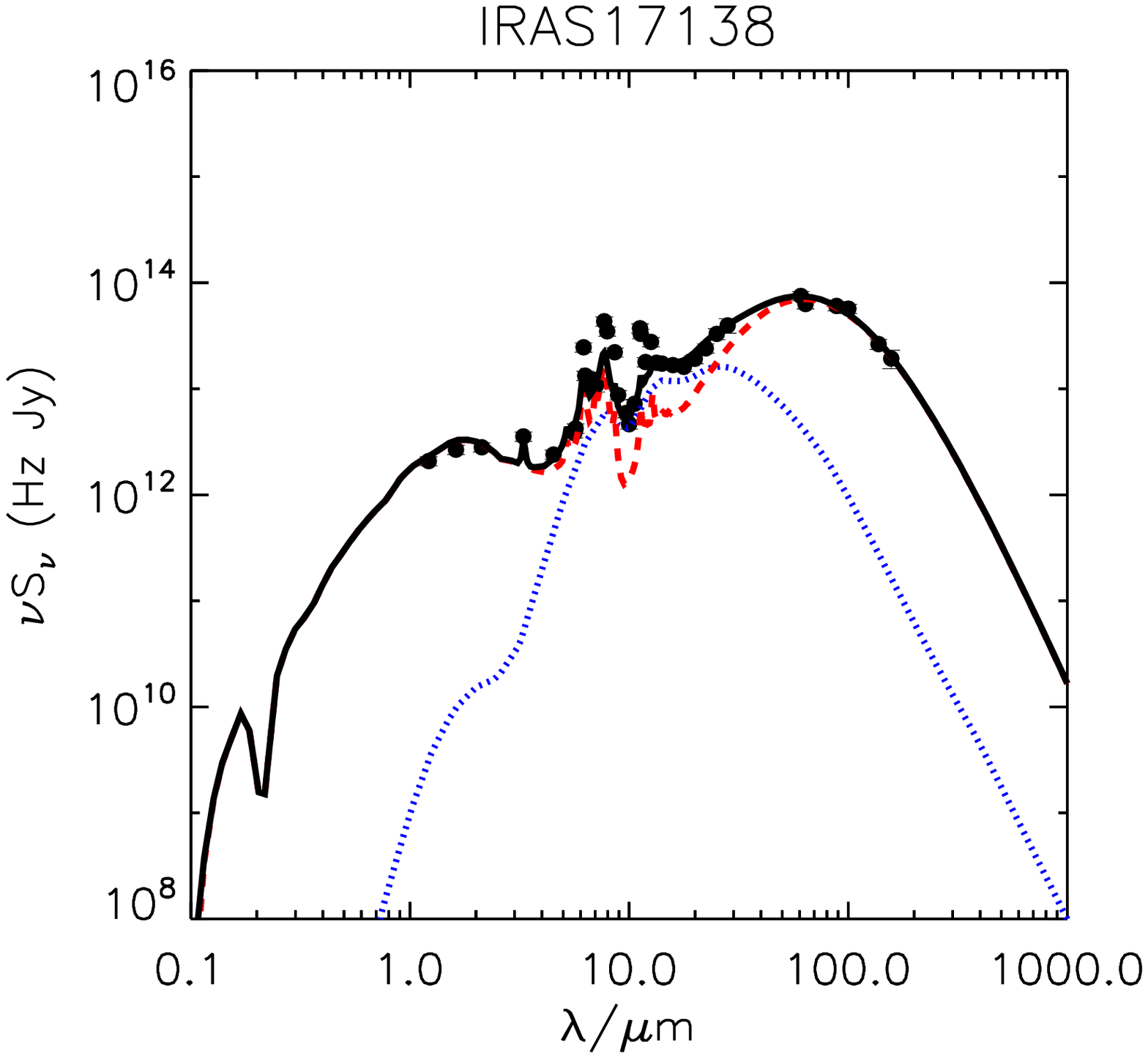}
\includegraphics[width=0.35\linewidth]{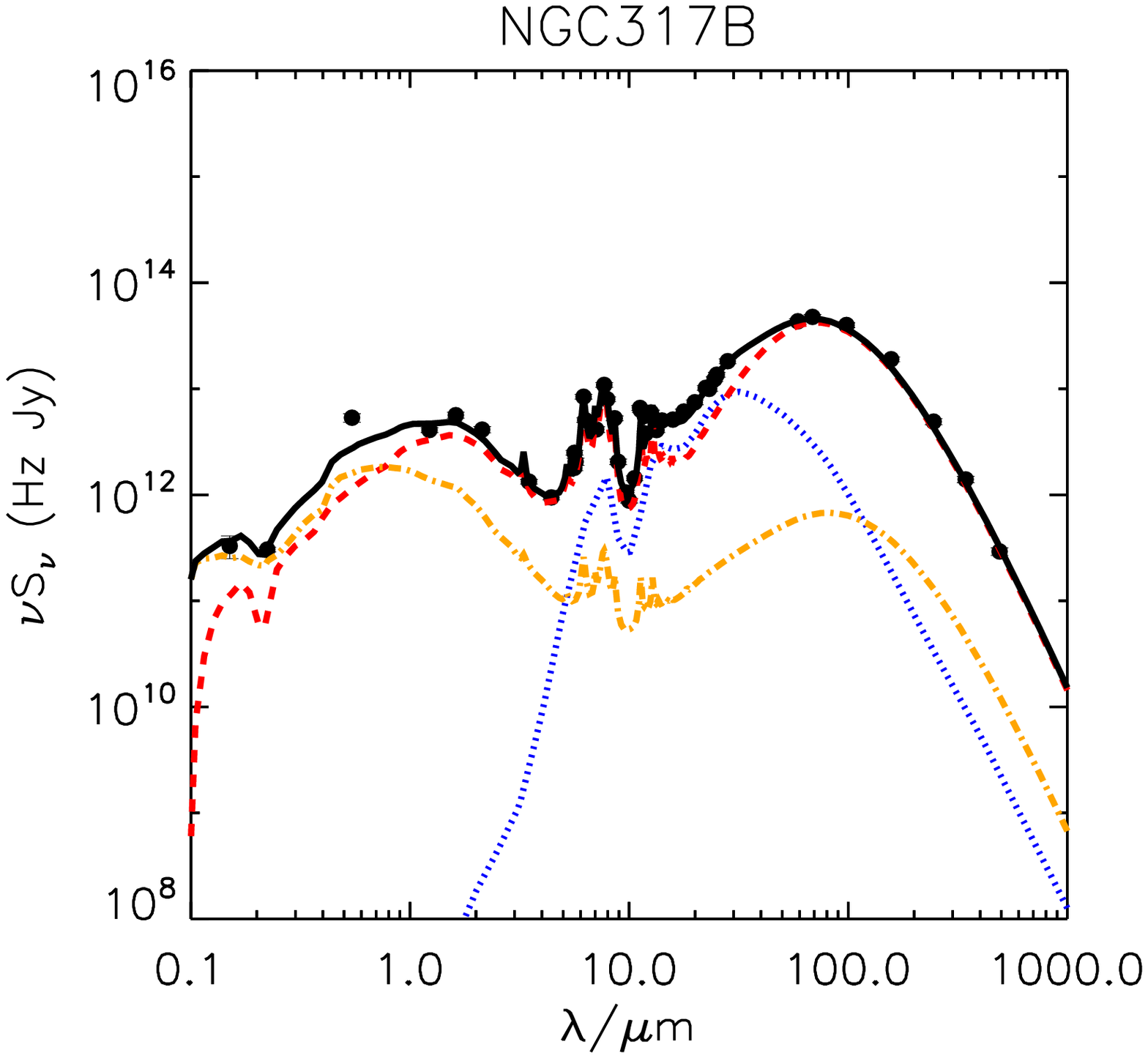}
\includegraphics[width=0.35\linewidth]{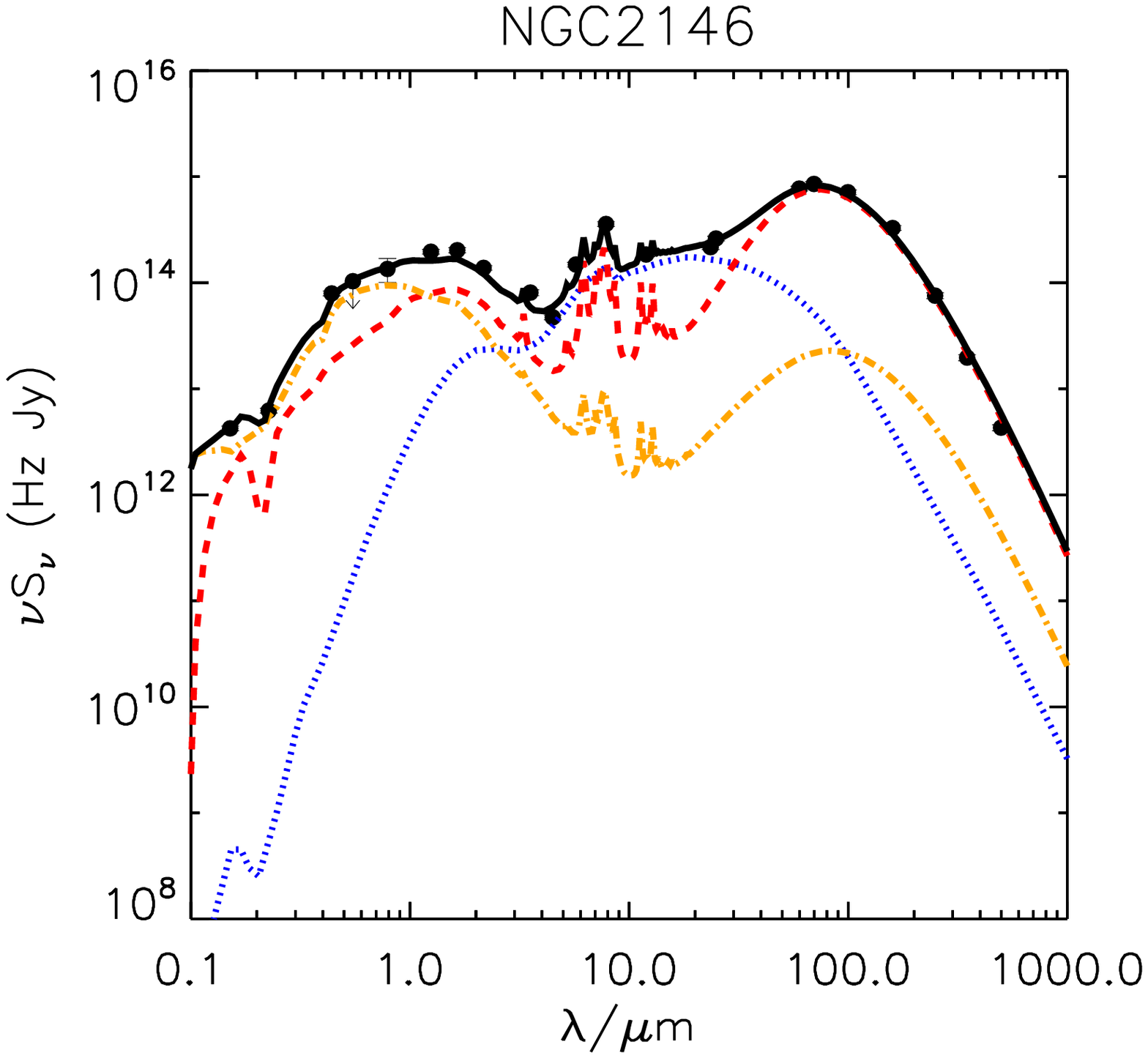}
\includegraphics[width=0.35\linewidth]{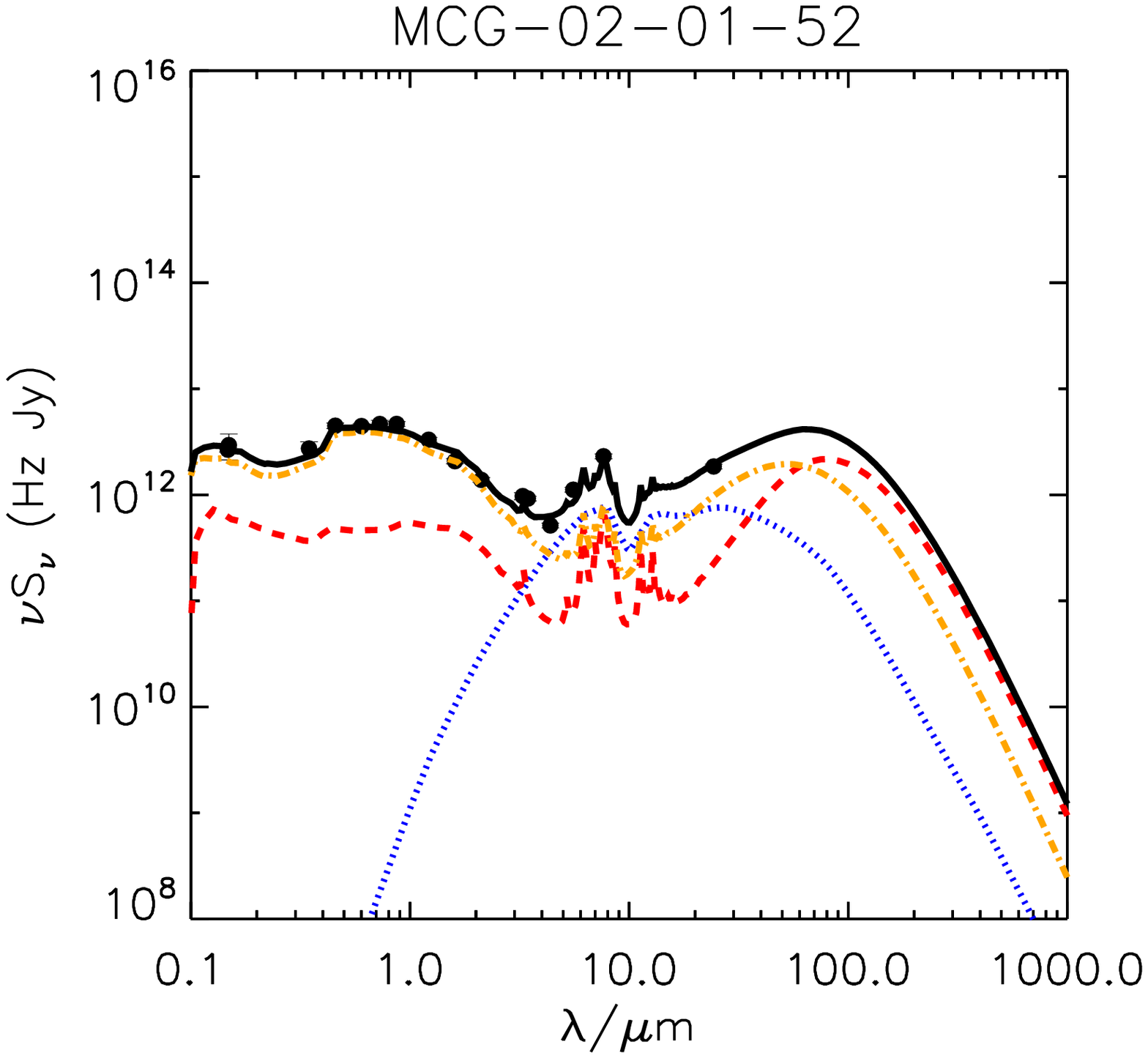}
\includegraphics[width=0.35\linewidth]{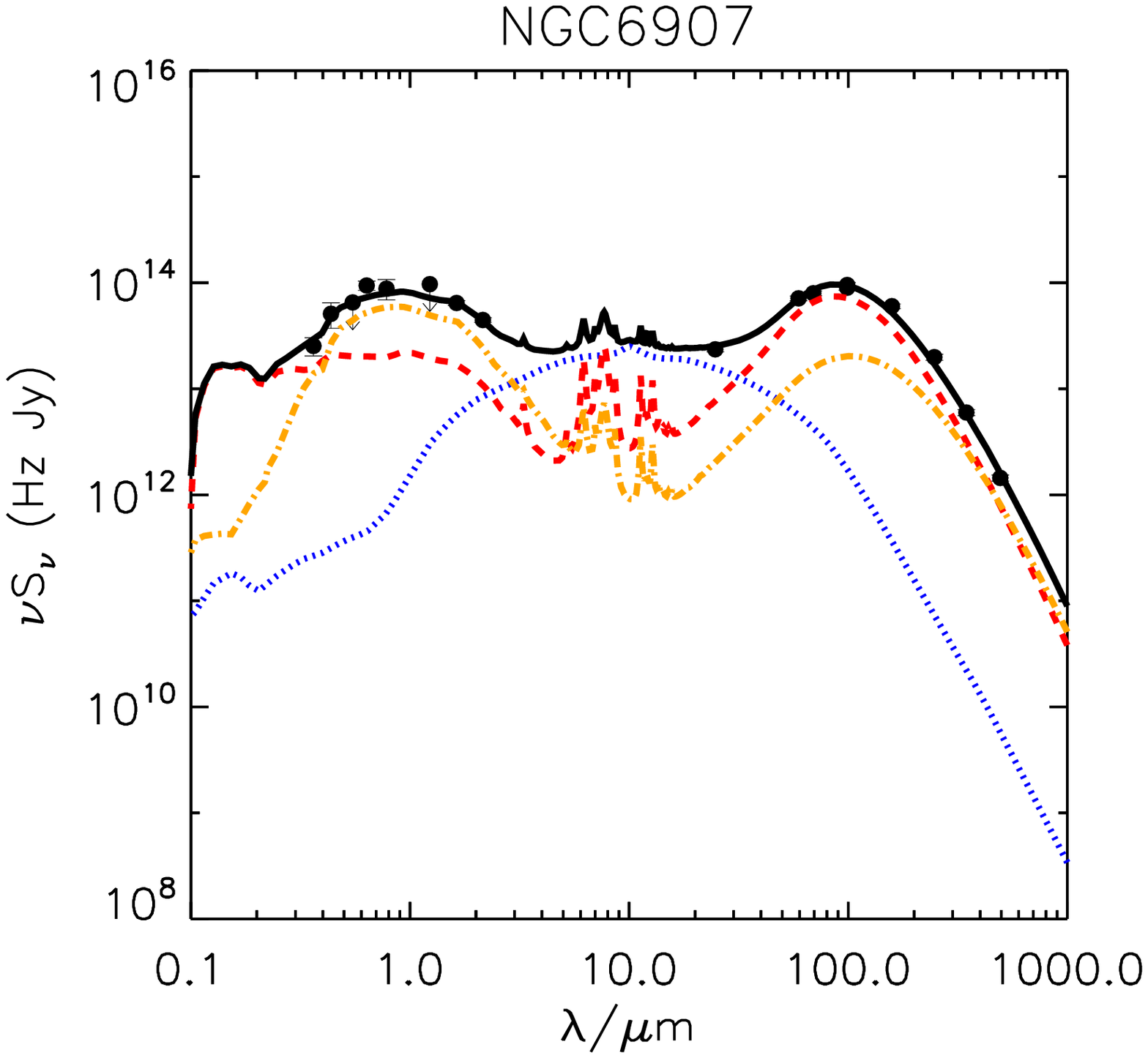}
\caption{Our SED models (solid black curves) for the observations (black points) of our initial sample of LIRGs for which a spherical host galaxy model is favoured. The model components include a spheroidal galaxy (dot-dashed orange), starburst contribution (dashed red), and an AGN (dotted blue).} 
\label{fig:sed}
\end{figure*}

\begin{figure*}[h]
\centering
\includegraphics[width=0.35\linewidth]{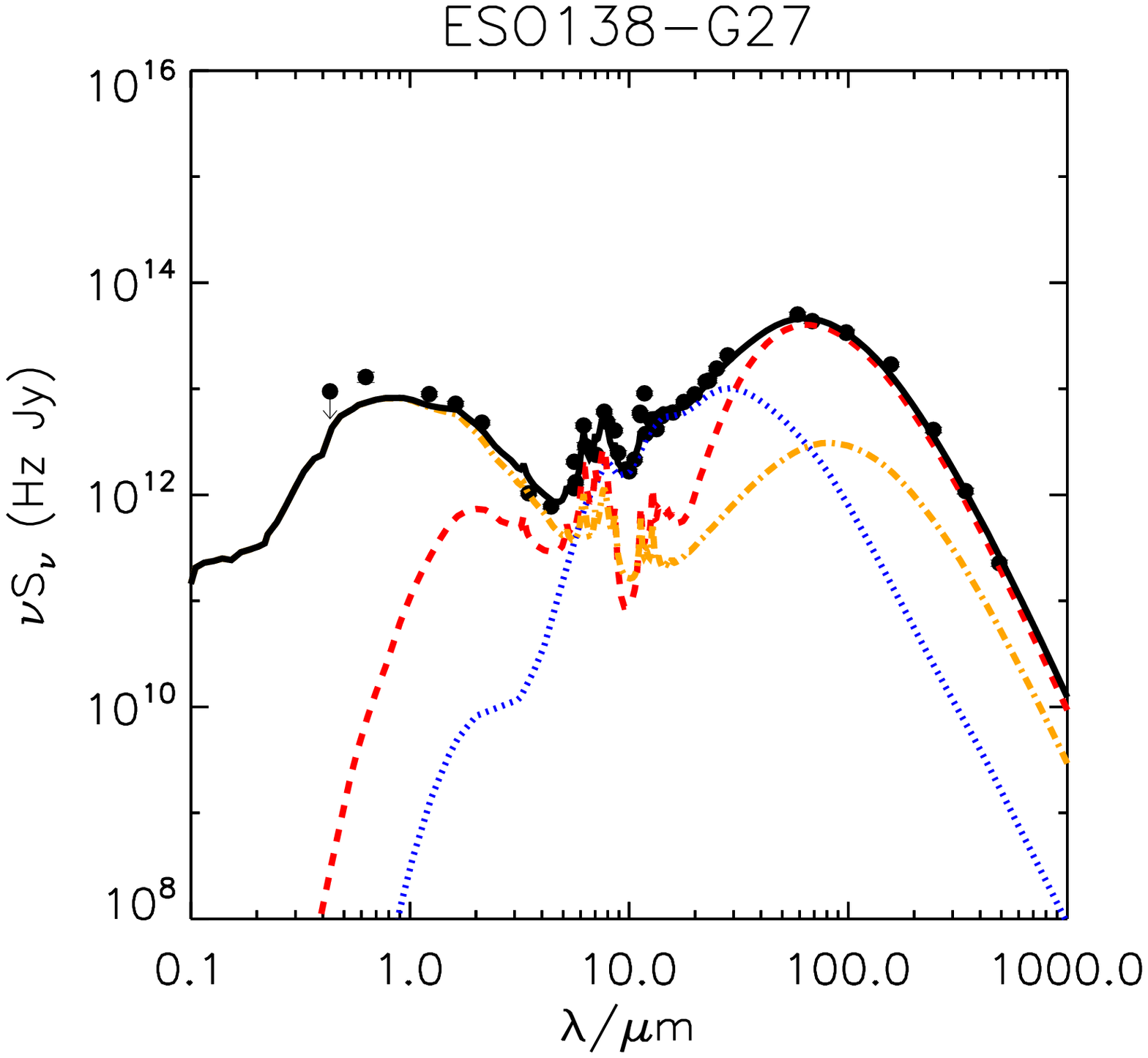}
\includegraphics[width=0.35\linewidth]{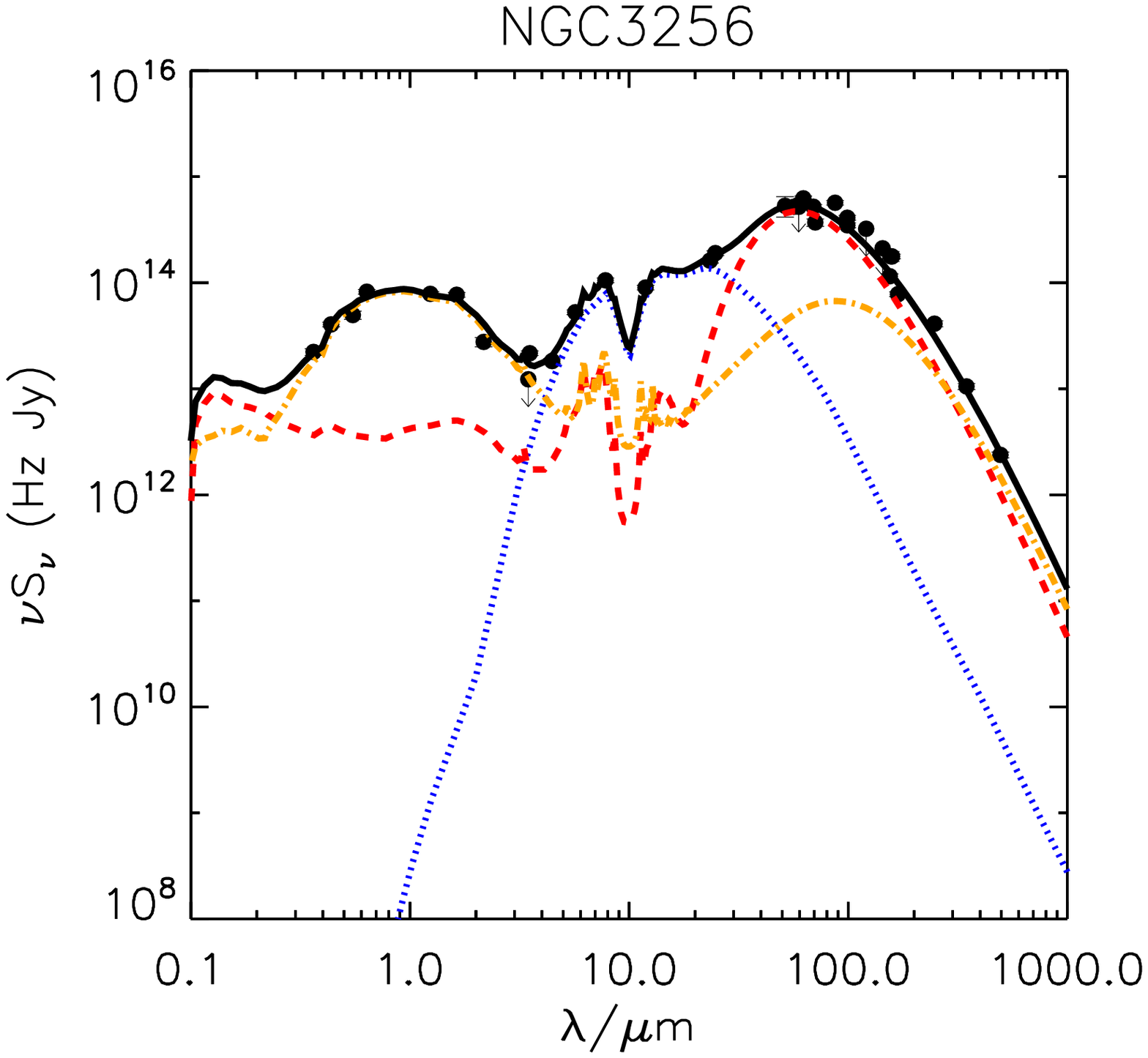}
\includegraphics[width=0.35\linewidth]{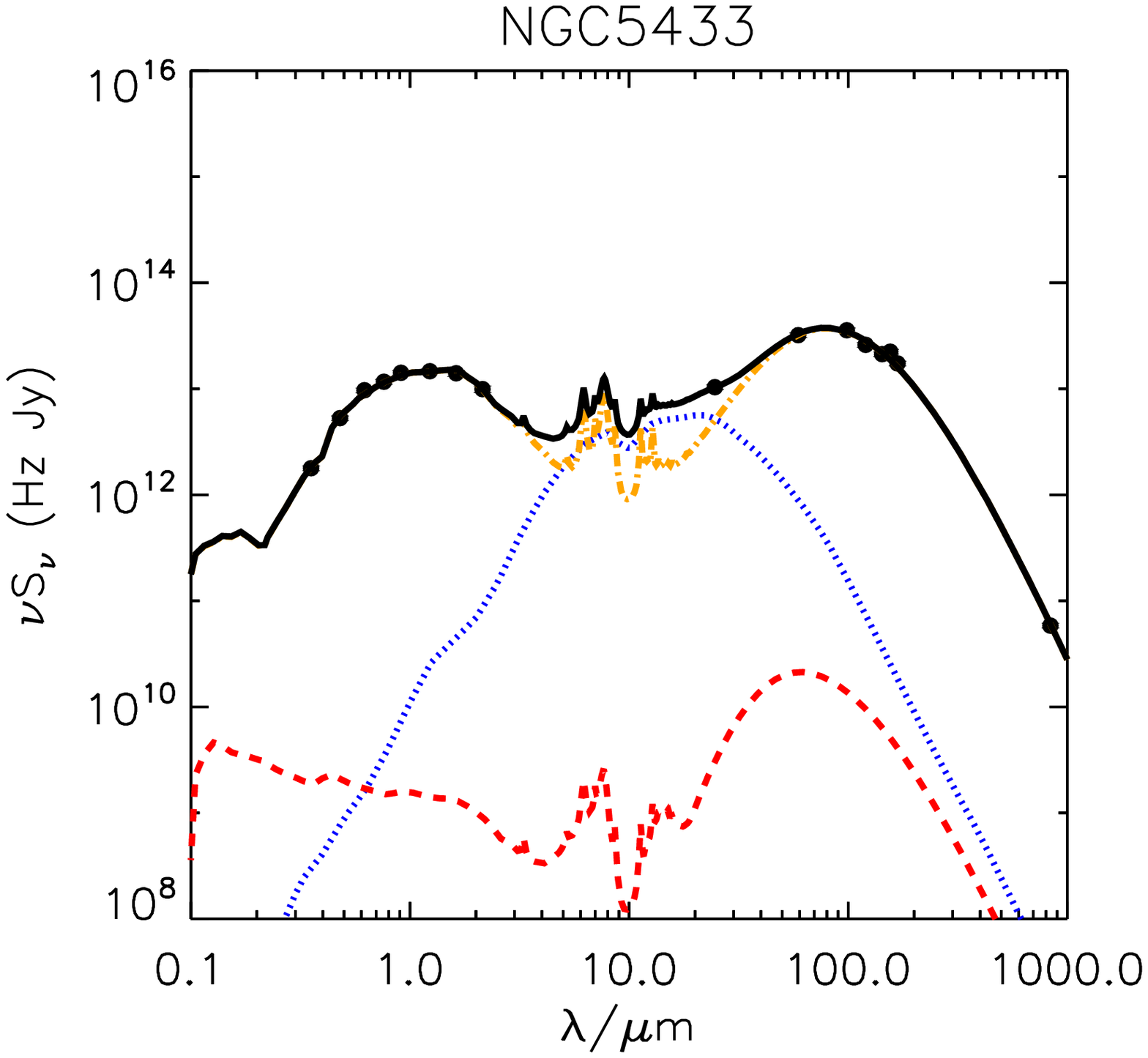}
\includegraphics[width=0.35\linewidth]{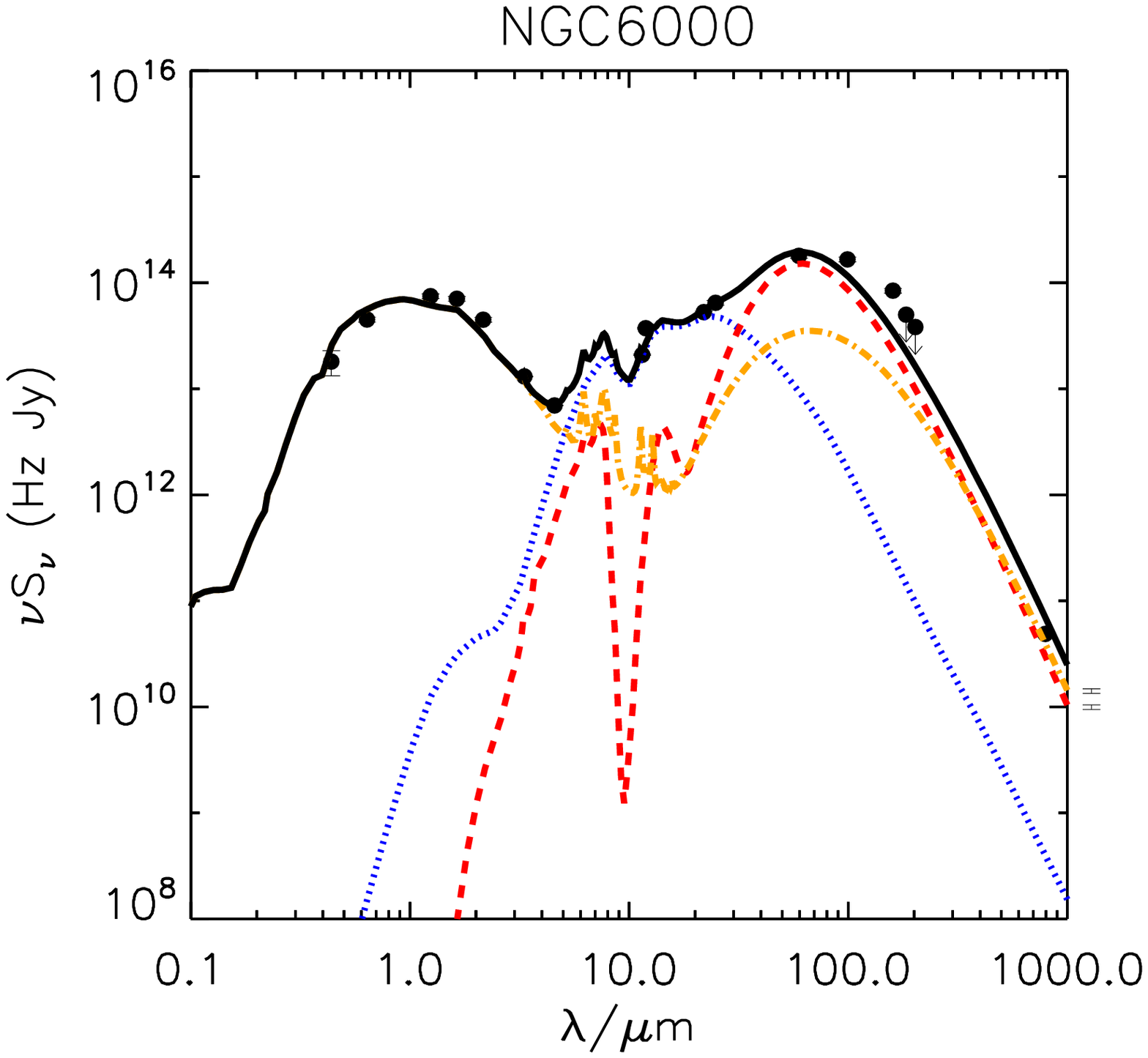}
{\\\textbf{Fig. A.3.} -- \textit{continued}}
\end{figure*}

\begin{figure*}[h]
\centering
\includegraphics[width=0.35\linewidth]{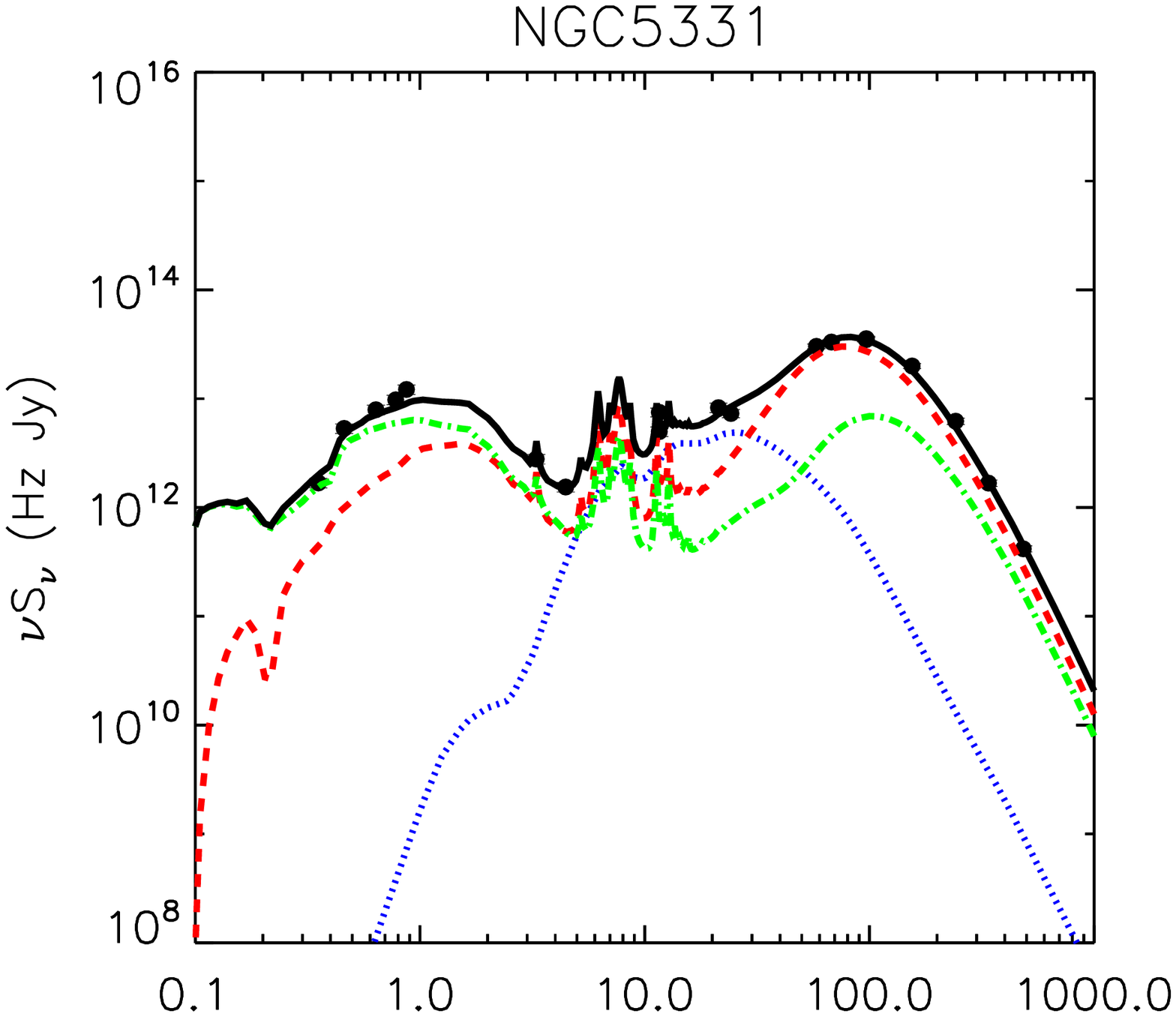}
\includegraphics[width=0.35\linewidth]{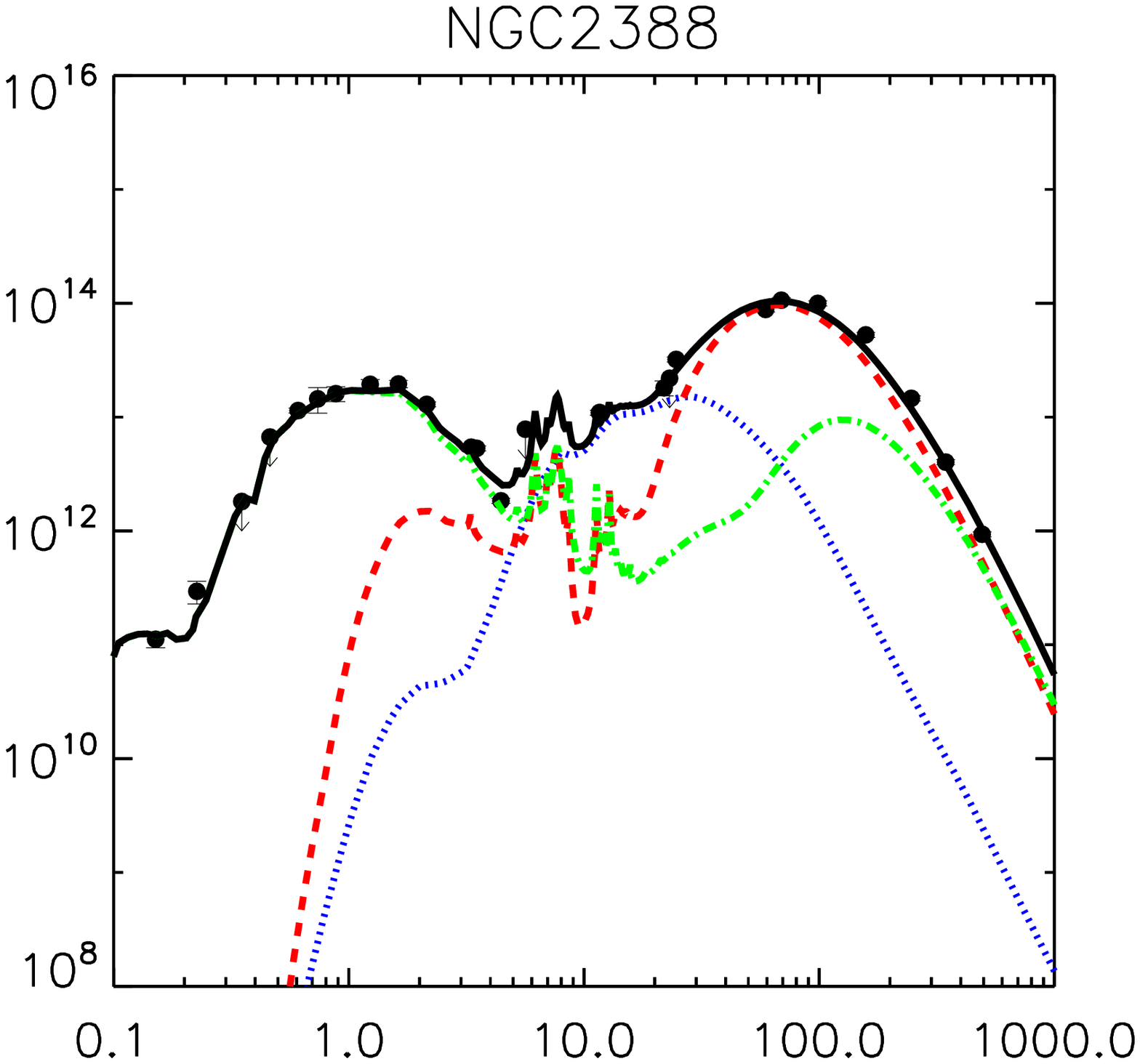}
\includegraphics[width=0.35\linewidth]{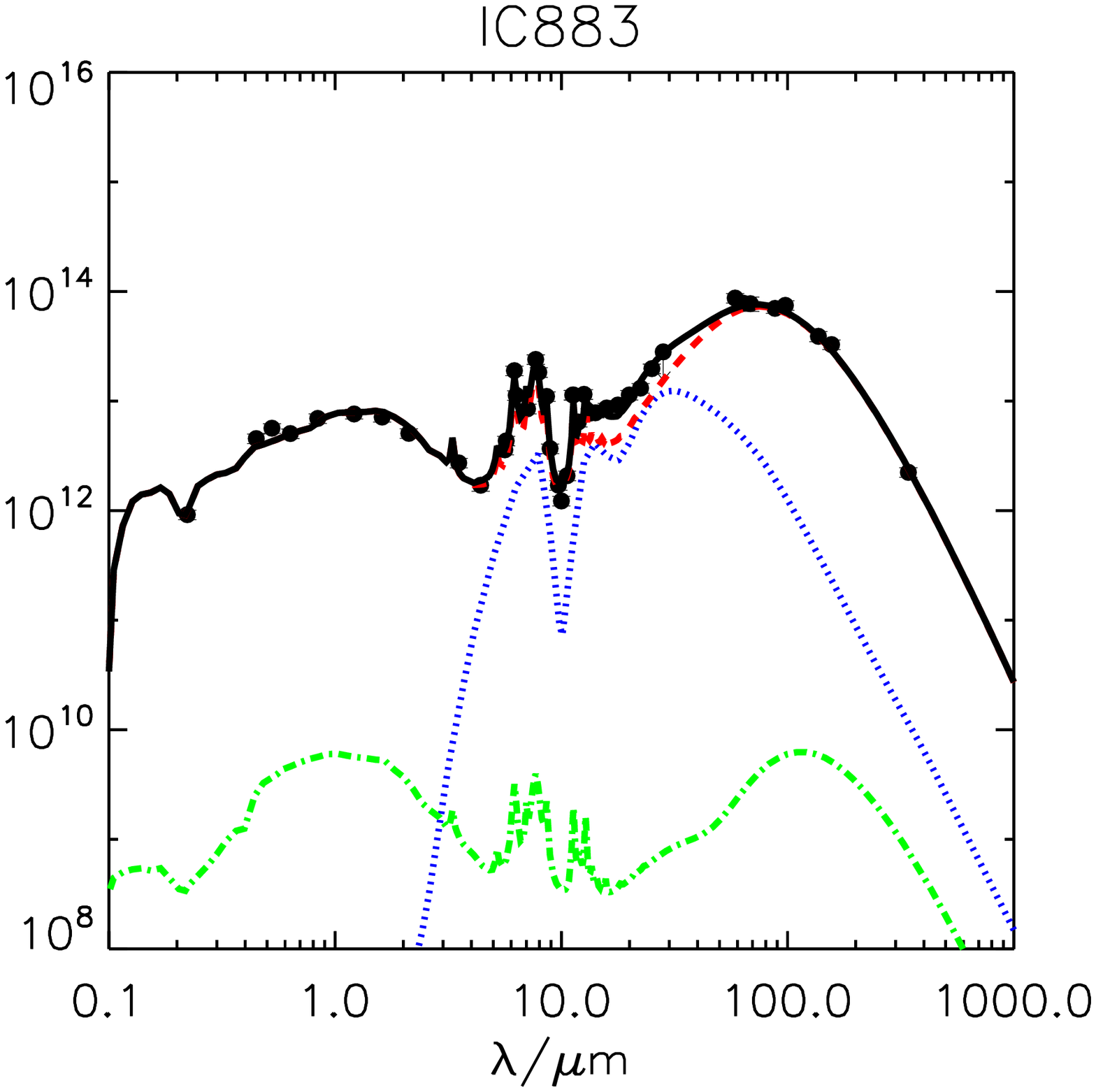}
\caption{Our SED models (solid black curves) for the observations (black points) of our initial sample of LIRGs for which a disc host galaxy model is favoured. The model components include a disc galaxy (dot-dashed green), starburst contribution (dashed red), and an AGN (dotted blue).}
\label{fig:sed2}
\end{figure*}

\begin{table*}[h]
\caption{Optical sequence star magnitudes in the field of SN 2018ec with the errors given in brackets.}
\centering
\begin{tabular}{cccccc}
\hline
\hline
\# & $m_{g}$ & $m_{r}$ & $m_{i}$ & $m_{z}$\\
 & (mag) & (mag) & (mag) & (mag) \\ 
\hline
1 & 18.538(0.014) & 17.920(0.018) & 17.703(0.019) & 17.603(0.042) \\ 
2 & 19.121(0.015) & 18.364(0.018) & 18.101(0.019) & 17.973(0.048) \\ 
3 & 18.741(0.014) & 18.115(0.017) & 17.862(0.019) & 17.786(0.059) \\ 
4 & 17.824(0.012) & 17.400(0.018) & 17.221(0.018) & 17.121(0.050) \\ 
5 & 17.557(0.014) & 16.782(0.017) & 16.482(0.018) & 16.232(0.023) \\ 
6 & 17.176(0.014) & 16.726(0.018) & 16.545(0.019) & 16.366(0.019) \\ 
7 & 18.022(0.015) & 17.170(0.018) & 16.840(0.019) & 16.594(0.022) \\ 
8 & 18.532(0.016) & 17.797(0.017) & 17.537(0.020) & 17.318(0.020) \\ 
9 & 18.180(0.014) & 17.644(0.017) & 17.429(0.019) & 17.253(0.022) \\ 
10 & 17.547(0.014) & 17.066(0.018) & 16.881(0.019) & 16.667(0.021) \\ 
11 & 18.892(0.016) & 18.376(0.018) & 18.163(0.020) & 18.026(0.030) \\ 
12 & 18.509(0.016) & 18.026(0.018) & 17.820(0.020) & 17.605(0.018) \\ 
13 & 19.572(0.021) & 18.125(0.018) & 16.861(0.020) & 16.284(0.018) \\ 
14 & 19.602(0.019) & 18.873(0.043) & 18.729(0.020) & 18.437(0.018) \\ 
15 & 18.521(0.014) & 18.114(0.018) & 17.943(0.020) & 17.671(0.022) \\ 
\hline
\end{tabular}
\label{table:field18ec}
\end{table*}

\begin{table*}[h]
\caption{Photometry for SN 2018ec with the errors given in brackets.}
\centering
\begin{tabular}{cccccccccc}
\hline
\hline
JD & $m_{g}$ & $m_{r}$ & $m_{i}$ & $m_{z}$ & $m_{J}$ & $m_{H}$ & $m_{K}$ & Telescope\\
(2400000+) & (mag) & (mag) & (mag) & (mag) & (mag) & (mag) & (mag) & \\ 
\hline
58121.86 & - & - & - & - & - & - & 15.13(0.05) & VLT \\
58124.86 & - & - & - & - & - & - & 15.22(0.04) & VLT \\
58130.76 & - & - & - & 17.15(0.03) & - & - & - & NTT \\
58134.77 & - & - & - & - & 16.35(0.04) & 15.79(0.03) & 15.72(0.05) & NTT \\
58135.71 & 20.25(0.03) & 18.79(0.06) & 18.02(0.01) & 17.40(0.03) & - & - & - & NTT \\
58154.82 & - & - & - & - & 17.04(0.09) & 16.30(0.03) & 16.37(0.06) & NTT \\
58162.83 & 20.74(0.07) & 19.16(0.07) & 18.48(0.02) & 17.85(0.03) & - & - & - & NTT \\
58164.86 & - & - & - & - & 17.38(0.10) & 16.51(0.04) & 16.62(0.06) & NTT \\
58169.80 & 20.89(0.06) & 19.31(0.10) & 18.70(0.03) & 18.04(0.03) & - & - & - & NTT \\
58186.61 & - & - & - & - & - & - & 17.08(0.16) & NTT \\
58186.79 & 20.97(0.17) & 19.64(0.15) & 19.02(0.03) & 18.32(0.08) & - & - & - & NTT \\
58201.70 & 21.21(0.07) & 19.67(0.12) & 19.20(0.03) & 18.62(0.04) & - & - & - & NTT \\
58204.61 & - & - & - & - & 18.64(0.20) & 17.51(0.06) & 17.44(0.10) & NTT \\
58215.72 & 21.38(0.23) & 20.01(0.17) & 19.41(0.05) & 18.98(0.07) & - & - & - & NTT \\
58216.53 & - & - & - & - & 18.71(0.41) & 17.52(0.10) & 17.76(0.11) & NTT \\
58228.57 & - & 20.29(0.20) & 19.72(0.05) & 19.12(0.09) & - & - & - & NTT \\
58230.58 & - & - & - & - & 18.84(0.29) & - & - & NTT \\
58243.55 & - & 20.11(0.20) & 19.94(0.04) & 19.37(0.06) & - & - & - & NTT \\
58250.65 & - & 20.72(0.45) & 20.09(0.04) & 19.67(0.06) & - & - & - & NTT \\
58251.58 & - & - & - & - & 19.04(0.25) & 18.47(0.15) & 18.42(0.30) & NTT \\
58260.46 & - & 20.15(0.27) & 20.17(0.05) & 19.85(0.08) & - & - & - & NTT \\
\hline
\end{tabular}
\label{table:phot18ec}
\end{table*}

\begin{table*}[h]
\caption{Spectroscopic log for SN 2018ec.}
\centering
\begin{tabular}{ccccccccc}
\hline
\hline
JD & Epoch & Grism & Slit & \textit{R} & PA & Exp. time & $\lambda$ & Telescope\\
(2400000+) & (d) & & (\arcsec) & ($\lambda/\Delta\lambda$) & ($\degr$) & (s) & (\AA) & \\ 
\hline
58130.8 & +28 & Gr13 & 1.0 & 355 & 121 & 1500 & 3645$-$9235 & NTT \\
58134.7 & +32 & GB & 1.0 & 550 & 0 & 12$\times$3$\times$90 & 9370$-$16460 & NTT \\
58135.7 & +33 & Gr13 & 1.0 & 355 & 83,91 & 2$\times$2700 & 3640$-$9230 & NTT \\
58154.7 & +52 & GB & 1.0 & 550 & 0 & 6$\times$3$\times$90 & 9370$-$16460 & NTT \\
58164.8 & +62 & GB & 1.0 & 550 & 0 & 12$\times$3$\times$90 & 9370$-$16460 & NTT \\
58165.8 & +63 & Gr13 & 1.0 & 355 & 228,249 & 2$\times$2700 & 3645$-$9235 & NTT \\ 
58187.8 & +85 & Gr13 & 1.0 & 355 & 236,252 & 2$\times$2400 & 3645$-$9235 & NTT \\ 
\hline
\end{tabular}
\label{table:spec18ec}
\end{table*}

\begin{table*}[h]
\caption{Photometry for AT 2018cux with the errors given in brackets.}
\centering
\begin{tabular}{cccccccccc}
\hline
\hline
JD & $m_{g}$ & $m_{r}$ & $m_{i}$ & $m_{z}$ & $m_{J}$ & $m_{H}$ & $m_{K}$ & Telescope\\
(2400000+) & (mag) & (mag) & (mag) & (mag) & (mag) & (mag) & (mag) & \\ 
\hline
58169.81 & $>$22.9 & $>$22.3 & $>$21.5 & $>$20.2 & - & - & - & NTT \\
58186.80 & $>$21.7 & $>$21.8 & $>$20.5 & $>$19.7 & - & - & - & NTT \\
58187.74 & - & - & - & $>$20.9 & - & - & - & NTT \\
58201.70 & 22.05(0.44) & 20.91(0.20) & 20.56(0.14) & 20.12(0.13) & - & - & - & NTT \\
58204.62 & - & - & - & - & 18.67(0.24) & 18.36(0.51) & 18.08(0.62) & NTT \\
58215.72 & 22.19(0.25) & 20.83(0.29) & 20.26(0.21) & 19.82(0.35) & - & - & - & NTT \\
58216.52 & - & - & - & - & 18.79(0.24) & 17.80(0.36) & $>$17.9 & NTT \\
58228.58 & - & 20.53(0.12) & 20.01(0.07) & 19.94(0.08) & - & - & - & NTT \\
58230.57 & - & - & - & - & 18.73(0.17) & - & - & NTT \\
58243.54 & - & 20.58(0.03) & 20.01(0.13) & 19.63(0.09) & - & - & - & NTT \\
58250.65 & - & 20.58(0.06) & 19.91(0.07) & 19.79(0.09) & - & - & - & NTT \\
58251.55 & - & - & - & - & 18.34(0.17) & 18.01(0.31) & 17.56(0.55) & NTT \\
58260.46 & - & 20.46(0.02) & 19.92(0.06) & 19.55(0.06) & - & - & - & NTT \\
\hline
\end{tabular}
\label{table:phot18cux}
\end{table*}

\begin{table*}[h]
\caption{Photometry for PSN102750 with the errors given in brackets.}
\centering
\begin{tabular}{ccccc}
\hline
\hline
JD & $m_{B}$ & $m_{R}$ & Telescope\\
(2400000+) & (mag) & (mag) & \\ 
\hline
56785.93 & - & 15.67(0.22) & T30 \\
56793.41 & - & 15.37(0.15) & C14 \\
56813.38 & - & 15.95(0.13) & C14 \\
56818.84 & 17.065(0.001) & - & HST \\
56975.16 & - & 19.116(0.004) & HST \\  
\hline
\end{tabular}
\label{table:photPSN}
\end{table*}

\begin{table*}[h]
\caption{Our photometry for SN 2019lqo with the errors given in brackets.}
\centering
\begin{tabular}{cccccccccc}
\hline
\hline
JD & $m_{U}$ & $m_{B}$ & $m_{V}$ & $m_{R}$ & $m_{I}$ & $m_{J}$ & $m_{H}$ & $m_{K}$ & Telescope\\
(2400000+) & (mag) & (mag) & (mag) & (mag) & (mag) & (mag) & (mag) & (mag) &  \\ 
\hline
58689.37 & - & - & - & 17.77(0.08) & - & - & - & - & NOT \\
58694.39 & 19.13(0.12) & 19.03(0.07) & 17.92(0.03) & 17.42(0.02) & 17.01(0.02) & - & - & - & NOT \\
58699.38 & 19.21(0.07) & 18.93(0.08) & 17.77(0.02) & 17.22(0.02) & 16.75(0.02) & - & - & - & NOT \\
58706.36 & 19.96(0.25) & 19.48(0.12) & 17.88(0.03) & 17.24(0.03) & 16.77(0.03) & - & - & - & NOT \\
58712.37 & - & 20.11(0.33) & 18.10(0.04) & 17.44(0.03) & 16.90(0.03) & - & - & - & NOT \\
58714.36 & - & - & - & - & - & 16.18(0.02) & 15.95(0.04) & - & NOT \\
58780.70 & - & - & - & - & - & 17.95(0.04) & 17.19(0.04) & 16.97(0.09) & NOT \\
58864.56 & - & - & 21.31(0.18) & 20.49(0.06) & 19.97(0.06) & - & - & - & NOT \\
\hline
\end{tabular}
\label{table:phot19lqo}
\end{table*}

\begin{table*}[h]
\caption{Photometry for SN 2019lqo based on public data including our measurements from the ZTF DR3 with the errors given in brackets.}
\centering
\begin{tabular}{ccccc}
\hline
\hline
JD & $m_{g}$ & $m_{G}$ & $m_{r}$ & Telescope\\
(2400000+) & (mag) & (mag) & (mag) &  \\ 
\hline
58679.68 & 20.27(0.15) & - & - & ZTF \\
58679.70 & - & - & 19.04(0.09) & ZTF \\
58680.73 & - & $>$21.5 & - & Gaia \\
58683.71 & - & - & 18.77(0.12) & ZTF \\
58685.82 & - & 18.33(0.20) & - & Gaia \\
58686.68 & 19.13(0.09) & - & - & ZTF \\
58766.01 & 20.13(0.15) & - & - & ZTF \\ 
58770.99 & - & - & 18.87(0.11) & ZTF \\ 
58777.01 & - & - & 19.05(0.12) & ZTF \\ 
58786.01 & - & - & 19.15(0.06) & ZTF \\ 
58791.01 & - & - & 19.20(0.11) & ZTF \\ 
58793.01 & - & - & 19.42(0.09) & ZTF \\ 
58794.02 & - & - & 19.42(0.07) & ZTF \\ 
58805.04 & - & - & 19.65(0.14) & ZTF \\ 
58807.05 & - & - & 19.53(0.10) & ZTF \\ 
58813.06 & - & - & 19.60(0.14) & ZTF \\ 
58828.92 & - & - & 19.87(0.20) & ZTF \\ 
58830.95 & - & - & 20.01(0.16) & ZTF \\ 
58833.93 & - & - & 19.91(0.23) & ZTF \\ 
58836.98 & - & - & 19.82(0.11) & ZTF \\ 
58846.97 & - & - & 20.36(0.30) & ZTF \\ 
\hline
\end{tabular}
\label{table:phot_19lqo}
\end{table*}

\begin{table*}[h]
\caption{Our optical photometry for SN 2020fkb with the errors given in brackets, and early public ZTF photometry listed for completeness.}
\centering
\begin{tabular}{ccccccccc}
\hline
\hline
JD & $m_{u}$ & $m_{B}$ & $m_{g}$ & $m_{V}$ & $m_{r}$ & $m_{i}$ & $m_{z}$ & Telescope\\
(2400000+) & (mag) & (mag) & (mag) & (mag) & (mag) & (mag) & (mag) &  \\ 
\hline
58936.66 & - & - & 17.83(0.08) & - & - & - & - & ZTF \\
58939.62 & - & - & - & - & 16.97(0.07) & - & - & ZTF \\
58939.83 & - & - & 17.33(0.07) & - & - & - & - & ZTF \\
58939.87 & - & - & - & - & 16.93(0.09) & - & - & ZTF \\
58942.33 & - & 17.53(0.11) & 17.17(0.04) & 17.00(0.07) & 16.85(0.09) & 17.04(0.05) & 17.00(0.11) & A1.82m \\
58943.30 & - & 17.38(0.04) & - & 16.80(0.06) & 16.72(0.05) & 16.86(0.03) & 16.99(0.07) & A1.82m \\
58944.44 & 17.98(0.11) & 17.51(0.24) & 16.88(0.04) & 16.75(0.06) & 16.56(0.04) & 16.75(0.03) & 16.79(0.07) & A1.82m \\
58945.31 & 18.61(0.57) & 17.44(0.11) & 16.88(0.04) & 16.87(0.06) & 16.63(0.03) & 16.78(0.04) & 16.85(0.04) & A1.82m \\
58947.32 & 18.26(0.41) & 17.41(0.09) & 16.94(0.09) & 16.69(0.04) & 16.45(0.08) & 16.36(0.09) & - & A67/92cm \\
58948.32 & 18.55(0.41) & 17.46(0.04) & 17.05(0.06) & 16.62(0.07) & 16.50(0.11) & 16.43(0.13) & - & A67/92cm \\
58949.33 & 18.54(0.49) & 17.48(0.07) & 17.02(0.08) & 16.74(0.06) & 16.37(0.09) & 16.36(0.18) & - & A67/92cm \\
58949.40 & 18.34(0.36) & 17.40(0.04) & 17.00(0.02) & 16.60(0.01) & 16.48(0.04) & 16.49(0.01) & 16.52(0.02) & LT \\
58950.37 & - & 17.63(0.06) & 17.10(0.10) & 16.74(0.08) & 16.41(0.08) & 16.47(0.13) & - & A67/92cm \\
58955.44 & - & 18.31(0.08) & 17.70(0.05) & 17.10(0.04) & 16.77(0.07) & 16.65(0.03) & 16.73(0.05) & A1.82m \\
58960.52 & - & 19.01(0.09) & 17.99(0.05) & 17.37(0.02) & 17.03(0.03) & 16.75(0.01) & 16.72(0.03) & NOT \\
58962.42 & - & 19.13(0.23) & 18.30(0.31) & 17.66(0.12) & 17.29(0.37) & 16.86(0.17) & - & A67/92cm \\
58969.42 & - & 19.42(0.08) & 18.58(0.07) & 17.93(0.03) & 17.63(0.06) & 17.23(0.02) & 17.02(0.04) & NOT \\
58976.42 & - & 19.57(0.05) & 18.82(0.08) & 18.13(0.02) & 18.01(0.07) & 17.56(0.04) & 17.17(0.05) & NOT \\
58989.45 & - & 20.37(0.16) & 18.95(0.12) & 18.57(0.04) & 18.24(0.07) & 17.82(0.03) & 17.33(0.07) & NOT \\
58999.38 & - & 20.03(0.10) & 19.05(0.12) & 18.74(0.25) & 18.48(0.08) & 18.06(0.03) & 17.38(0.06) & NOT \\
59011.40 & - & - & 19.03(0.14) & 18.86(0.18) & 18.64(0.14) & 18.27(0.09) & 17.59(0.08) & NOT \\
59021.42 & - & 20.59(0.21) & 19.32(0.19) & 18.94(0.04) & 18.76(0.07) & 18.45(0.05) & 17.73(0.08) & NOT \\
59033.43 & - & - & 19.55(0.20) & 19.16(0.07) & 19.17(0.21) & 18.79(0.07) & 17.92(0.11) & NOT \\
59049.38 & - & - & 19.53(0.21) & 19.38(0.11) & 19.38(0.19) & 19.15(0.16) & 18.18(0.10) & NOT \\
\hline
\end{tabular}
\label{table:phot20fkb}
\end{table*}

\begin{table*}[h]
\caption{Our near-IR photometry for SN 2020fkb with the errors given in brackets.}
\centering
\begin{tabular}{ccccc}
\hline
\hline
JD & $m_{J}$ & $m_{H}$ & $m_{K}$ & Telescope\\
(2400000+) & (mag) & (mag) & (mag) &  \\ 
\hline
58945.37 & - & 15.95(0.05) & - & LT \\
58964.45 & - & 15.69(0.08) & - & LT \\
58970.38 & - & 15.88(0.04) & - & LT \\
58974.36 & - & 15.99(0.04) & - & LT \\
58978.44 & 16.43(0.04) & - & 15.88(0.13) & NOT \\
58979.45 & - & 16.31(0.05) & - & LT \\
59000.39 & 17.09(0.06) & 16.68(0.08) & 16.60(0.06) & NOT \\
59025.47 & 17.95(0.16) & 17.55(0.14) & 17.34(0.35) & NOT \\
\hline
\end{tabular}
\label{table:phot_20fkb}
\end{table*}

\begin{table*}[h]
\caption{Spectroscopic log for SN 2020fkb.}
\centering
\begin{tabular}{ccccccccc}
\hline
\hline
JD & Epoch & Grism & Slit & \textit{R} & PA & Exp. time & $\lambda$ & Telescope\\
(2400000+) & (d) & & (\arcsec) & ($\lambda/\Delta\lambda$) & ($\degr$) & (s) & (\AA) & \\ 
\hline
58942.3 & $-7$ & Gr4 & 1.7 & 310 & 259 & 2700 & 3400$-$8200 & A1.82m \\
58943.3 & $-6$ & Gr4 & 1.7 & 310 & 265 & 1800 & 3400$-$8200 & A1.82m \\
58945.4 & $-4$ & Gr4 & 1.7 & 310 & 203 & 1800 & 3400$-$8200 & A1.82m \\
58956.4 & +7 & Gr4 & 1.7 & 310 & 126 & 3600 & 3400$-$8175 & A1.82m \\
58968.5 & +20 & Gr4 & 1.3 & 280 & 124 & 1800 & 3500$-$9300 & NOT \\ 
59021.4 & +72 & Gr4 & 1.3 & 280 & 122 & 2400 & 3400$-$9300 & NOT \\ 
59036.4 & +87 & Gr4 & 1.3 & 280 & 101 & 2400 & 3500$-$9300 & NOT \\ 
\hline
\end{tabular}
\label{table:spec20fkb}
\end{table*}

\end{appendix}

\end{document}